\documentclass[draftclsnofoot,onecolumn]{IEEEtran}

\usepackage[flushleft]{threeparttable} 
\usepackage{booktabs}       
\usepackage{hhline}
\usepackage{graphicx}
\usepackage{mathtools}
\usepackage{amsmath,amsthm,amsfonts,latexsym}    
\usepackage{stfloats}       
\usepackage{balance}
\usepackage{siunitx}
\usepackage{icomma}
\usepackage{xcolor}
\usepackage[english]{babel}
\usepackage{comment}
\usepackage{amssymb}
\usepackage[T1]{fontenc}
\usepackage{xspace}
\usepackage{tikz}
\usetikzlibrary{arrows.meta,positioning,fit,calc}

\usepackage{cite}
\DeclareUnicodeCharacter{2212}{-}

\usepackage[switch]{lineno}
\usepackage[named]{algo}
\usepackage{algpseudocode}
\usepackage{algorithm}
\algnewcommand{\LineComment}[1]{\State \(\triangleright\) #1}

\ifCLASSOPTIONcompsoc
  \usepackage[caption=false,font=normalsize,labelfont=sf,textfont=sf]{subfig}
\else
  \usepackage[caption=false,font=footnotesize]{subfig}
\fi

\newtheorem{theorem}{Theorem}
\newtheorem{lemma}{Lemma}
\newtheorem{proposition}[theorem]{Proposition}

\newtheorem*{remark}{Remark}

\def\Den{\mathrm{D}}
\mathchardef\mhyphen="2D

\newcommand{\atanTwo}[2]{\operatorname{atan2}\left(#1,\,#2\right)}

\newcommand{\SDR}{\ensuremath{\mathrm{SDR}}}

\newcommand{\SNR}{\ensuremath{\mathrm{SNR}}}
\newcommand{\SNRs}{\ensuremath{\mathrm{SNRs}}}

\newcommand{\dBval}[1]{\ensuremath{#1\,\mathrm{dB}}}

\graphicspath{{./plts/}{./figs/}}

\begin{document}

\title{Low-Latency Joint Source-Channel Coding with Multiple Constant Curvature Curves}

\author{Anders M. Buvarp,~\IEEEmembership{Senior Member,~IEEE} and Stefan Werner,~\IEEEmembership{Fellow,~IEEE},
        \thanks{This work was supported by the Research Council of Norway.}
        \thanks{Anders Buvarp and Stefan Werner are with the Department of Electronic Systems, Norwegian University of Science and Technology, 7491 Trondheim, Norway (e-mail: \{anders.buvarp,stefan.werner\}@ntnu.no). Stefan Werner is also with the Department of Information and Communications Engineering, Aalto University, 00076, Finland.}}

\maketitle

\begin{abstract}
We propose constant-curvature curve multicode (C3M), a multi-branch extension
of constant-curvature curve tube (C3T) analog joint source--channel coding for
one-shot transmission over fixed isotropic AWGN channels. Instead of mapping a
scalar source to a single curve on the power sphere, C3M partitions the source
interval across a union of constant-speed branch curves in even-dimensional
Euclidean space. This construction increases the available geometric degrees
of freedom while preserving a low-blocklength analog encoding structure. We
derive the induced mixture output distribution, posterior representation, MAP
and MMSE estimators, and mutual-information decomposition into
branch-identification and along-curve components. We then introduce geometry-
and classification-aware branch allocation rules based on tube density, curve
length, and branch-classification reliability. The resulting design procedure
fixes the optimal C3T curve as a starting branch and sequentially selects
additional branches, thereby reducing the combinatorial cost of exhaustive
multi-curve search. We use NORM-inspired branch-conditioned spectral decoders
based on Fourier-feature lifting of the received ambient vector. Numerical experiments in four and six dimensions show that adding candidate branch curves improves the finite-dimensional end-to-end signal-to-distortion
ratio, with selected two- and three-candidate C3M systems outperforming the
one-curve baseline at \dBval{+10} \SNR{}. The results show that multi-curve
analog mappings provide a practical finite-dimensional design route for
low-blocklength joint source--channel coding.
\end{abstract}


\section{Introduction}
\label{sec:intro}

Sixth-generation wireless technology is expected to combine communication,
sensing, distributed learning, and non-terrestrial connectivity in a single
networked infrastructure~\cite{Letaief2022,Letaief2019,Liu2020}. Integrated
sensing and communication, edge intelligence, radio-access-network slicing, and
large numbers of heterogeneous IoT devices will create many short-packet,
low-latency, and energy-constrained links~\cite{Pulkkinen2024,Ravi2022,
Gogineni2022}. In such settings, the main difficulty is not only to increase
data rate, but also to transmit analog-valued sensor information with small
delay, limited complexity, and robustness to interference, impulsive
disturbances, and model mismatch~\cite{Matthaiou2021,Clavier2021,Anders2025}.

Shannon's separation theorem states that, in the infinite-block-length limit,
a source can be compressed to a rate below the channel capacity and then
transmitted reliably using a channel code~\cite{Shannon1948}. This separated
architecture, commonly referred to as separate source--channel coding
(SSCC), is asymptotically optimal under the classical assumptions, but it can
require long source and channel blocks to approach the asymptotic limit
\cite{Kostina2012}. The resulting buffering and computational delay are poorly
matched to ultra-reliable low-latency communication and real-time sensing
applications. At finite block length and under implementation constraints,
joint source--channel coding (JSCC) provides an alternative in which the source
is mapped directly to channel inputs.

Analog JSCC, also known in this context as Shannon--Kotel'nikov (SK) mapping,
encodes continuous-alphabet source samples directly into channel vectors without
a separate quantization and channel-coding stage~\cite{Hekland2009}. For scalar
sources with bandwidth expansion by a factor of $n$, a common construction maps
one source sample to a point on a one-dimensional curve embedded in
$\mathbb R^n$. The curve geometry determines the tradeoff between local
resolution and noise robustness. Longer curves improve tangential resolution,
whereas excessive folding or small separation between curve segments increases
the risk of threshold effects. Curve-based analog JSCC mappings on spheres,
tori, and related manifolds therefore provide a natural finite-dimensional
design problem for one-shot and low-blocklength communication
~\cite{Vaishampayan1999,Vaishampayan2003,AndersC3T,AndersR4C}. Related
nonlinear analog JSCC schemes with MMSE decoding have also been studied as
practical finite-dimensional reference points~\cite{Hu2011}.

The present work is also adjacent to finite-blocklength JSCC, hybrid
digital--analog JSCC, and learned JSCC. Finite-blocklength JSCC characterizes
distortion under short-block constraints where asymptotic separation arguments
are not directly decisive~\cite{Kostina2012}. Learned JSCC instead optimizes
encoder--decoder pairs end to end for structured sources, especially images
\cite{Bourtsoulatze2019}. C3M differs from these lines by focusing on one-shot
scalar analog mappings whose signal sets are explicit unions of
constant-curvature curves, so that branch geometry, source allocation, and
branch identification can be studied directly.

The present paper builds on constant-curvature curve tube (C3T) mappings, where
a scalar source is transmitted through a single constant-speed curve embedded in
an even-dimensional Euclidean space~\cite{AndersC3T}. C3T provides a
geometrically structured one-shot analog code, but a single curve also limits
the available design degrees of freedom. This motivates a multi-branch
extension in which the source interval is partitioned across several
constant-curvature branch curves. The resulting construction can be viewed as a
discrete-index/continuous-position analog code: the branch index selects a
geometric component, while the position along the selected branch carries the
analog refinement.

Multi-curve analog mappings have appeared previously in related forms. Hekland
et al.\ studied a two-curve analog code based on a double Archimedes spiral in a
$2{:}1$ bandwidth-reducing JSCC scheme~\cite{Hekland2005}. In that setting, two
intertwined spiral curves form the codebook for mapping pairs of i.i.d.
Gaussian source samples to a single channel input. Campello, Torezzan, and
Costa proposed another multi-curve construction based on a set of $m$ curves
foliating the surface of an $n$-dimensional sphere~\cite{Campello2012}. These
works show that multi-curve mappings can increase geometric design flexibility,
but they also introduce inter-curve separation and branch-identification
constraints that are absent in the single-curve case.

These observations motivate the constant-curvature curve multicode (C3M)
developed in this paper. C3M partitions the source interval across multiple
constant-speed branch curves and uses a branch-conditioned decoder to estimate
the transmitted local coordinate. The resulting design problem combines
intra-branch resolution, inter-branch separation, source allocation across
branches, and receiver-side branch classification.

This paper derives the induced mixture output distribution, posterior
representation, MAP and MMSE estimators, mutual-information expressions, and
end-to-end and branch-known \SDR{} references for the C3M construction. It then
evaluates learned branch-conditioned decoders under nominal AWGN and
non-Gaussian robustness tests. The learned receiver includes a NORM-inspired
branch-conditioned spectral decoder that applies Fourier-feature lifting to the
received ambient vector and uses circular Huber training for local-coordinate
regression.

\begin{figure}[b!]
\begin{center}
  \includegraphics[width=.65\columnwidth]{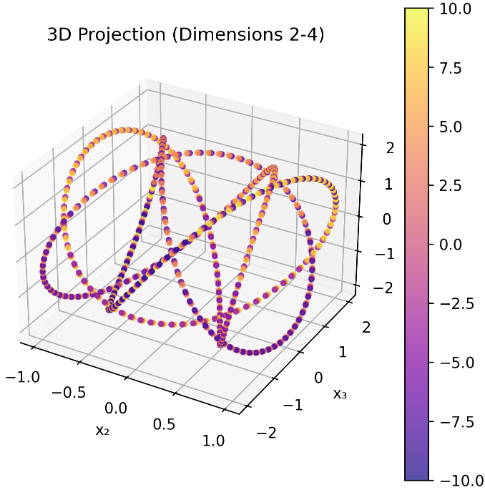}
    \caption{Three-dimensional projection of the C3T manifold for dimensions $2$, $3$, and $4$ with $n=4$. The colors denote the source parameter $\alpha$.}
  \label{fig:proj}
\end{center}
\vspace{-15pt}
\end{figure}

Prior work on analog curve codes suggests that performance is governed by the
geometry of the signal locus, including curve length, local stretch, and
minimum separation between distinct curve segments~\cite{Vaishampayan2003}.
In multi-curve analog mappings, these geometric constraints become central
because intra-curve reconstruction and inter-curve branch identification must
be handled simultaneously~\cite{Campello2012}.

Our previous work developed constant-curvature curve tube (C3T) mappings for
$1{:}n$ bandwidth-expansion analog JSCC using tube-packing and ropelength
ideas~\cite{AndersC3T,Gonzalez1999}. In C3T, a uniformly distributed scalar
source is mapped to a single constant-curvature curve in $\mathbb R^n$.
Fig.~\ref{fig:proj} shows an example of such a C3T trajectory projected into
three dimensions. The corresponding receivers used geometric feature extraction,
including averaged torus projections, together with learned MLP decoders. A
subsequent extension studied complex-valued and quaternion-valued neural
decoders for related curve-based analog mappings~\cite{AndersR4C}.

The present work extends this single-curve construction to a multi-branch
constant-curvature code. Related multi-curve analog mappings were studied by
Campello et al.~\cite{Campello2012}, who used multiple curves on spherical
geometries and length-based allocation principles for non-overlapping curves.
Here, the branch curves are derived from the C3T geometry, and the receiver is
explicitly branch conditioned. The source interval is therefore not allocated
solely according to geometric length. Instead, the allocation also accounts for
receiver-side branch-identification reliability. The relevant design question is
whether the receiver can first identify the correct branch and then accurately
estimate the local coordinate along that branch.

In this work, our main contributions are as follows:\\
\emph{1) Multi-branch C3M construction:} We extend C3T from a single
constant-curvature curve to a multi-branch analog JSCC construction, denoted
C3M. Each branch is a C3T-type constant-speed constant-curvature curve embedded
in an even-dimensional Euclidean space and associated with a flat-torus
parameterization~\cite{Costa1990}. The resulting encoder is a piecewise
constant-curvature map in which each source symbol activates one branch curve
and one local coordinate on that branch.\\
\emph{2) Analytical model and \SDR{} references:} We derive the induced mixture
output distribution, posterior representation, MAP and MMSE estimators, and
mutual-information expressions for the C3M construction. We also distinguish
between the ordinary end-to-end Gaussian \SDR{} reference and the genie-aided
branch-known Gaussian \SDR{} reference. This distinction is essential because
the practical receiver must infer the branch index from the noisy channel
output.\\
\emph{3) Interval-allocation rules and diagnostics:} We formulate
equal-splitting, length-only, tube-density-only, and density--reliability
allocation rules for partitioning the $\mathcal{U}[0,1)$ source interval
across branch curves. The density--reliability rule accounts for both the local
geometry of each branch and the probability that the receiver selects the
correct branch before performing along-curve decoding. The allocation ablation
separates the effect of source-interval allocation from the effect of
branch-geometry selection.\\
\emph{4) NORM-inspired branch-conditioned spectral receiver:} We propose a
hierarchical receiver that first estimates the branch index and then applies a
branch-conditioned decoder. Motivated by neural operators on Riemannian
manifolds (NORM)~\cite{Chen2024}, the implemented decoder applies
Fourier-feature lifting to the received ambient vector and then uses residual
and dense layers to estimate the local branch coordinate. A separate decoder is
trained for each branch. A circular Huber loss is used for periodicity-aware
local-coordinate regression under Gaussian-mixture impulsive noise, $S\alpha S$
perturbations, and branch-confusion events. \\
\emph{5) Numerical evaluation and design tradeoff:} We evaluate C3M in four and
six dimensions under fixed isotropic AWGN and compare the learned decoders with
the appropriate \SDR{} references. The results show that additional branches
can improve finite-dimensional end-to-end \SDR{}, but they also introduce
branch-confusion errors. The resulting tradeoff links curve geometry, source
allocation, branch priors, and classifier reliability.

The remainder of this paper is organized as follows.
Section~\ref{sec:bg} reviews the background material underlying the proposed
C3M construction, including C3T, flat tori, and geometry-based learning methods
such as neural operators. Section~\ref{sec:sys} introduces the system model,
including the source, encoder, channel, noise, and receiver models.
Section~\ref{sec:multi} presents the proposed multi-curve design and develops
equal-splitting, length-only, tube-density-only, and density--reliability
allocation rules.
Section~\ref{sec:norm} presents the NORM-inspired branch-conditioned spectral
decoder, including the Fourier-feature lifting, decoder architecture, and
circular Huber-loss training objective. Section~\ref{sec:analysis} develops the analytical model,
posterior estimators, mutual-information expressions, and \SDR{} references.
Section~\ref{sec:numexp} presents the numerical results, and
Section~\ref{sec:conclusion} concludes the paper.

\noindent \emph{Notation:}
$\mathcal{O}(\cdot)$ denotes Bachmann--Landau big-$O$ notation, and
${}^{\circ}$ denotes degrees. The set of real numbers is denoted by $\mathbb R$. The unit sphere in
$\mathbb R^n$ is denoted by
$\mathbb S^{n-1}=\{\mathbf x\in\mathbb R^n:\|\mathbf x\|_2=1\}$. The operators
$\mathbb{E}[\cdot]$ and $\Pr(\cdot)$ denote expectation and probability, and
$\times$ denotes the Cartesian product. The Euclidean norm is denoted by
$\lVert\cdot\rVert$, and $\mathbf{I}_n$ denotes the $n\times n$ identity
matrix. The notation $x\sim\mathcal{U}[a,b)$ denotes a continuous uniform random variable on $[a,b)$, and
$\mathbf{x}\sim\mathcal{N}(\boldsymbol{\mu},\mathbf{\Sigma})$ denotes a
Gaussian random vector with mean $\boldsymbol{\mu}$ and covariance matrix
$\mathbf{\Sigma}$. A symmetric alpha-stable random variable is denoted by
$x\sim\mathrm{S\alpha S}(\alpha,c,\mu)$, where $\alpha$ is the stability
parameter, $c$ is the scale, and $\mu$ is the location.
Uppercase letters denote random variables, lowercase letters denote their
realizations, and hatted variables denote estimates. For example, $S$, $K$, and
$Z$ denote random variables, $s$, $k$, and $z$ denote realizations, and
$\widehat{S}$, $\widehat{K}$, and $\widehat{Z}$ denote estimates. The
transmitted and received vectors are denoted by $\mathbf{X}$ and $\mathbf{Y}$,
respectively. The integer $m$ denotes the number of branch curves in the C3M
construction. The interval endpoints are denoted by
$a_0,a_1,\ldots,a_m$, the corresponding subintervals by $I_k$, and their
lengths by $p_k$. The random variable $K$ denotes the transmitted branch index.
The $k$th branch curve is denoted by $\gamma_k(\cdot)$, where
$\gamma_k:[0,1)\to\mathcal{M}_k\subset\mathcal{T}_k\subset\mathbb{R}^n$.
Here, $\mathcal{M}_k$ denotes the associated one-dimensional branch manifold,
$\mathcal{T}_k$ the associated flat torus, and
$\theta_k=(\mathbf{r}_k,\boldsymbol{\omega}_k)$ the corresponding parameter
vector, with $\mathbf{r}_k$ and $\boldsymbol{\omega}_k$ denoting the radii and
frequency vectors, respectively. The signal set is
$\mathcal{X}\triangleq\bigcup_{k=1}^m\gamma_k([0,1))$. The length and
protection radius of the $k$th curve are denoted by $L_k$ and $\rho_k$, and
the minimum separation between curves $k$ and $\ell$ is denoted by
$d_{k\ell}$.

\section{Background}
\label{sec:bg}

Section~\ref{ss:c3t} reviews constant-curvature curves and geodesics on flat
tori, which form the geometric basis for the C3T and C3M constructions.
Section~\ref{ss:principles} discusses geometric principles for multi-curve code
design, including curve length, separation, packing, and Grassmannian-frame
viewpoints. Section~\ref{ss:nop} reviews feature extraction on non-Euclidean
data structures and motivates the NORM-inspired branch-conditioned spectral
decoder used later in the paper.

\subsection{Constant-curvature curves and geodesics on flat tori}
\label{ss:c3t}

A manifold is a space that is locally homeomorphic to Euclidean space. In
particular, a one-dimensional manifold $\mathcal{M}$ may be represented by a
parametric curve embedded in $\mathbb{R}^n$. Thus, for a suitable fundamental
parameter interval $I\subset\mathbb{R}$, we view a branch curve as a bijective
map $\gamma:I\to\mathcal{M}\subset\mathbb{R}^n$, where $\gamma$ is a smooth
parametric curve and $\mathcal{M}$ is the associated embedded one-dimensional
manifold.

In our setting, the relevant one-dimensional manifolds are the images of
parametric constant-curvature curves embedded in $\mathbb R^n$. The C3T and
C3M curves considered in this paper are constructed as geodesics on flat tori.
In even dimensions, these constant-curvature curves have constant generalized
curvatures and are geodesics on flat tori $T^{n/2}$ embedded in
$\mathbb S^{n-1}\subset\mathbb R^n$. These curves are represented by the
vector~\cite{Costa1990}
\begin{align}
\mathbf{x}(\alpha)&=[r_1\cos(\omega_1 \alpha), r_1\sin(\omega_1 \alpha),  
r_2 \cos(\omega_2 \alpha), r_2 \sin(\omega_2 \alpha), 
\ldots, r_{n/2} \cos(\omega_{n/2} \alpha), r_{n/2} \sin(\omega_{n/2} \alpha)]^T,
\label{equ:parametric_curve}
\end{align}
where $r_1,r_2,\ldots,r_{n/2}$ are the radii and
$\omega_1,\omega_2,\ldots,\omega_{n/2}$ are the frequency parameters. These
harmonic coordinates satisfy a number of useful structural properties,
summarized in the following lemma.

\begin{lemma}
\label{lem:const_speed}
Consider the even-dimensional parametric curve in
Eq.~\eqref{equ:parametric_curve}, with $\alpha\in[0,2\pi)$, radii $r_j\ge 0$,
and frequencies $\omega_j\in\mathbb{R}$ for $j=1,\ldots,n/2$. Then:
\begin{enumerate}
    \item The squared norm is constant and satisfies
    $\|\mathbf{x}(\alpha)\|^2=\sum_{j=1}^{n/2} r_j^2$.
    Hence, under the power constraint $\sum_{j=1}^{n/2} r_j^2=P$, the curve
    lies on the sphere $\mathbb{S}^{n-1}(\sqrt{P})$.
    
    \item The squared speed is constant and satisfies
    $\|\mathbf{x}'(\alpha)\|^2=\sum_{j=1}^{n/2}\omega_j^2 r_j^2$.
    Therefore the natural parametrization is constant-speed,
    \begin{equation}
        \label{eq:speed}
        v=\|\mathbf{x}'(\alpha)\|
        =
        \sqrt{\sum_{j=1}^{n/2}\omega_j^2 r_j^2}.
    \end{equation}
    
    \item The curve length over the parameter interval $\alpha\in[0,2\pi)$ is
    \[
    L=2\pi v
    =
    2\pi\sqrt{\sum_{j=1}^{n/2}\omega_j^2 r_j^2}.
    \]
    When the frequencies are compatible with a $2\pi$-periodic closed
    parametrization, this is the length over one period.
\end{enumerate}
\end{lemma}

\begin{proof}
Using $\cos^2(\theta)+\sin^2(\theta)=1$ for each harmonic pair gives
\[
\|\mathbf{x}(\alpha)\|^2
=
\sum_{j=1}^{n/2} r_j^2
\bigl(\cos^2(\omega_j\alpha)+\sin^2(\omega_j\alpha)\bigr)
=
\sum_{j=1}^{n/2} r_j^2,
\]
which proves the first claim. Differentiating component-wise yields
\begin{equation}
\begin{aligned}
\mathbf{x}'(\alpha)
=
[
-r_1\omega_1\sin(\omega_1\alpha),\,
 r_1\omega_1\cos(\omega_1\alpha),\,
\ldots, 
-r_{n/2}\omega_{n/2}\sin(\omega_{n/2}\alpha),\,
 r_{n/2}\omega_{n/2}\cos(\omega_{n/2}\alpha)
]^T .
\end{aligned}
\end{equation}
Hence
\[
\|\mathbf{x}'(\alpha)\|^2
=
\sum_{j=1}^{n/2}\omega_j^2 r_j^2
\bigl(\sin^2(\omega_j\alpha)+\cos^2(\omega_j\alpha)\bigr)
=
\sum_{j=1}^{n/2}\omega_j^2 r_j^2,
\]
which is independent of $\alpha$. This proves the constant-speed claim. Since
the speed is constant, the length over $\alpha\in[0,2\pi)$ is
\[
L
=
\int_0^{2\pi}\|\mathbf{x}'(\alpha)\|\,d\alpha
=
2\pi\,\|\mathbf{x}'(\alpha)\|
=
2\pi\sqrt{\sum_{j=1}^{n/2}\omega_j^2 r_j^2}.
\]
This proves the third claim.
\end{proof}

In the present C3T/C3M construction, we typically choose integer frequencies,
and in particular harmonic choices such as $\omega_j=j$, so that the resulting
torus geodesic is closed over one $2\pi$ parameter interval. Each coordinate
pair in Eq.~\eqref{equ:parametric_curve} lies on a radius-$r_i$ circle
$S_i^1(r_i)$, so the curve lies on the flat torus
\[
T_{\mathbf r}^{n/2}
=
S_1^1(r_1)\times S_2^1(r_2)\times\cdots\times S_{n/2}^1(r_{n/2}).
\]
Defining the angular coordinates $\theta_i=\omega_i\alpha$ gives
$(\theta_1,\ldots,\theta_{n/2})\in T^{n/2}$. Thus, the C3T/C3M branch curve
corresponds to the linear flow $\theta(\alpha)
=
(\omega_1\alpha,\ldots,\omega_{n/2}\alpha)$ on the flat torus, where
$\boldsymbol{\omega}=(\omega_1,\ldots,\omega_{n/2})$ is the slope vector of the
geodesic. A flat torus can be represented as
$T^{n/2}=\mathbb{R}^{n/2}/(2\pi\mathbb{Z}^{n/2})$. Under the Euclidean
embedding with radii $\mathbf r$, the induced metric is the constant diagonal 
metric $ds^2
=
\sum_{i=1}^{n/2} r_i^2\,d\theta_i^2$. For $i=1,\ldots,n/2$, geodesics of this flat metric satisfy $\frac{d^2\theta_i}{d\alpha^2}=0$. Since the C3T/C3M flow is given by $\theta_i=\omega_i\alpha$, it follows that
the curve is a geodesic on the corresponding flat torus. Mapping these torus
coordinates into the Euclidean embedding via
$(r_i\cos\theta_i,r_i\sin\theta_i)$ yields the harmonic coordinate
representation in Eq.~\eqref{equ:parametric_curve}.
\begin{figure}[t!]
\begin{center}
  \includegraphics[width=\columnwidth]{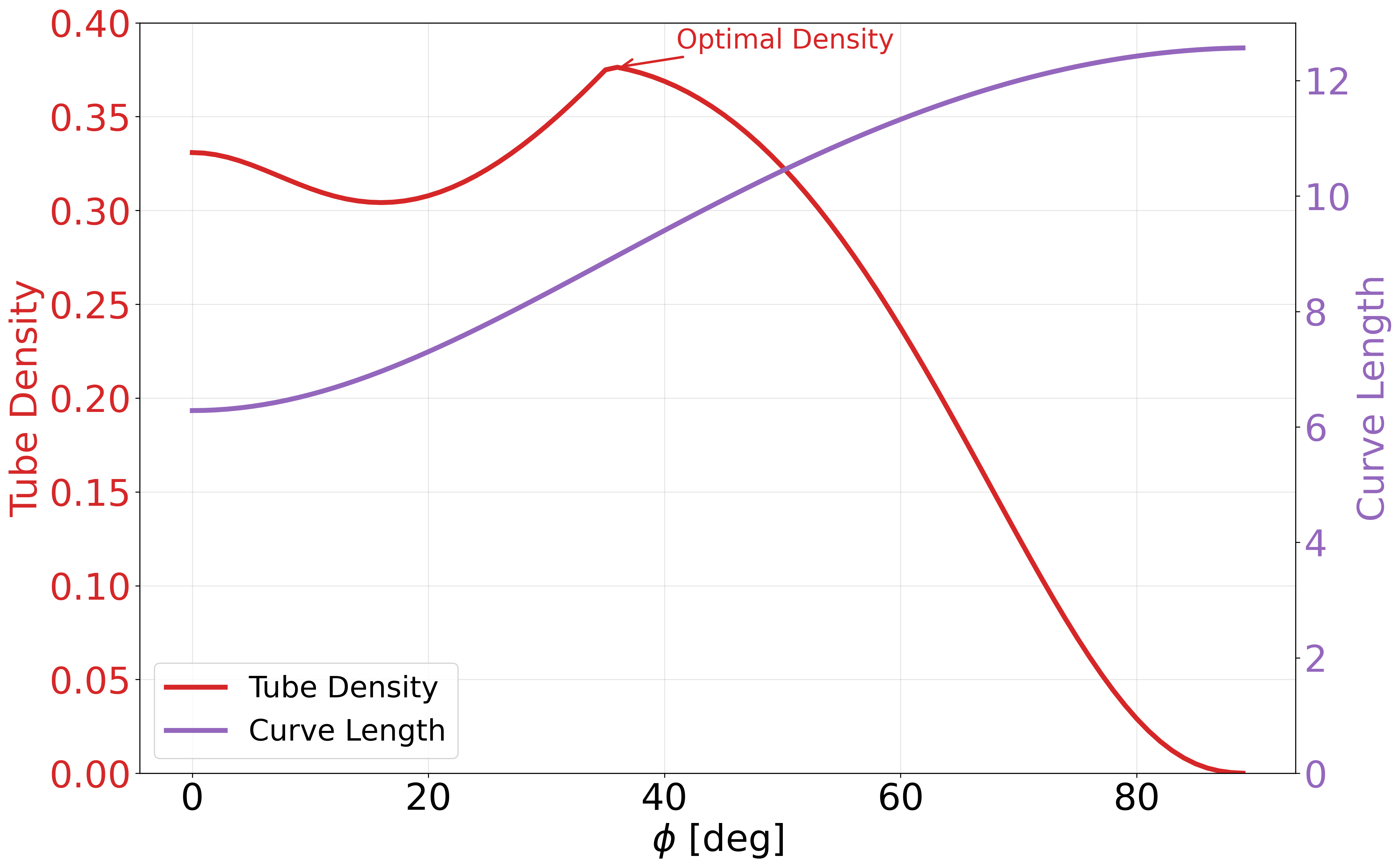}
  \caption{Tube density (left axis) and curve length (right axis) versus the radius angle $\phi=\atanTwo{r_2}{r_1}$ for the $n=4$ radii parameters in Eq.~\eqref{equ:parametric_curve}. The SPSA-optimized radii reported in~\cite[Table I]{AndersC3T}, $(r_1,r_2)\approx(0.8165,0.5774)$, correspond to $\phi=\ang{35.27}$, which coincides with the maximum tube density. For $\omega=(1,2)$ and $r_1=\cos\phi$, $r_2=\sin\phi$, the curve length is $L(\phi)=2\pi\sqrt{\cos^2\phi+4\sin^2\phi}$, ranging from $2\pi$ to $4\pi$.}
  \label{fig:density}
\end{center}
\end{figure}

It is known that packing density is important for the performance of
continuous codes~\cite{Kravtsov2007}, and the radii are therefore selected to
increase the usable tube density within the ambient power sphere. Consider the
curve length in Lemma~\ref{lem:const_speed}, the tube cross-sectional volume
$V_{cs}=\kappa_{n-1}\rho^{n-1}$, the bounding $n$-ball volume
$V_b=\kappa_n(1+\rho)^n$, and the radii vector $\mathbf r$. The tube density
is defined as
\begin{align}
  \Den(\mathbf r) = L\frac{V_{cs}(\rho)}{V_b(\rho)},
  \label{equ:den}
\end{align}
where $\kappa_n$ denotes the volume of the unit $n$-ball,
\begin{align}
  \kappa_n
  =
  \frac{\pi^{n/2}}{\Gamma \left(\frac{n}{2}+1\right)} .
\end{align}
Here $\Gamma(t)=\int_{0}^{\infty} x^{t-1}e^{-x}\,dx$ is the Gamma
function~\cite{Gonzalez1999}. For even $n$, this reduces to
$\kappa_n=\pi^{n/2}/(n/2)!$. The protection radius $\rho$ is obtained from the
global radius of curvature~\cite{Gonzalez1999}. For two parameter values
$\alpha_1$ and $\alpha_2$ on the curve in Eq.~\eqref{equ:parametric_curve},
let $\Delta=\alpha_1-\alpha_2$ and define
\begin{align}   
  \label{equ:sq_radius}
  \rho^2(\Delta)&=\frac{\left[\sum_{i=1}^{n/2}r_i^2(1-\cos(\omega_i\Delta))\right]^2}
      {2\sum_{i=1}^{n/2}r_i^2(1-\cos(\omega_i\Delta))-\frac{\left[2\pi\sum_{i=1}^{n/2}\omega_i r_i^2\sin(\omega_i\Delta)\right]^2}{L^2}} .
\end{align}
The global protection radius used in the density calculation is then obtained
by minimizing $\rho(\Delta)$ over the admissible nonzero parameter separations.
For a C3T curve on $\mathbb S^3$, Fig.~\ref{fig:density} shows tube densities and
curve lengths for a range of $(r_1,r_2)$ pairs, where all pairs satisfy the
same power constraint $\|\mathbf{x}(\alpha)\|^2=P$.

\begin{figure}[t!]
\centering
\begin{tikzpicture}[scale=0.85, line cap=round, line join=round]

\draw[line width=18pt, gray!25]
  plot[smooth, tension=0.9] coordinates {
    (-3.2,0.0) (-2.2,1.15) (-1.1,0.25) (0.0,-0.75)
    (1.1,0.25) (2.2,1.15) (3.2,0.0)
  };

\draw[line width=1.4pt]
  plot[smooth, tension=0.9] coordinates {
    (-3.2,0.0) (-2.2,1.15) (-1.1,0.25) (0.0,-0.75)
    (1.1,0.25) (2.2,1.15) (3.2,0.0)
  };

\draw[<->, line width=0.9pt]
  (-2.2,1.15) -- (-2.2,1.75);
\node[above] at (-2.2,1.75) {$\rho_G$};

\fill (-1.05,0.22) circle (1.7pt);
\fill (1.05,0.22) circle (1.7pt);
\draw[<->, dashed, line width=0.9pt] (-1.05,0.22) -- (1.05,0.22);
\node[above] at (0,0.22) {$d_{\min}$};


\node at (-3.15,-0.45) {curve centerline};
\node at (2.75,-0.55) {protection tube};

\end{tikzpicture}
\caption{Schematic illustration of a protection tube around a constant-curvature
curve. The tube radius is determined by the global protection radius
$\rho_G$, while the minimum separation $d_{\min} \ge 2\rho_G$ between distinct folds
controls the onset of threshold errors when noise moves a received point toward
a neighboring fold.}
\label{fig:folds}
\end{figure}
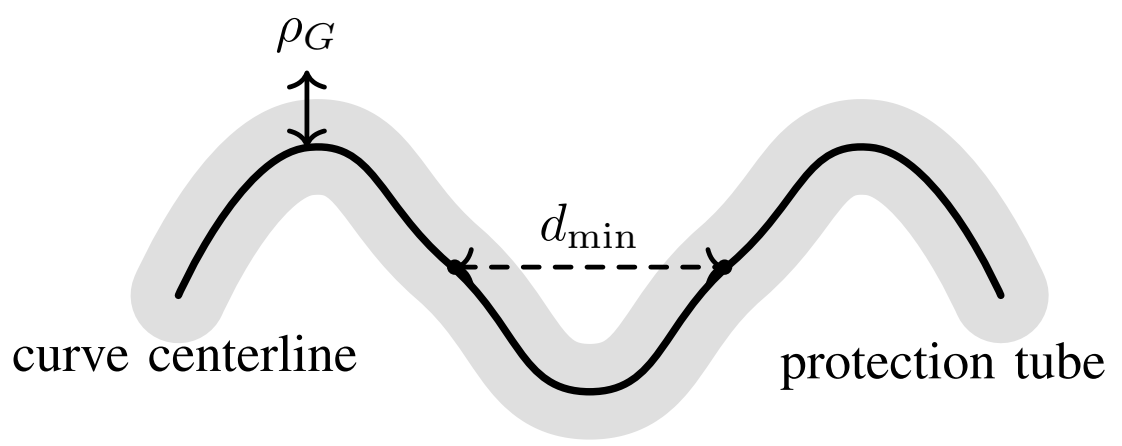

In our C3T construction, we used harmonic integer frequencies,
$\omega_i=i$, so that the parametrized curve is closed over one $2\pi$
interval and may be interpreted as a torus-knot-type curve on the associated
flat torus. In knot theory and ropelength analysis, the global radius of
curvature determines the radius of an embedded tube around the curve
centerline~\cite{Gonzalez1999}. It is defined as the minimum circumradius over
all triplets of points on the curve and therefore determines the admissible tube
radius, see Fig.~\ref{fig:folds} for an illustration. This tube enforces a
minimum separation between distinct points on the curve, thereby reducing
confusion between nearby source values at the receiver. The corresponding
global protection radius is $\rho_G= \sqrt{\min_{\Delta\in [0,\pi]}\rho^2(\Delta)}$.

\subsection{Geometric principles for multi-curve code design}
\label{ss:principles}

The design of multi-curve analog codes can be viewed as a geometric
optimization problem. The relevant design variables include curve shape,
orientation, spacing, and source allocation across branches. These variables
affect both along-curve resolution and the probability of confusing nearby
curve segments or nearby branch curves. Consequently, ideas from manifold
geometry, Riemannian optimization, frame design, and packing theory provide
useful guiding principles for C3M.

Vaishampayan and Costa showed that the performance of analog codes is governed
by the geometry of the signal locus~\cite{Vaishampayan2003}. In their curved
modulation construction, they derived expressions for arc length and minimum
distance along the curve and related these quantities to distortion
performance. This reinforces the central design tradeoff in curve-based analog
JSCC: increasing curve length improves local resolution, while maintaining
sufficient separation between distinct curve segments reduces threshold errors
at high \SNR{}. In the multi-curve setting, the same principle applies both
within each branch and across different branches.

A related viewpoint is provided by Grassmannian frames. Strohmer and Heath
studied frames that minimize the maximum pairwise correlation, thereby linking
frame design to packings in Grassmannian spaces and spherical codes
~\cite{Strohmer2003}. Although their focus was on discrete vector sets rather
than continuous curves, the underlying design principle is relevant here:
different signaling components should be spread apart in the ambient space so
that mutual interference is reduced. For C3M, this suggests choosing branch
curves whose dominant directions, occupied regions, or local tangent structures
are sufficiently separated. This is analogous to placing lines or subspaces far
apart in a Grassmannian packing, but applied here to continuous branch
manifolds rather than to finite frame elements.

These geometric viewpoints motivate the C3M design problem considered in this
paper. The branch curves should provide sufficient along-curve resolution while
maintaining enough inter-branch separation for reliable classifier-assisted
decoding. Thus, the relevant design criteria combine packing geometry,
manifold structure, source allocation, and receiver-side branch identification.

\subsection{Feature extraction on non-Euclidean data structures}
\label{ss:nop}

The geometric viewpoint above also motivates learning architectures that
respect manifold structure. The receiver problem in C3M is not merely a
Euclidean regression problem: the transmitted point lies on one of several
structured branch manifolds, and the receiver must first identify the active
branch and then estimate a local coordinate on that branch. This motivates
feature-extraction methods that exploit non-Euclidean structure rather than
treating the received vector only as an unstructured point in $\mathbb{R}^n$.

Masci et al.\ generalized convolutional neural networks (CNNs) to
non-Euclidean manifolds by introducing geodesic CNNs (GCNNs)~\cite{Masci2015}.
Their construction defines convolutions through local geodesic coordinate
patches on curved surfaces, thereby learning filters that exploit intrinsic
manifold geometry. This illustrates how neural networks can be adapted to data
that lie on curved domains and provides part of the motivation for decoder
architectures that are adapted to branch geometry.

In our previous work on C3T, decoding was performed using a conventional
multilayer perceptron (MLP). Although effective, such a model does not
explicitly exploit the geometry of the underlying branch manifold. A broader
class of approaches for non-Euclidean feature extraction is provided by
geometric deep learning. One representative example is MoNet~\cite{Monti2017},
which introduces mixture-model convolution kernels for graphs and manifolds.
MoNet defines convolution as a template-matching operation on local intrinsic
patches using learnable pseudo-coordinates and parametric kernels, typically
based on Gaussian mixtures. In this way, MoNet unifies several earlier
geometric deep-learning constructions, including geodesic CNNs and anisotropic
CNNs~\cite{Masci2015,Yu2016}. The main relevance for C3M is that such methods
show how feature extraction can be made geometry-aware by operating on local
manifold structure rather than on a fixed Euclidean grid.

A related line of work moves from discrete graph processing toward
continuous-domain models. Continuous graph neural networks (CGNNs)
~\cite{Poli2021,Ma2024} extend geometric learning from finite graph
neighborhoods to continuous domains, where learned kernels can be evaluated at
arbitrary points on the underlying space. This is appealing in the present
setting because the signal geometry is fundamentally continuous, even when the
receiver is implemented through sampled or discretized representations.

Neural operators provide a further step in this direction by learning mappings
between function spaces rather than only mappings between finite-dimensional
vectors. Instead of approximating a continuous function through a fixed
discrete graph or mesh, a neural operator learns the underlying continuous
operator. In Euclidean domains, the Fourier neural operator realizes this idea
by parameterizing operator kernels in the Fourier domain, which gives a
resolution-independent representation of the learned map.

Neural operators have subsequently been extended to Riemannian manifolds
through NORM~\cite{Chen2024}. NORM represents functions in truncated subspaces
spanned by Laplace--Beltrami eigenfunctions. Since the Laplace--Beltrami
operator plays a role on manifolds analogous to the Fourier transform in
Euclidean space, it provides a natural spectral representation for learning
maps on curved domains. This viewpoint motivates the decoder used in this
paper: each C3M branch is associated with an embedded one-dimensional manifold
contained in a flat torus, and the implemented receiver uses
branch-conditioned spectral features to estimate the corresponding local
coordinate.

This progression motivates the NORM-inspired receiver adopted in this work. The
geometric-design viewpoints explain why the branch manifolds and their relative
placement matter, while geometric deep-learning methods explain how feature
extraction can be adapted to non-Euclidean domains. NORM provides a suitable
endpoint for the present setting because it learns maps on Riemannian manifolds
using spectral structure, allowing each branch-conditioned decoder to exploit
the geometry associated with the selected C3M branch. Section~\ref{ss:receiver}
introduces the receiver model, and Section~\ref{sec:norm} presents the
corresponding NORM-inspired decoder and circular Huber training objective.

\begin{figure}[t!]
\begin{center}
  \includegraphics[width=\columnwidth]{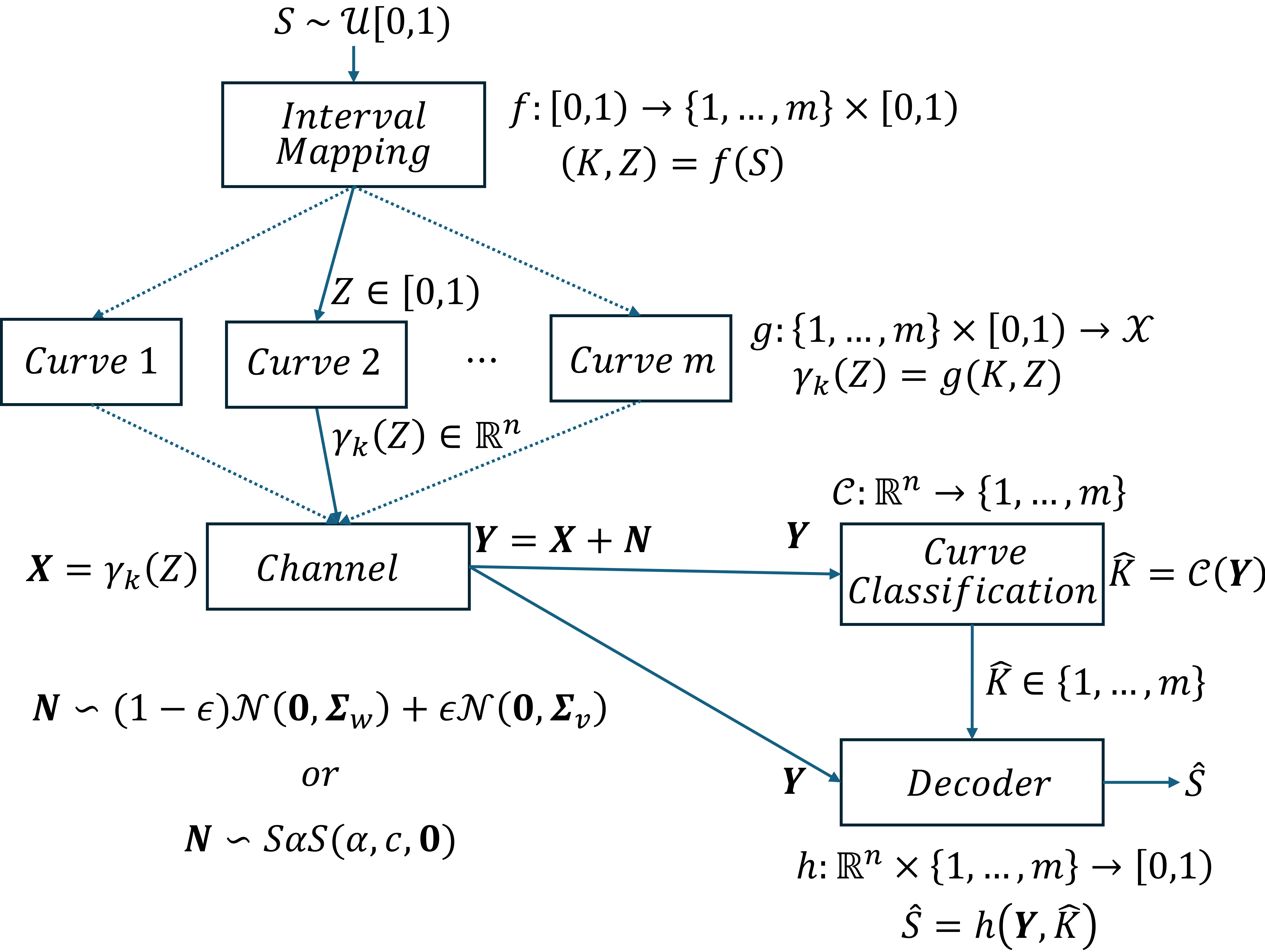}
  \caption{Block diagram of the proposed C3M system. The source symbol satisfies $S\sim\mathcal{U}[0,1)$ and is first mapped to a branch index and a normalized local coordinate, which determine the active constant-curvature branch curve for transmission. The nominal channel model is fixed isotropic AWGN. Gaussian-mixture impulsive noise and symmetric alpha-stable noise are used as robustness test cases. At the receiver, a curve classifier estimates the transmitted branch, after which a branch-conditioned decoder reconstructs the source symbol.}
\label{fig:sys}
\end{center}
\end{figure}

\section{System Model}
\label{sec:sys}

Our system model is illustrated in Fig.~\ref{fig:sys} for several additive-noise settings, including symmetric alpha-stable noise, Gaussian mixture noise, and Gaussian noise, with additive white Gaussian noise (AWGN) arising as a special case when $\epsilon = 0$ or $\alpha = 2$. The proposed C3M framework extends the single-curve C3T construction to a multi-branch setting in which the source interval is partitioned across $m$ branches, each associated with its own constant-curvature curve on a nearby flat torus. In this way, the encoder operates as a piecewise constant-curvature map from the scalar source domain to an $n$-dimensional signal set. For each realization of the scalar memoryless source, the encoder first determines the active branch and a normalized local coordinate within that branch, and then maps this pair to a point on the corresponding branch curve. The resulting channel input is transmitted over an $n$-dimensional additive-noise channel. At the receiver, the noisy observation is first processed by a curve classifier that estimates the transmitted branch index, after which a branch-conditioned decoder reconstructs the source symbol. The system performance is measured through the end-to-end mean-squared distortion $D\triangleq \mathrm{E}\!\left[(S-\widehat{S})^2\right]$, which is used throughout the paper to compare interval allocations, branch geometries, and decoder configurations.

\subsection{Source and encoder model}

Consider a scalar memoryless source $S \sim \mathcal{U}(0,1)$, which is partitioned into $m$ disjoint source intervals
\begin{equation}
    I_k=[a_{k-1},a_k), \qquad k=1,\ldots,m,
\end{equation}
with
\begin{equation}
    0=a_0\le a_1\le\cdots\le a_m=1,
    \qquad
    p_k \triangleq a_k-a_{k-1}\ge 0.
\end{equation}
Branches with $p_k>0$ are active source-carrying branches. Let $\mathcal A=\{k:p_k>0\}$ denote the active branch set. All interval maps below are understood to be restricted to branches in $\mathcal A$. A candidate branch
with $p_k=0$ is interpreted as pruned by the allocation rule and does not carry
source mass in the final encoder. The corresponding branch-index random variable is defined by
\begin{equation}
    K=k \quad \Longleftrightarrow \quad S\in I_k,
\end{equation}
so that, for the uniform source, $\Pr(K=k)=p_k$.

For an active branch with $p_k>0$ and a realization $s\in I_k$, define the
normalized local coordinate
\begin{equation}
    z=\frac{s-a_{k-1}}{p_k}\in[0,1).
\end{equation}
Accordingly, the corresponding local-coordinate random variable is
\begin{equation}
    Z \triangleq \frac{S-a_{K-1}}{p_K}\in[0,1).
\end{equation}

The first stage of the encoder is the interval map
\begin{equation}
\label{eq:f_map}
    f:[0,1)\to\{1,\ldots,m\}\times[0,1),
    \qquad
    f(s)=(k,z),
\end{equation}
or, equivalently,
\begin{equation}
\label{eq:f_explicit}
    f(s)=\left(k,\frac{s-a_{k-1}}{p_k}\right),
    \qquad s\in I_k.
\end{equation}
At the random-variable level, this becomes $(K,Z)=f(S)$. Because each source symbol belongs to exactly one active interval and can be
recovered uniquely from the pair $(k,z)$ with $k\in\mathcal A$, the map $f$ is bijective on the active branch set, with inverse
$f^{-1}(k,z)=a_{k-1}+p_k z$.

For each branch $k\in\{1,\ldots,m\}$, let
\begin{equation}
    \gamma_k:[0,1)\to\mathbb{R}^n
\end{equation}
denote a constant-curvature branch curve of the form reviewed in Section~\ref{ss:c3t}, with branch-specific parameter vector $\theta_k=(\mathbf{r}_k,\boldsymbol{\omega}_k)$. Define the branch signal set $\mathcal{X}_k \triangleq \gamma_k([0,1))$ and the full signal set
\begin{equation}
\label{eq:signal_set}
    \mathcal{X}\triangleq\bigcup_{k=1}^m \mathcal{X}_k
    =\bigcup_{k=1}^m \gamma_k([0,1)).
\end{equation}

The second stage of the encoder is geometric. Given a branch index $k$ and local coordinate $z$, the map $g$ selects the corresponding branch curve and generates the transmitted point on that curve:
\begin{equation}
\label{eq:g_map}
    g:\{1,\ldots,m\}\times[0,1)\to\mathcal{X},
    \qquad
    g(k,z)=\gamma_k(z).
\end{equation}
Since the branch curves are pairwise disjoint and each parameterization $z\mapsto\gamma_k(z)$ is one-to-one on $[0,1)$, the map $g$ is bijective onto $\mathcal{X}$. Consequently, the overall noiseless encoder
\begin{equation}
    e \triangleq g\circ f:[0,1)\to\mathcal{X}
\end{equation}
is also bijective. The transmitted vector random variable is therefore
\begin{equation}
\label{eq:x_encoder}
    \mathbf{X}=e(S)=g(f(S))=\gamma_K(Z)\in\mathbb{R}^n.
\end{equation}
Fig.~\ref{fig:sys} illustrates the resulting C3M transmitter, channel, classifier, and decoder.
Throughout, the branch curves are normalized to satisfy the transmit-power constraint $\mathrm{E}\!\left[\|\mathbf{X}\|^2\right]=P$ and, unless otherwise stated, we use the normalization $P=1$.

\subsection{Channel model and noise assumptions}

Typically, wireless channels are modeled using AWGN together with some form of
fading. However, the Gaussian assumption can be inadequate in practical
implementations, where the channel may be perturbed by impulsive or heavy-tailed
disturbances~\cite{Shhab2017,Clavier2021b}. In such settings, the noise
distribution is no longer well described by a single Gaussian law. Common
non-Gaussian models include Gaussian-mixture impulsive-noise models~\cite{Shhab2017,Clavier2021b,NikiasShao1995} and
symmetric alpha-stable models~\cite{NikiasShao1995,Nolan2020}, which provide useful test cases for robustness
evaluation.

The Gaussian-mixture impulsive-noise model is particularly attractive because it
connects naturally to robust estimation theory. In this interpretation, most
samples follow a nominal Gaussian distribution, while a fraction of the
observations behave as outliers. This viewpoint is well aligned with
impulsive-noise channels, in which rare but large perturbations contaminate an
otherwise nominal background-noise process. To also capture heavier-tailed behavior, we consider symmetric alpha-stable
noise~\cite{NikiasShao1995,Nolan2020}.\footnote{A distribution is stable if a linear combination of two
independent random variables with that distribution has the same distribution,
up to location and scale.} Stable distributions are heavy-tailed and
leptokurtotic, except for the Gaussian case obtained at $\alpha=2$. For
$\beta=0$, the distribution is symmetric about its location parameter and is
referred to as a symmetric alpha-stable ($S\alpha S$) distribution. Thus, the
two non-Gaussian noise models considered here provide complementary robustness
test cases: Gaussian-mixture impulsive noise captures finite-variance
contamination by rare large perturbations, while $S\alpha S$ noise captures
more general heavy-tailed behavior.

After encoding, the transmitted vector is corrupted by additive channel noise.
Fig.~\ref{fig:sys} illustrates the proposed C3M system. The nominal channel
model is fixed isotropic AWGN, and the Gaussian-mixture impulsive-noise and
symmetric alpha-stable ($S\alpha S$) models are used as robustness test cases.
Accordingly, the channel is modeled as
\[
\mathbf{Y}=\mathbf{X}+\mathbf{N},
\]
where the noise vector $\mathbf{N}$ is independent of the source $S$. Each
source symbol is transmitted using one $n$-dimensional channel use,
corresponding to a bandwidth-expansion factor of $n$.
In the Gaussian case, we write $\mathbf{N}\sim\mathcal{N}(\mathbf{0},\mathbf{\Sigma}_n)$, where $\mathbf{\Sigma}_n$ is a positive-definite covariance matrix. The isotropic AWGN case is recovered as the special case $\mathbf{\Sigma}_n=\sigma_n^2\mathbf{I}_n$. For impulsive Gaussian-mixture noise, we write the noise vector $\mathbf{N}$ as
\begin{equation}
\label{eq:gmm_noise}
    \mathbf{N}\sim (1-\epsilon)\,\mathcal{N}(\mathbf{0},\mathbf{\Sigma}_w)
    + \epsilon\,\mathcal{N}(\mathbf{0},\mathbf{\Sigma}_v),
\end{equation}
where $\epsilon\in[0,1]$ is the contamination probability and $\mathbf{\Sigma}_w,\mathbf{\Sigma}_v\in\mathbb{R}^{n\times n}$ are covariance matrices. The AWGN case is recovered when $\epsilon=0$ and $\mathbf{\Sigma}_w=\sigma_n^2\mathbf{I}_n$.
For symmetric alpha-stable noise, we write the $j$th component of $\mathbf{N}$ as
\begin{equation}
\label{eq:sas_noise}
    N_j \overset{\mathrm{i.i.d.}}{\sim} \mathrm{S\alpha S}(\alpha,c,0),
    \qquad j=1,\ldots,n,
\end{equation}
where $\alpha\in(0,2]$ is the stability parameter, and $c>0$ is the scale parameter. The Gaussian case is recovered at $\alpha=2$.
Unless otherwise stated, we normalize the transmit power to
\begin{equation}
    P=\mathrm{E}\!\left[\|\mathbf{X}\|^2\right]=1,
    \label{eq:power}
\end{equation}
and hence in the isotropic AWGN case, i.e., when
$\mathbf{N}\sim\mathcal{N}(\mathbf{0},\sigma_n^2\mathbf{I}_n)$, we define the total signal-to-noise ratio as
\begin{equation}
    \SNR\triangleq \frac{P}{\sigma_n^2}=\frac{1}{\sigma_n^2}.
    \label{eq:snr}
\end{equation}

Unless explicitly stated otherwise, the analytical posterior expressions, MAP
and MMSE estimators, mutual-information calculations, capacity references, and
\SDR{} reference curves in Sections~\ref{sec:analysis}--\ref{sec:numexp} are
evaluated under the fixed isotropic AWGN model. In
Fig.~\ref{fig:sdr_vs_snr}, the one-curve panels use the end-to-end Gaussian
\SDR{} reference, while the multi-curve panels use the branch-known Gaussian
\SDR{} reference. The Gaussian-mixture and $S\alpha S$ models are used only as
robustness test cases for the Huber-trained branch-conditioned decoder and for
evaluating deviations from the nominal Gaussian channel.

\subsection{Receiver model}
\label{ss:receiver}

The receiver is hierarchical. The first stage estimates the transmitted branch index from the noisy observation, and the second stage performs branch-conditioned decoding within the selected branch. This structure mirrors the encoder: first a coarse branch decision, then a fine along-curve reconstruction.

The first stage is a curve classifier
\begin{equation}
\label{eq:classifier}
    \widehat{K}=\mathcal{C}(\mathbf{Y}),
    \qquad
    \mathcal{C}:\mathbb{R}^n\to\{1,\ldots,m\},
\end{equation}
which estimates the transmitted branch index.

Equivalently, the receiver may be viewed as a bank of $m$ branch-conditioned decoders. Each decoder observes the same channel output $\mathbf{Y}\in\mathbb{R}^n$ and is associated with one candidate branch. The classifier estimate $\widehat{K}$ selects the active decoder, which then produces the local-coordinate estimate $\widehat{Z}$ and, through the inverse interval map, the final scalar estimate $\widehat{S}$. In the absence of side information, the receiver must infer the transmitted branch identity from the channel output. From a detection perspective, this can be formulated as a maximum a posteriori (MAP) branch-selection problem, followed by branch-conditioned decoding within the selected branch.

Conditioned on $\widehat{K}$, the decoder produces the reconstructed source symbol
\begin{equation}
\label{eq:overall_decoder}
    h:\mathbb{R}^n\times\{1,\ldots,m\}\to[0,1),
    \qquad
    \widehat{S}=h(\mathbf{Y},\widehat{K}).
\end{equation}
In the implementation considered in this paper, $h(\cdot,\cdot)$ is realized
by a NORM-inspired branch-conditioned spectral decoder, see
Section~\ref{sec:norm}. Conceptually, the decoder may be viewed as first estimating a branch-local coordinate and then mapping it back to the source interval through the inverse interval map.

\section{Multi-curve Design}
\label{sec:multi}

This section describes how the branch curves used in C3M are constructed once a baseline C3T design has been fixed. The main idea is to retain the constant-curvature branch structure while introducing additional nearby branches through new radii vectors. Throughout this section, the frequency parameters are kept fixed, and the branch design is performed through the choice of the radii parameters. 

Because the measured SDR depends on the noise normalization, all main C3M design sweeps are evaluated using a fixed channel-noise covariance that is independent of the candidate curve radii. This distinguishes the communication-channel model from radius-dependent normalizations, where coordinate pairs with small radii also receive proportionally smaller noise. The latter can mask near-degenerate radius choices and is therefore treated only as an auxiliary normalization rather than as the main C3M channel model.
Unless otherwise stated, all C3M design sweeps are performed under fixed isotropic AWGN, $\mathbf{N}\sim\mathcal{N}(\mathbf{0},\sigma_n^2\mathbf{I}_n)$,  where the noise variance is independent of the branch radii and candidate curve parameters. 

The decoders are trained using a mixed-SNR{} data set with samples drawn from $0$, $+3$, $+6$, and \dBval{+10}. Unless otherwise stated, the \SNRs{} are sampled uniformly, so that the learned receiver is exposed to a range of moderate-noise conditions rather than a single operating point.
\begin{figure*}[t!]
  \centering

  \begin{minipage}{\textwidth}
    \centering
    \subfloat[\label{fig:sdr_vs_phi}]{
      \includegraphics[width=0.45\textwidth]{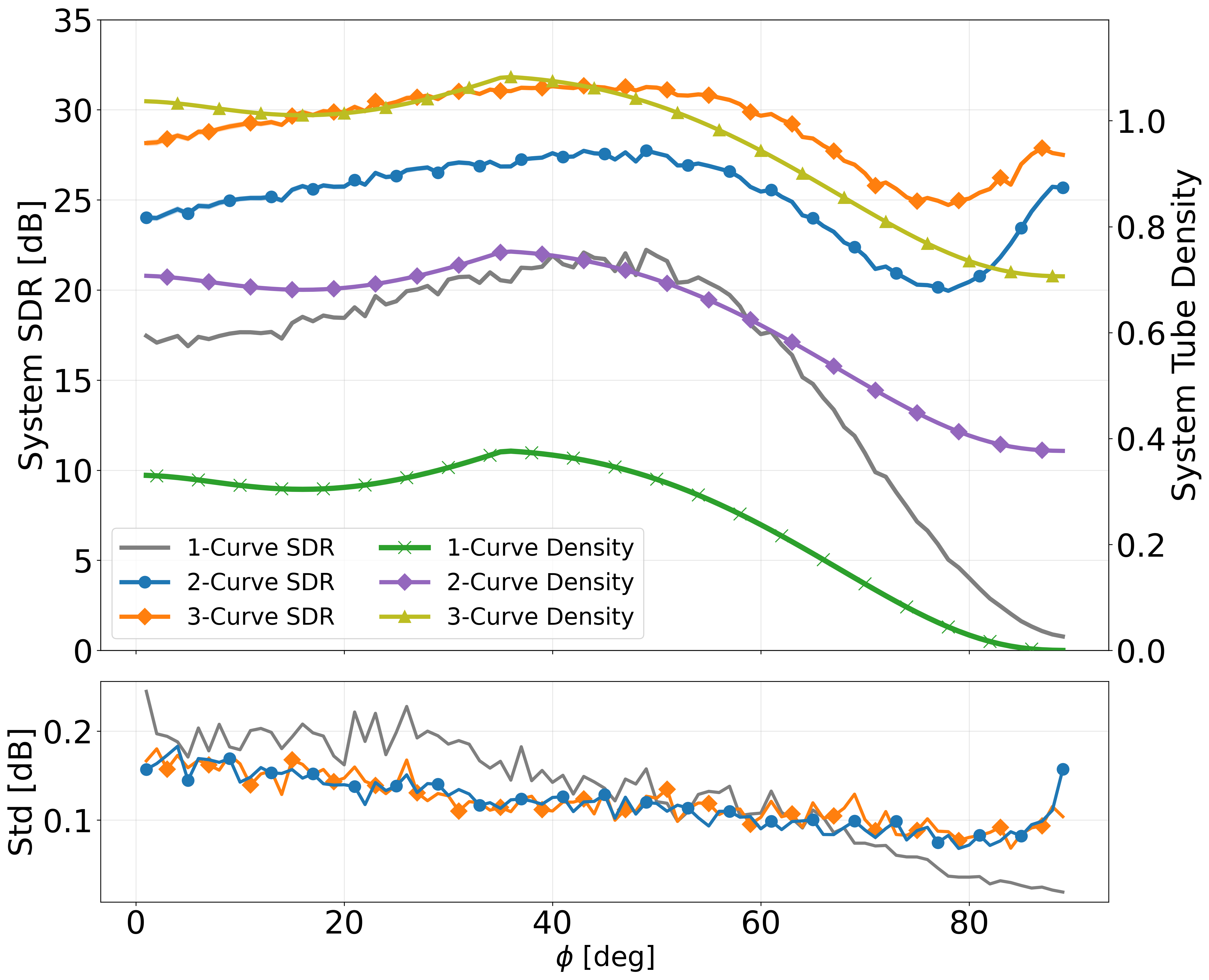}}\hfil
    \subfloat[\label{fig:accu_vs_phi}]{
      \includegraphics[width=0.45\textwidth]{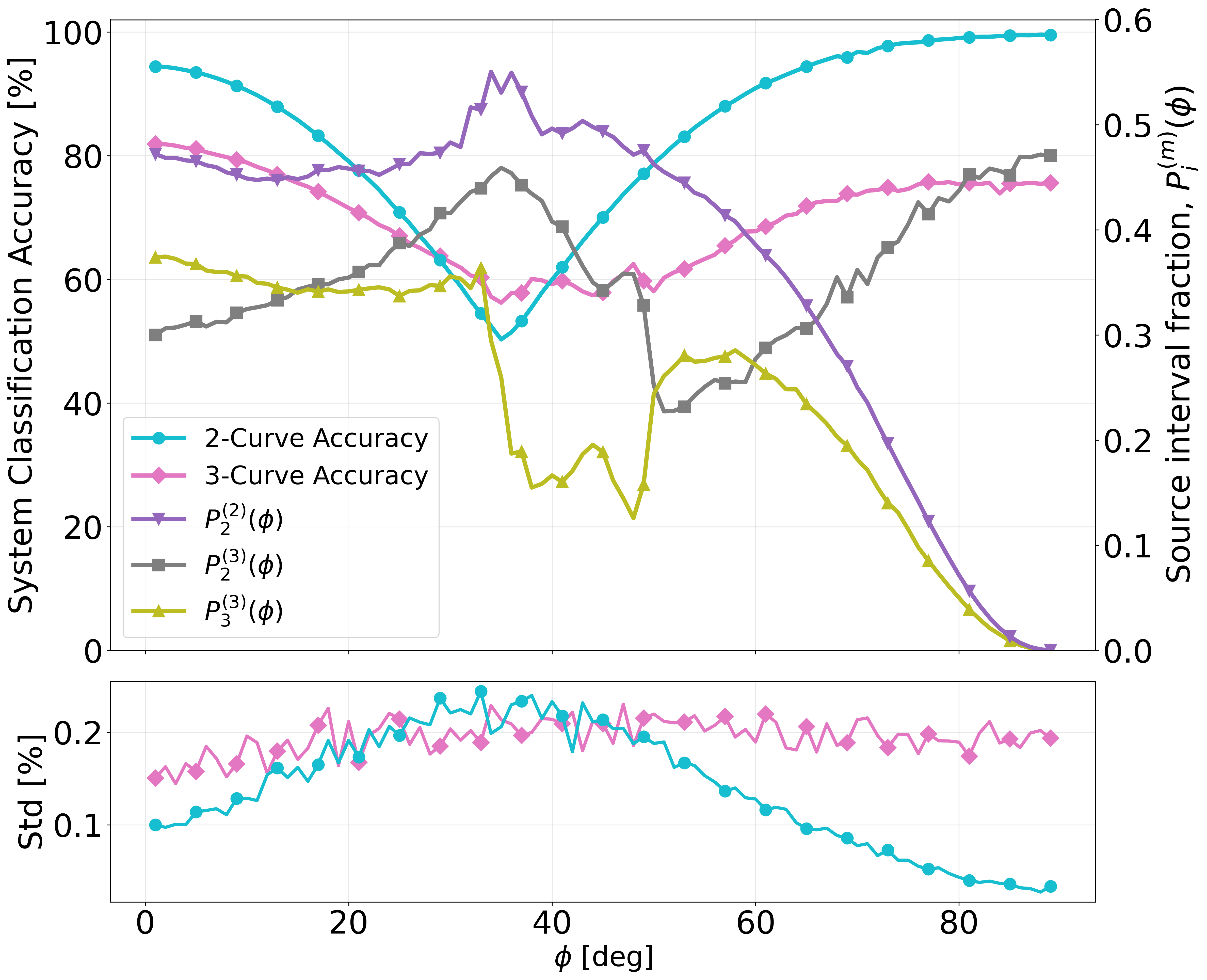}}
    \caption{Sequential branch-selection results for the $n=4$ C3M design at \dBval{+10} \SNR{}. (a) System \SDR{} (left axis) and system tube density (right axis) versus $\phi$ for the one-, two-, and three-candidate systems. (b) Classification accuracy (left axis) and interval fraction $P_i^{(m)}(\phi)$ (right axis) versus $\phi$ for the corresponding $m=2$ and $m=3$ sweeps.}  
    \label{fig:phi_block}
  \end{minipage}
\end{figure*}
\begin{figure*}[t!]
  \centering
    \centering
    \subfloat[\label{fig:a}]{
      \includegraphics[width=0.32\textwidth]{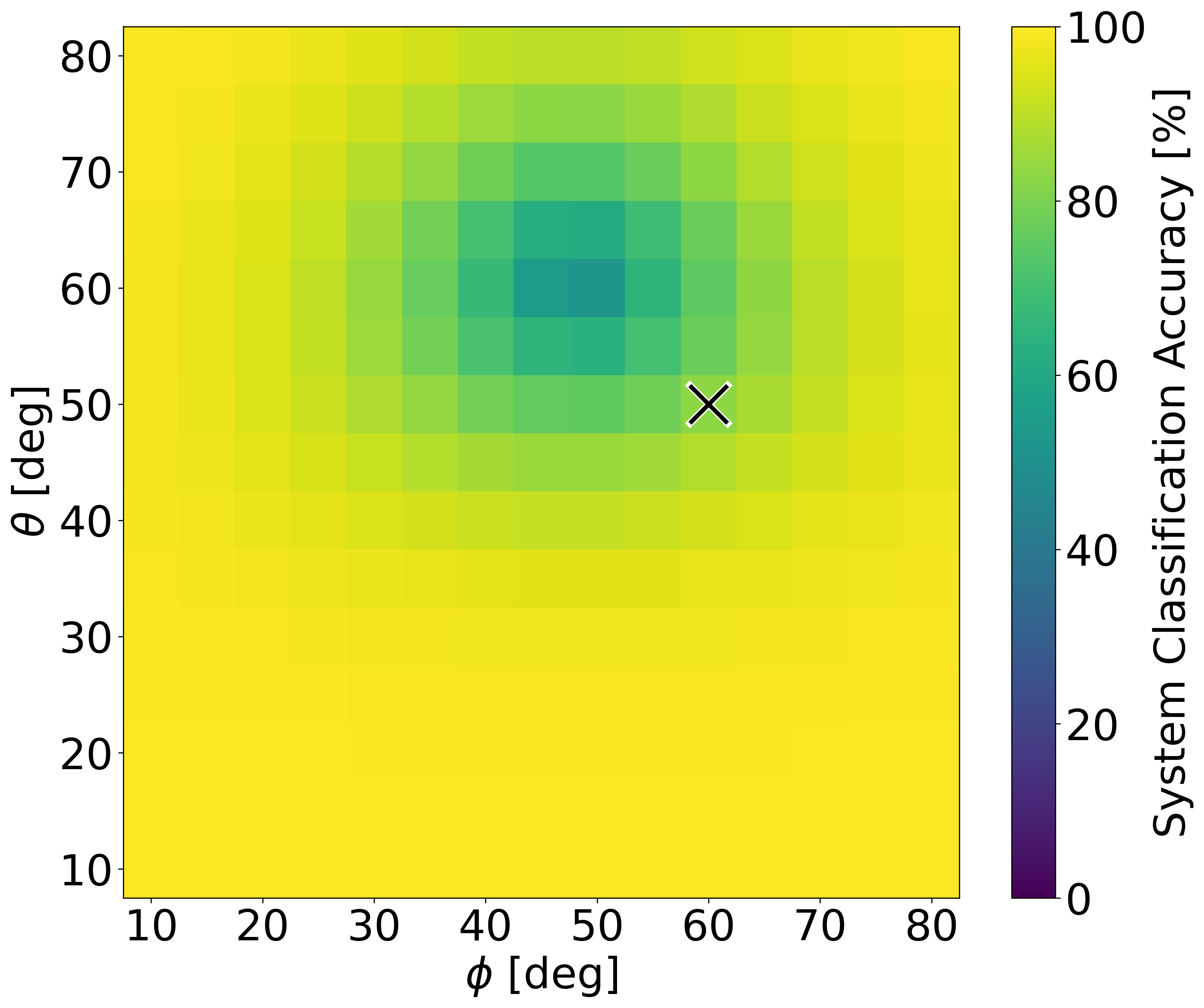}}\hfil
    \subfloat[\label{fig:b}]{
      \includegraphics[width=0.32\textwidth]{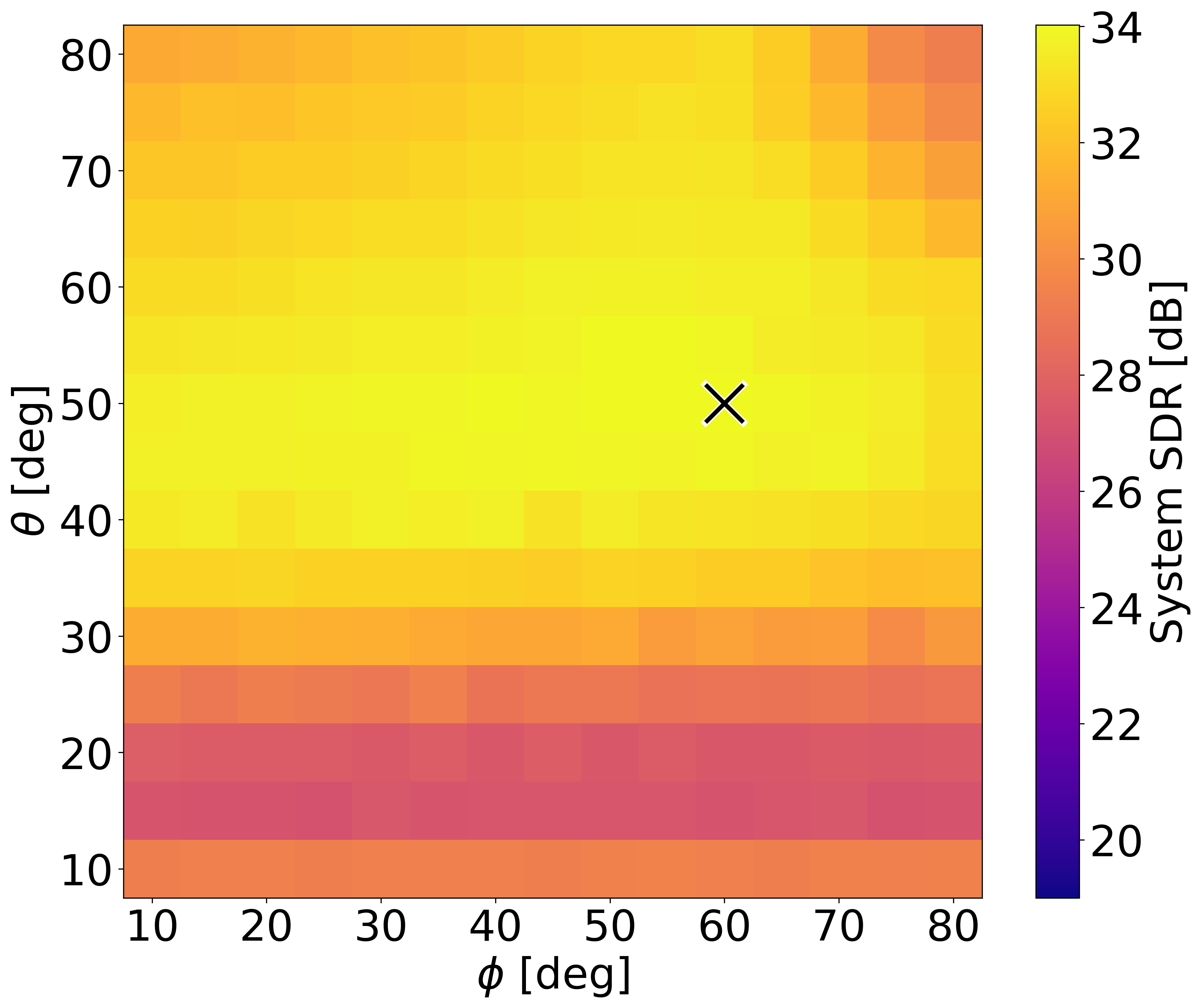}}\hfil
    \subfloat[\label{fig:c}]{
      \includegraphics[width=0.32\textwidth]{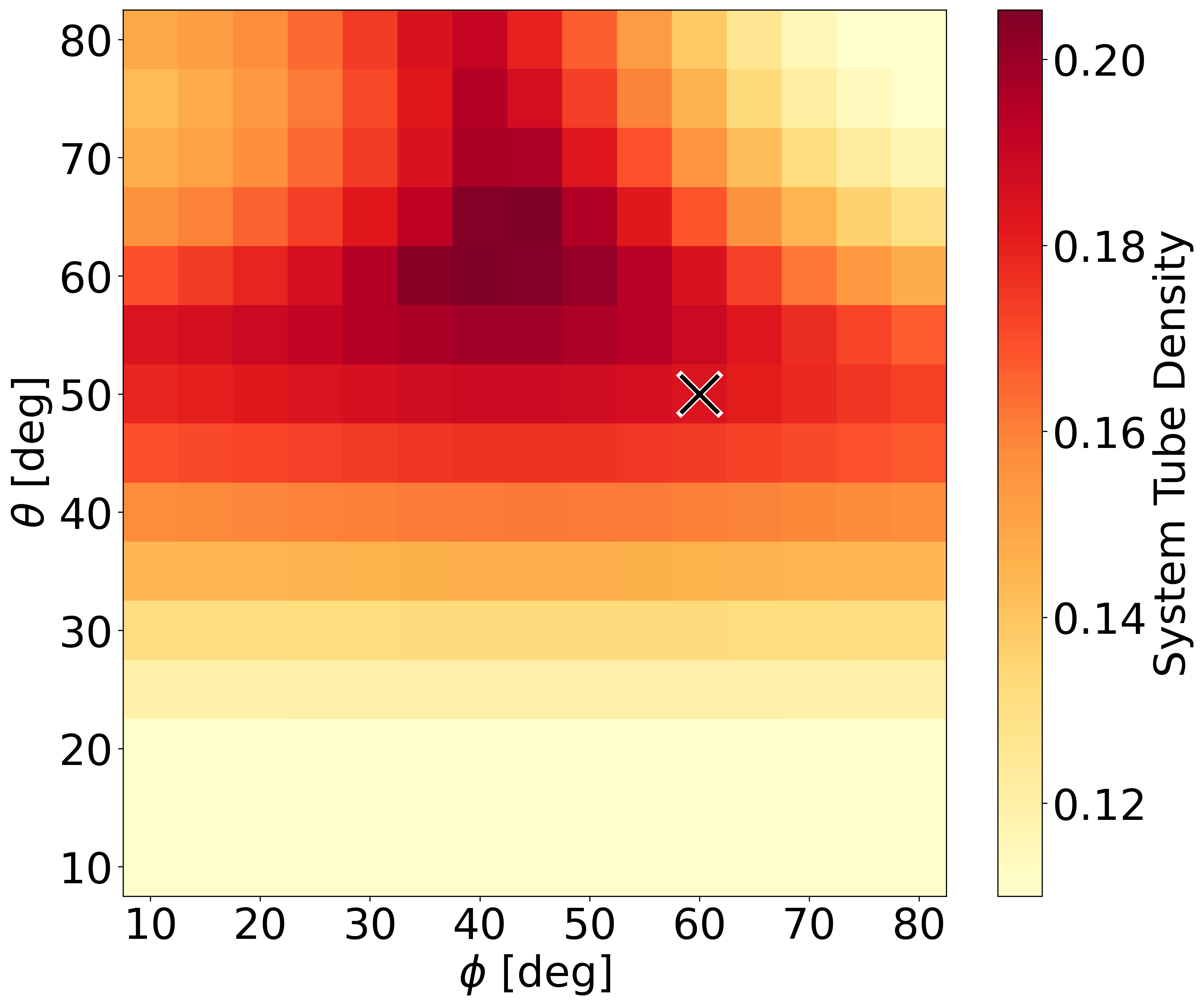}}\\
    \subfloat[\label{fig:d}]{
      \includegraphics[width=0.32\textwidth]{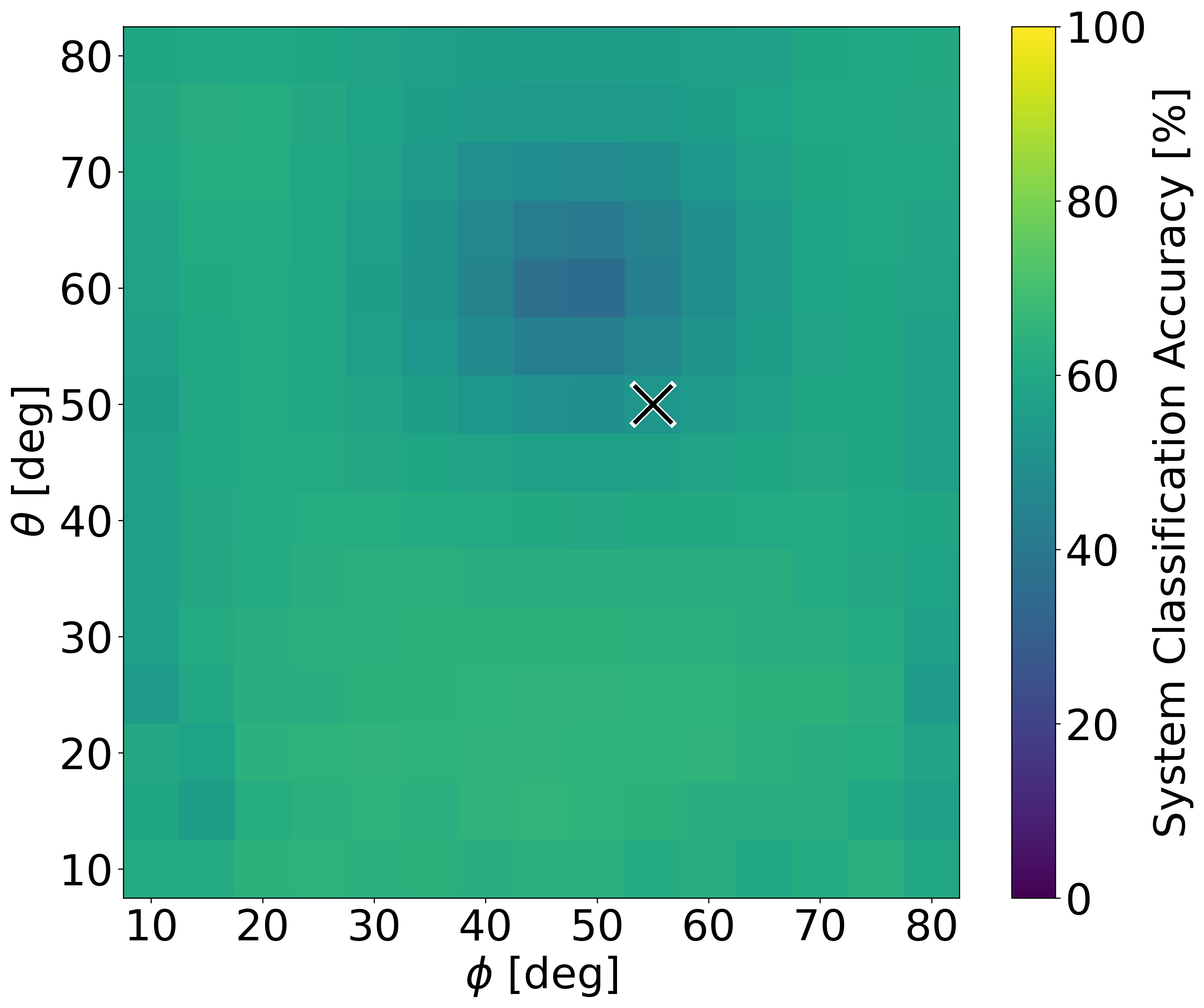}}\hfil
    \subfloat[\label{fig:e}]{
      \includegraphics[width=0.32\textwidth]{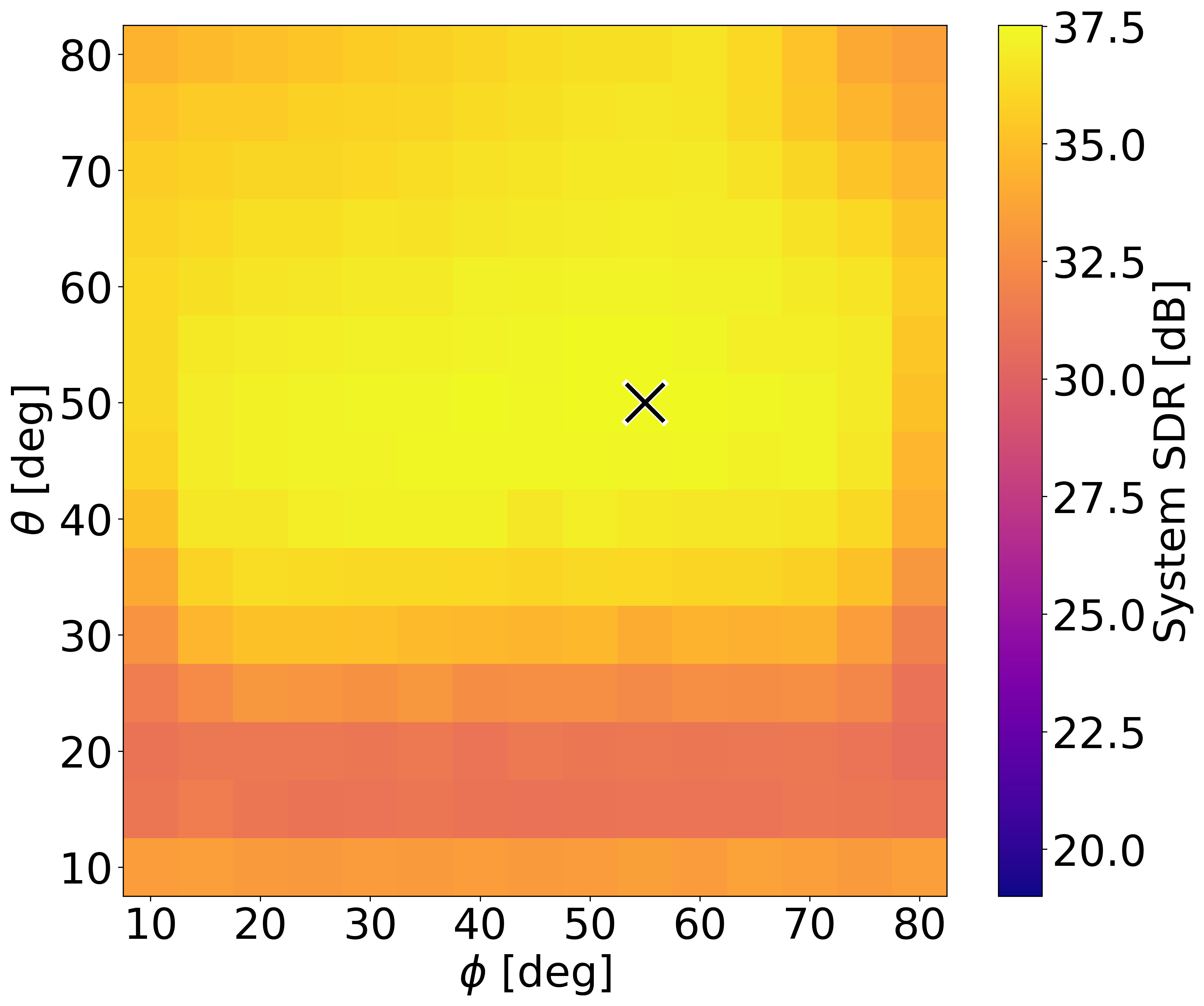}}
    \caption{Six-dimensional C3M candidate sweeps at \dBval{+10} \SNR{}, using the optimal C3T curve from~\cite[Table I]{AndersC3T} as the starting branch. Panels (a)--(c) show the $15\times15$ second-curve sweep for the $m=2$ system: (a) classification accuracy, (b) system \SDR{}, and (c) system tube density. The selected $m=2$ point is $(\theta,\phi)=(50^\circ,60^\circ)$. Panels (d)--(e) show the corresponding $15\times15$ third-curve sweep for the $m=3$ system: (d) classification accuracy and (e) system \SDR{}. The selected third-curve candidate is $(\theta,\phi)=(50^\circ,55^\circ)$.} 
    \label{fig:accu}
\end{figure*}
\begin{table}[b!]
\caption{Four-dimensional C3M candidate-branch parameters at \dBval{+10} \SNR{}.}
\label{tab:curves_n4}
\centering
\begin{threeparttable}
\begin{tabular}{lccc}
\toprule
Quantity & Starting C3T Curve\tnote{a} & Second Curve & Third Curve \\
\midrule
$\phi=\atanTwo{r_2}{r_1}$ [deg] & $35.27$ & $47.00$ & $44.00$ \\
$r_1$ & $0.8165$ & $0.6820$ & $0.7193$ \\
$r_2$ & $0.5774$ & $0.7314$ & $0.6947$ \\
Length & $8.8860$ & $10.1404$ & $9.8300$ \\
Tube density & $0.3767$ & $0.3411$ & $0.3554$ \\
\bottomrule
\end{tabular}
\begin{tablenotes}[flushleft]
    \item[a] The optimal C3T curve from~\cite[Table I]{AndersC3T} is used as
    the starting branch for the multi-branch C3M construction. All curve
    parameters use the radii convention in Eq.~\eqref{equ:parametric_curve};
    lengths are computed using Lemma~\ref{lem:const_speed}, and tube densities
    are computed using Eq.~\eqref{equ:den}.
\end{tablenotes}
\end{threeparttable}
\end{table}
Under the power normalization used in this work, the radii vector lies on the positive orthant of the unit sphere, i.e., $\sum_{i=1}^{n/2} r_i^2 = 1$, $r_i \ge 0$. Thus, the multi-curve design problem may be viewed as the selection of multiple radii vectors, each defining a constant-curvature branch curve on its own nearby flat torus.

For $n=4$, let $\phi$ denote the angle associated with the radii pair $(r_1,r_2)$ in Eq.~\eqref{equ:parametric_curve}, i.e., $\phi=\atanTwo{r_2}{r_1}$. The optimal pair $(r_{1,\mathrm{opt}},r_{2,\mathrm{opt}})$ maximizing the tube density in Eq.~\eqref{equ:den} is reported in~\cite[Table I]{AndersC3T}. Note that this optimal pair satisfies $r_{1,\mathrm{opt}} \approx \frac{4}{5}$, $r_{2,\mathrm{opt}} \approx \frac{3}{5}$, corresponding approximately to the Pythagorean triplet $(3,4,5)$.

Given this optimal radii pair, and keeping the frequency parameters fixed, the multi-curve design objective is to identify additional branch curves $(r_1,r_2)$ that maximize the resulting SDR while remaining geometrically distinguishable from the baseline curve. For $n=4$, the candidate radii pairs are parameterized as
\begin{equation}
\label{eq:radii_n4}
\begin{cases}
r_1=\cos\phi,\\
r_2=\sin\phi,
\end{cases}
\qquad 0^\circ \le \phi \le 90^\circ.
\end{equation}
Thus, the search is restricted to the first quadrant. For the case $n=4$ and $m=2$, Fig.~\ref{fig:sdr_vs_phi} illustrates the dependence of both the system SDR and the tube density of the second curve on $\phi$, when the first curve is fixed to the optimal C3T design from~\cite[Table I]{AndersC3T}. The resulting multi-curve design details, including radii, tube densities, curve lengths, interval fractions, and classification accuracies for $n=4$, are reported in Table~\ref{tab:curves_n4}.

For $n=6$, the candidate radii triples are parameterized by two angles as
\begin{equation}
\label{eq:radii_n6}
\begin{cases}
r_1=\sin\theta\cos\phi,\\
r_2=\sin\theta\sin\phi,\\
r_3=\cos\theta,
\end{cases}
\qquad
0^\circ \le \theta \le 90^\circ,
\quad
0^\circ \le \phi \le 90^\circ.
\end{equation}
Again, the search is restricted to the first quadrant of the corresponding angular parameterization. The radii vector $\mathbf{r}=(r_1,r_2,r_3)$ satisfies the power constraint $r_1^2+r_2^2+r_3^2=1$, so the admissible radii lie on the unit sphere $\mathbb S^2\subset\mathbb{R}^3$. Since the harmonic pairs occupy mutually orthogonal two-dimensional subspaces of $\mathbb{R}^6$, the corresponding encoder curve lies on the unit sphere $\mathbb S^5\subset\mathbb{R}^6$, while its geometry is determined by the choice of $\mathbf{r}$. Thus, once the frequencies are fixed, designing an additional branch in $\mathbb{R}^6$ may be viewed as selecting an additional point on the positive orthant of $\mathbb S^2$. The multi-curve design details for $n=6$ are shown in Table~\ref{tab:curves_n6} and the corresponding two-curve and three-curve sequential branch-selection procedures are illustrated in Fig.~\ref{fig:accu}.  
For $n=8$, the candidate radii quadruples $(r_1,r_2,r_3,r_4)$ are parameterized analogously by three angles in the first quadrant. More generally, for an $n$-dimensional space with even $n$, the radii design involves $n/2-1$ angular parameters. Thus, the branch-design problem reduces to a search over the positive orthant of the unit sphere in $\mathbb{R}^{n/2}$. 
\begin{table}[t!]
\caption{Six-dimensional C3M candidate-branch parameters at \dBval{+10} \SNR{}.}
\label{tab:curves_n6}
\centering
\begin{threeparttable}
\begin{tabular}{lccc}
\toprule
Quantity & Starting C3T Curve\tnote{a} & Second Curve & Third Curve \\
\midrule
$\theta$ [deg] & $60.54$ & $50.00$ & $50.00$ \\
$\phi$ [deg] & $47.92$ & $60.00$ & $55.00$ \\
$r_1$ & $0.5835$ & $0.3830$ & $0.4394$ \\
$r_2$ & $0.6463$ & $0.6634$ & $0.6275$ \\
$r_3$ & $0.4918$ & $0.6428$ & $0.6428$ \\
Length & $12.86$ & $14.90$ & $14.72$ \\
Tube density & $0.1014$ & $0.0825$ & $0.0848$ \\
\bottomrule
\end{tabular}
\begin{tablenotes}[flushleft]
    \item[a] The optimal C3T curve from~\cite[Table I]{AndersC3T} is used as
    the starting branch for the multi-branch C3M construction. All curve
    parameters use the radii convention in Eq.~\eqref{equ:parametric_curve};
    lengths are computed using Lemma~\ref{lem:const_speed}, and tube densities
    are computed using Eq.~\eqref{equ:den}.
\end{tablenotes}
\end{threeparttable}
\end{table}
\begin{table*}[b!]
\caption{Final interval allocations and branch-classification reliabilities at
\dBval{+10} \SNR{} for the selected C3M configurations. The reliabilities
$q_k$ are the diagonal entries of the row-normalized branch-confusion matrix,
and the overall branch-classification accuracy is $\sum_k p_k q_k$.}
\label{tab:alloc_reliability}
\centering
\begin{tabular}{ccccccccc}
\toprule
$n$ & $m$ & $p_1$ & $p_2$ & $p_3$ & $q_1$ [\%] & $q_2$ [\%] & $q_3$ [\%] & Overall [\%] \\
\midrule
$4$ & $2$ & $0.5207$ & $0.4793$ & --       & $73.16$ & $74.23$ & --      & $73.67$ \\
$4$ & $3$ & $0.4729$ & $0.3234$ & $0.2037$ & $67.19$ & $50.69$ & $30.88$ & $54.45$ \\
$6$ & $2$ & $0.5483$ & $0.4517$ & --       & $82.38$ & $83.33$ & --      & $82.81$ \\
$6$ & $3$ & $0.5413$ & $0.0000$ & $0.4587$ & $77.67$ & $0.00$  & $78.77$ & $78.17$ \\
\bottomrule
\end{tabular}
\end{table*}

Table~\ref{tab:alloc_reliability} shows that the selected
six-dimensional three-candidate configuration contains one candidate branch
with zero allocated source mass. Thus, this case should be interpreted as a
three-candidate branch set in which the allocation rule effectively prunes the
second candidate branch. The reported overall branch-classification accuracy is
the interval-weighted value $\sum_k p_k q_k$, not the unweighted average over
candidate branches. This distinction is important for the $n=6$, $m=3$ case
because the second candidate branch has $p_2=0$.

For even ambient dimension $n$, the branch geometry admits a convenient radius-vector interpretation. Writing $N=n/2$, the radii vector of branch $k$ is $\mathbf{r}_k=(r_{1,k},\ldots,r_{N,k})\in\mathbb{R}^{N}$, $\|\mathbf{r}_k\|^2=\sum_{i=1}^{N} r_{i,k}^2 = 1$, so that $\mathbf{r}_k$ lies on the unit sphere $\mathbb{S}^{N-1}$. Thus, once the frequency parameters are fixed, selecting multiple branch curves may be viewed as selecting multiple points on $\mathbb{S}^{N-1}$, restricted to the positive orthant by the nonnegativity of the radii. In this sense, the multi-curve design problem reduces to a spherical-code-like packing problem in radius-vector space: one seeks branch radii vectors that are sufficiently well separated to support reliable classification while still yielding geometrically efficient curves.

This viewpoint is particularly useful for interpreting the low-dimensional cases considered in this paper. For $n=4$, one has $N=2$, so the admissible radii vectors lie on $\mathbb{S}^{1}$ and may be parameterized by a single angle $\phi$. For $n=6$, one has $N=3$, so the admissible radii vectors lie on $\mathbb{S}^{2}$ and may be parameterized by the angular pair $(\theta,\phi)$. More generally, for even $n$, the branch-design problem may be viewed as the selection of points on $\mathbb{S}^{n/2-1}$, followed by evaluation of the resulting geometric metrics, classification reliability, and end-to-end SDR.

It is useful to separate multi-curve design into two stages. The first stage
selects the branch curves. The second stage determines how the source interval
is divided among the selected branches. In the geometry stage, the objective is
to find a branch set with favorable geometric scores and sufficient separation.
In the allocation stage, the source interval is assigned according to the
resulting per-branch utilities. Sections~\ref{ss:allocation},
\ref{ss:dens_only}, and~\ref{ss:eq_interval} focus on the allocation stage,
where the selected branches are already fixed and the remaining task is to
divide the source interval across them.

\subsection{Classification-aware interval allocation}
\label{ss:allocation}

Once the branch curves have been selected, the next step is to determine how
the source interval should be divided among them. Consider $m$ candidate branch
curves in $\mathbb{R}^n$. The source interval $[0,1)$ is partitioned into
subintervals of lengths $p_1,p_2,\ldots,p_m$, with $p_k\ge 0$ and
$\sum_{k=1}^m p_k=1$. Branches with $p_k>0$ are active source-carrying
branches, while branches with $p_k=0$ are pruned by the allocation rule. The
$k$th active subinterval is mapped to branch $k$. Let $\delta_k$ denote the
tube-density measure associated with branch $k$, and let
$q_k \triangleq \Pr(\widehat{K}=k \mid K=k)=1-\epsilon_k$ denote the
corresponding probability of correct branch classification, where $\epsilon_k$
is the misclassification probability for branch $k$, that is,
$\epsilon_k=\Pr(\widehat K\ne k\mid K=k)$.

To account jointly for geometric efficiency and branch-identification
reliability, we allocate the interval lengths proportionally to the effective
branch weights $\delta_k q_k$. Under the adopted approximation that branch
utility is proportional to the product of tube density and correct-classification
probability, the resulting interval allocation is
\begin{equation}
\label{eq:interval_alloc}
    p_k
    =
    \frac{\delta_k q_k}{\sum_{\ell=1}^m \delta_\ell q_\ell}
    =
    \frac{\delta_k(1-\epsilon_k)}
    {\sum_{\ell=1}^m \delta_\ell(1-\epsilon_\ell)},
    \qquad k=1,\ldots,m.
\end{equation}
Thus, each branch is weighted by both its tube density and its probability of
correct classification. In the special case of two branches, the allocation
rule becomes
\[
p_1=
\frac{\delta_1(1-\epsilon_1)}
{\delta_1(1-\epsilon_1)+\delta_2(1-\epsilon_2)},
\qquad
p_2=1-p_1.
\]
If the source is parameterized over $[0,2\pi)$, the corresponding split point
is $\theta=2\pi p_1$.

The rule in Eq.~\eqref{eq:interval_alloc} is a classification-aware extension
of density-based source allocation. A branch with larger tube density can
support finer along-curve resolution and improved noise tolerance. In a
multi-curve receiver, however, this geometric advantage is useful only if the
branch can also be identified reliably before along-curve decoding. The factor
$\delta_k$ captures the geometric contribution, and
$q_k=1-\epsilon_k$ captures branch-identification reliability. Their product
defines the effective branch utility used for proportional allocation. The rule
should therefore be viewed as a geometry- and reliability-aware allocation
surrogate, not as a finite-\SNR{} optimality claim.

For the case $n=4$, Fig.~\ref{fig:accu_vs_phi} visualizes the sequential branch-selection procedure. In the $m=2$ sweep, the second-curve candidate is varied over $\phi$, and the plotted interval fraction is $P_2^{(2)}(\phi)$; the corresponding starting-branch fraction is $P_1^{(2)}(\phi)=1-P_2^{(2)}(\phi)$. In the $m=3$ sweep, the second curve is fixed while the third-curve candidate is varied. Since the interval allocation is recomputed for each candidate, the figure shows both the fixed second-branch fraction $P_2^{(3)}(\phi)$ and the third-candidate fraction $P_3^{(3)}(\phi)$, with $P_1^{(3)}(\phi)=1-P_2^{(3)}(\phi)-P_3^{(3)}(\phi)$. Together, these curves illustrate the tradeoff between system \SDR{}, geometric admissibility, and branch-identification reliability that underlies the proposed allocation rule.

\subsection{Density-only allocation rule}
\label{ss:dens_only}

A useful special case of Eq.~\eqref{eq:interval_alloc} arises when the branch-classification reliabilities are identical, i.e., $\epsilon_1=\epsilon_2=\cdots=\epsilon_m$. Then the interval allocation reduces to $p_k=\frac{\delta_k}{\sum_{\ell=1}^m \delta_\ell}$, $k=1,\ldots,m$, which is the density-only allocation rule. In this regime, the interval division depends only on the geometric efficiency of the branch curves, as measured by their tube densities, and no longer distinguishes among branches through classification reliability.

\begin{algorithm}[t!]
\caption{Sequential C3M candidate-branch selection for $n\in\{4,6\}$ and
$m\in\{2,3\}$.}
\label{alg:c3m_design}
\textbf{Input:} Design case $(n,m)$ and the optimal C3T curve
from~\cite[Table I]{AndersC3T}\\
\textbf{Output:} Selected candidate branches
$\mathcal B_m=\{\mathrm{Curve}\;1,\ldots,\mathrm{Curve}\;m\}$, selected
parameters $\{\psi_j^\star\}_{j=2}^{m}$, interval allocation
$\{p_k\}_{k=1}^{m}$, and final performance metrics
\begin{algorithmic}[1]
\State Initialize $\mathcal B_1$ with the optimal C3T branch
\State Compute $\delta_1$ and $L_1$ for the starting branch
\For{$j=2$ \textbf{to} $m$}
    \State $SDR_{\max}\gets -\infty$
    \If{$n=4$}
        \State $\Psi \gets$ finite sweep grid for $\phi\in[0^\circ,90^\circ]$
    \Else
        \State $\Psi \gets$ finite sweep grid for
        $(\theta,\phi)\in[0^\circ,90^\circ]^2$
    \EndIf
    \For{\textbf{each} $\psi\in\Psi$}
        \State Form the candidate branch set
        $\mathcal B_j(\psi)=\mathcal B_{j-1}\cup\{\mathrm{Curve}\;j(\psi)\}$
        \State Compute $\delta_k(\psi)$ and $L_k(\psi)$ for
        $k=1,\ldots,j$
        \State Estimate branch reliabilities
        $q_k(\psi)=1-\epsilon_k(\psi)$ for $k=1,\ldots,j$
        \State Compute interval fractions
        $p_k(\psi)\propto\delta_k(\psi)q_k(\psi)$ for $k=1,\ldots,j$
        \State Train and evaluate the classifier and branch-conditioned
        decoders for $\mathcal B_j(\psi)$
        \State Compute the end-to-end $\SDR(\psi)$
        \If{$\SDR(\psi)>SDR_{\max}$}
            \State $SDR_{\max}\gets \SDR(\psi)$
            \State $\psi_j^\star\gets\psi$
            \State $\mathcal B_j^\star\gets\mathcal B_j(\psi)$
            \State Store the corresponding allocation and performance metrics
        \EndIf
    \EndFor
    \State $\mathcal B_j\gets\mathcal B_j^\star$
\EndFor
\State Re-evaluate the final selected system $\mathcal B_m$ and record the
reported \SDR{}, inference accuracy, interval allocation, and branch
reliabilities
\State \Return $\mathcal B_m$, $\{\psi_j^\star\}_{j=2}^{m}$,
$\{p_k\}_{k=1}^{m}$, and the final performance metrics
\end{algorithmic}
\end{algorithm}
\subsection{Equal-splitting allocation rule}
\label{ss:eq_interval}

A useful baseline is obtained by dividing the source interval uniformly across
all branches. For a source $S\sim\mathcal{U}(0,1)$, the source variance is
$\sigma_\mathbb S^2=1/12$. If the source interval is partitioned uniformly, then
\[
I_k=\left[\frac{k-1}{m},\frac{k}{m}\right),
\qquad
k=1,\ldots,m,
\]
so that $p_k=1/m$. Conditioned on $S\in I_k$, the source is uniform on
$I_k$, and the conditional variance within each interval is
\[
\mathrm{Var}(S\mid S\in I_k)=\frac{1}{12m^2}.
\]
This conditional variance reduction is a property of the source partition and
does not by itself imply an end-to-end $m^2$ information-theoretic improvement,
since the branch index must still be inferred at the receiver. This
equal-splitting construction provides a natural allocation baseline against
which the classification-aware allocation rule in Section~\ref{ss:allocation}
can be compared.

Equal splitting also serves as the natural allocation rule when the branches are
geometrically identical. In particular, if all branches have the same geometric
score and the same branch-reliability score, then a geometry- and
classification-aware allocation assigns equal mass to each branch. Thus, in the
absence of branch-dependent geometric or classification advantages, the uniform
choice $p_k=1/m$ is the baseline allocation.

\begin{figure*}[b!]
    \centering
    \includegraphics[width=\textwidth]{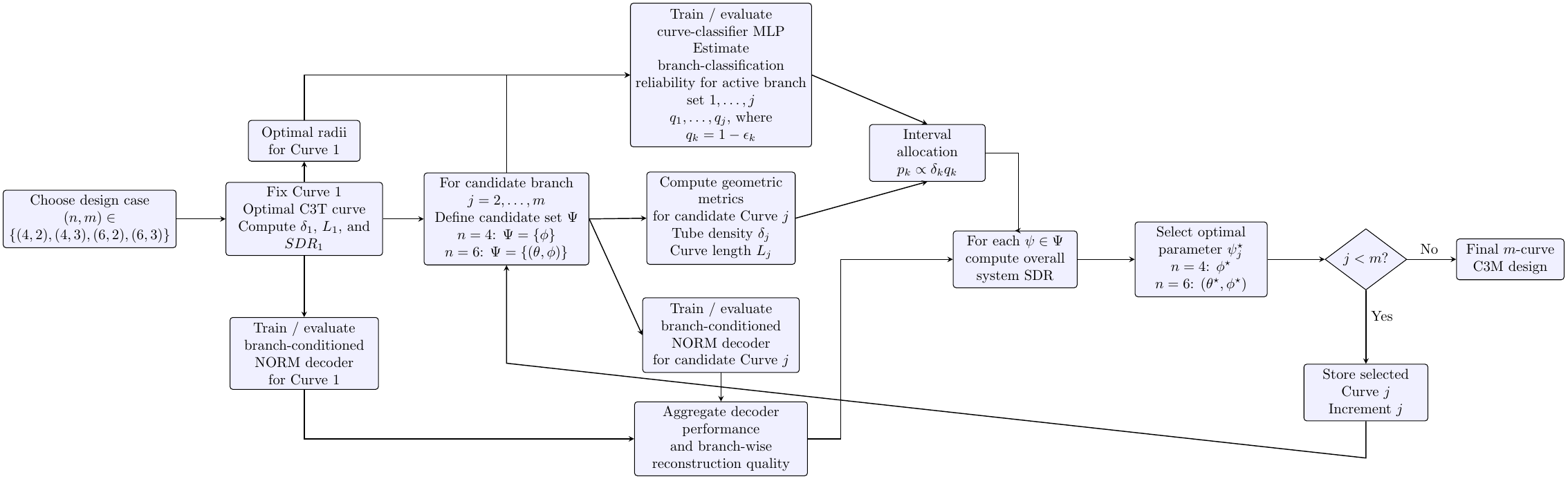}
    \caption{Multi-curve selection procedure for $(n,m)\in\{(4,2),(4,3),(6,2),(6,3)\}$. At each branch-selection stage, a candidate parameter set $\Psi$ is defined, where $\Psi=\{\phi\}$ for $n=4$ and $\Psi=\{(\theta,\phi)\}$ for $n=6$. Geometric quantities and branch-classification reliability determine the interval allocation, while decoder performance is evaluated in parallel. These quantities are then combined to select the parameter choice that maximizes the end-to-end SDR.}
    \label{fig:design}
\end{figure*}
\subsection{Algorithmic selection procedure}
\label{ss:workflow}

The proposed C3M multi-curve selection procedure combines geometric branch construction, classification-aware interval allocation, and branch-conditioned decoder evaluation, see Fig.~\ref{fig:design} and Algorithm~\ref{alg:c3m_design}. The first branch is fixed to the optimal single-curve C3T design from~\cite[Table I]{AndersC3T}, and the remaining branches are selected sequentially. For each branch-selection stage, a candidate parameter set $\Psi$ is defined. For $n=4$, $\Psi$ consists of values of $\phi$ according to Eq.~\eqref{eq:radii_n4}, while for $n=6$, $\Psi$ consists of angular pairs $(\theta,\phi)$ according to Eq.~\eqref{eq:radii_n6}. For each candidate parameter choice in $\Psi$, the geometric branch quantities and classification reliability are evaluated, the interval fractions are computed, and the decoder performance is assessed. Among the geometrically admissible candidates, the parameter choice yielding the largest end-to-end \SDR{} is selected as the next branch. This procedure is repeated until the desired multi-curve design with $m$ branches is obtained.

For the $n=4$ case in Fig.~\ref{fig:phi_block}, the second-curve candidate is
swept over $\phi$, yielding the selected second branch at $\phi=47^\circ$ for
the final NORM C3M configuration. This second curve is then fixed, and a third-curve candidate is swept over $\phi$; among the admissible third-curve candidates, the final branch is selected according to the highest system \SDR{}. Fig.~\ref{fig:sdr_vs_phi} shows the resulting system \SDR{} and tube-density trends, while Fig.~\ref{fig:accu_vs_phi} shows the corresponding classification accuracy and interval-fraction curves. The interval-fraction curves are denoted by $P_i^{(m)}(\phi)$, where $i$ is
the candidate-branch index and $m$ is the number of candidate branches in the sweep. Thus, the $m=2$ sweep shows $P_2^{(2)}(\phi)$, while the $m=3$ sweep shows both the recomputed second-branch fraction $P_2^{(3)}(\phi)$ and the third-candidate fraction $P_3^{(3)}(\phi)$.

Although the above description focuses on the cases $n\in\{4,6\}$ for concreteness, the same workflow extends directly to higher even dimensions. In general, the radii vector in $\mathbb{R}^{n/2}$ is parameterized by $n/2-1$ angular variables in the positive orthant of the unit sphere, and the corresponding candidate set is denoted by $\Psi$. For example, the case $n=8$ requires three angular parameters, so that each candidate branch corresponds to one element of a three-dimensional parameter set $\Psi$. Once $\Psi$ is specified, the remainder of the procedure is unchanged: for each candidate $\psi\in\Psi$, the geometric branch quantities and classification reliability are evaluated, the interval fractions are computed via the classification-aware allocation rule, the branch-conditioned decoder performance is assessed, and the final branch is selected by maximizing the resulting end-to-end SDR. In this way, the proposed selection procedure provides a unified framework for C3M design in arbitrary even dimensions.

The ambient dimension $n$ alone does not impose a hard upper bound on the
number of admissible branch curves. Without robustness constraints, arbitrarily
many distinct curves can be placed in the ambient sphere by allowing the
corresponding tubes or inter-curve separations to become sufficiently small. A
meaningful design limit appears only after imposing geometric robustness
requirements, including a minimum tube radius, a minimum inter-branch
separation, or a target branch-classification reliability. The practical
multi-curve design problem is therefore a constrained packing problem. The
number of branches must be balanced against tube robustness, branch
separability, and decoding performance.

\begin{remark}[Sequential search complexity and design rationale]
One could in principle perform a full unordered two-curve search over
$\phi_1,\phi_2\in\{1^\circ,\ldots,89^\circ\}$ in the four-dimensional case.
Excluding identical pairs, this gives the binomial coefficient $\binom{89}{2}
=
\frac{89\cdot 88}{2}
=
3916$, i.e., the number of unordered distinct pairs. By contrast, the sequential
procedure fixes the first branch to the optimal C3T curve and sweeps only the
second branch, reducing the second-curve selection stage to $89$ candidates, a
factor of approximately $44$. This reduction is significant because each candidate is not only a geometric
lookup because it requires interval allocation, branch-classifier evaluation, decoder
training/evaluation, and end-to-end \SDR{} estimation under Monte Carlo
sampling. 

Moreover, the cost grows combinatorially when adding further branches. On the same four-dimensional grid, an unordered three-candidate exhaustive search would require $\binom{89}{3}=113564$ distinct triples before accounting for training seeds, \SNR{} sweeps, and decoder hyperparameter choices.

The sequential strategy is therefore not merely a computational convenience. It
is a scalable code-design rule that anchors the construction at the known
optimal single-curve C3T design and then measures the marginal benefit of adding
each new branch. This makes the procedure reproducible, extensible to higher
dimensions and larger branch counts, and aligned with the practical objective of
constructing finite-dimensional multi-curve analog codes without an intractable
global combinatorial search.
\hfill$\blacksquare$
\end{remark}

\begin{figure*}[t!]
\centering
\resizebox{0.98\textwidth}{!}{%
\begin{tikzpicture}[
    >=Latex,
    font=\small,
    node distance=6mm and 5mm,
    box/.style={
        draw=blue!70!black,
        rounded corners=2pt,
        thick,
        align=center,
        fill=blue!3,
        inner sep=4pt,
        minimum height=12mm
    },
    smallbox/.style={
        draw=blue!70!black,
        rounded corners=2pt,
        thick,
        align=center,
        fill=blue!3,
        inner sep=4pt,
        minimum height=10mm
    },
    groupbox/.style={
        draw=blue!70!black,
        dashed,
        rounded corners=3pt,
        thick,
        inner sep=7pt
    },
    note/.style={
        draw=blue!70!black,
        dashed,
        rounded corners=2pt,
        align=center,
        fill=blue!3,
        inner sep=4pt
    },
    line/.style={->, thick}
]


\node[box, text width=25mm] (recv)
{Received vector\\[1mm]
$\mathbf{Y}\in\mathbb{R}^n$};

\node[box, text width=28mm, above=9mm of recv] (branch)
{Estimated branch\\[1mm]
$\widehat{K}=C(\mathbf{Y})$};

\node[box, text width=28mm, right=9mm of recv] (cond)
{Branch conditioning\\[1mm]
select $d_{\widehat K}$};

\node[box, text width=40mm, right=6mm of cond] (lift)
{Fourier-feature lifting\\[1mm]
$\Gamma_{\widehat K}(\mathbf{Y})$\\[1mm]
$\big[\sin(2\pi\mathbf{Y}\mathbf{B}_{\widehat K}),
\cos(2\pi\mathbf{Y}\mathbf{B}_{\widehat K})\big]$};

\node[box, text width=30mm, right=6mm of lift] (net)
{Residual / dense network\\[1mm]
$F_{\widehat K}(\Gamma_{\widehat K}(\mathbf{Y}))$};

\node[box, text width=34mm, right=6mm of net] (head)
{Scalar regression head\\[1mm]
$\widehat{Z}
=
q_{\widehat K}\!\left(
F_{\widehat K}(\Gamma_{\widehat K}(\mathbf{Y}))
\right)$};

\node[box, text width=29mm, right=6mm of head] (src)
{Source reconstruction\\[1mm]
$\widehat{S}
=
a_{\widehat K-1}
+
p_{\widehat K}\widehat{Z}$};

\draw[line] (recv) -- (cond);
\draw[line] (branch) -- (cond);
\draw[line] (cond) -- (lift);
\draw[line] (lift) -- (net);
\draw[line] (net) -- (head);
\draw[line] (head) -- (src);


\node[smallbox, text width=25mm, below=18mm of cond] (resid)
{Angular residual\\[1mm]
$\Delta=2\pi(\widehat{Z}-Z)$};

\node[smallbox, text width=31mm, right=6mm of resid] (wrap)
{Circular wrapping\\[1mm]
$\Delta_{\mathrm{circ}}
=
\operatorname{atan2}(\sin\Delta,\cos\Delta)$};

\node[smallbox, text width=25mm, right=6mm of wrap] (mag)
{Angular magnitude\\[1mm]
$\delta=|\Delta_{\mathrm{circ}}|$};

\node[smallbox, text width=34mm, right=6mm of mag] (huber)
{Robust Huber loss\\[1mm]
$\displaystyle
L(\delta)=
\begin{cases}
\frac{1}{2}\delta^2, & \delta\le 1,\\[1mm]
\delta-\frac{1}{2}, & \delta>1 .
\end{cases}$};

\node[note, text width=25mm, right=7mm of huber] (note1)
{Quadratic for small errors,\\
linear for large errors,\\
robust to impulsive outliers};

\draw[line] (resid) -- (wrap);
\draw[line] (wrap) -- (mag);
\draw[line] (mag) -- (huber);
\draw[line, dashed] (note1.west) -- (huber.east);

\draw[line] (head.south) |- (huber.north);


\node[groupbox, fit=(cond)(lift)(net)(head)(src), inner sep=7pt] (topbox) {};

\node[
    fill=white,
    text=blue!70!black,
    font=\bfseries,
    inner sep=2pt,
    anchor=south east
] at ([xshift=-2mm,yshift=1mm]topbox.north east)
{NORM-inspired branch-conditioned spectral decoder};

\node[groupbox, fit=(resid)(wrap)(mag)(huber), inner sep=7pt] (botbox) {};

\node[
    fill=white,
    text=blue!70!black,
    font=\bfseries,
    inner sep=2pt,
    anchor=south west
] at ([xshift=2mm,yshift=-6mm]botbox.north west)
{Training objective};

\end{tikzpicture}%
}
\caption{NORM-inspired branch-conditioned spectral decoder used in the proposed
C3M receiver. The curve classifier first estimates the transmitted branch
$\widehat K=C(\mathbf Y)$. The estimated branch selects one decoder from a bank
of branch-conditioned decoders. The selected decoder takes the received ambient
vector $\mathbf Y\in\mathbb R^n$ as input, applies a Fourier-feature lifting
$\Gamma_{\widehat K}(\mathbf Y)$, processes the lifted representation through
residual and dense layers, and outputs a normalized local-coordinate estimate
$\widehat Z\in[0,1)$. The source estimate is then obtained using the inverse
interval map. During training, the local-coordinate error is converted to an
angular residual, wrapped onto the principal interval, and penalized using a
robust Huber loss.}
\label{fig:norm}
\end{figure*}
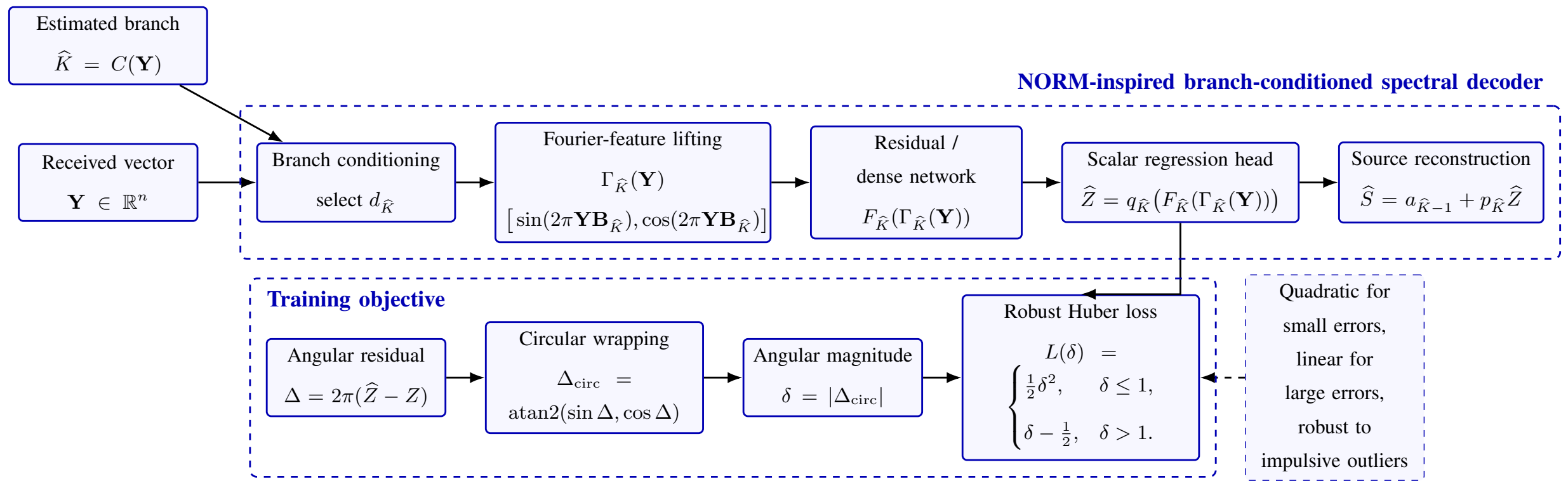

\section{NORM-Inspired Branch-Conditioned Decoder}
\label{sec:norm}

As motivated in Section~\ref{ss:nop}, decoding on curved signal geometries
should ideally exploit the geometric structure of the underlying branch set.
The receiver described in Section~\ref{sec:sys} is hierarchical: after branch
classification, decoding is performed conditionally on the estimated branch
index. This motivates a branch-dependent decoder rather than a single regressor
shared across all branch curves.

The decoder used in this work is motivated by the spectral viewpoint underlying
neural operators on Riemannian manifolds, but it is not a full
Laplace--Beltrami neural-operator implementation. The implemented decoder does
not construct a sampled function on the branch manifold, compute a
nearest-point projection onto the curve, or form an explicit
Laplace--Beltrami coefficient field. Instead, after the classifier selects
branch $\widehat K=k$, the corresponding branch-conditioned decoder takes the
ambient noisy channel vector $\mathbf Y\in\mathbb R^n$ as its input and
estimates the local coordinate on that branch:
\[
    \widehat Z = d_{\widehat K}(\mathbf Y),
    \qquad
    d_k:\mathbb R^n\to[0,1).
\]

The overall decoder structure is illustrated in Fig.~\ref{fig:norm}. The
estimated branch $\widehat K$ selects one decoder from a bank of
branch-conditioned decoders. Each decoder applies a Fourier-feature lifting of
the received vector, followed by residual and dense layers, and produces a
scalar local-coordinate estimate. During training, the estimate is compared
with the true local coordinate using the circular Huber loss described in
Section~\ref{ss:Huber}.

\subsection{Branch-conditioned decoder architecture}
\label{ss:decoder}

For each branch $k\in\{1,\ldots,m\}$, training samples are generated according
to
\begin{equation}
    \label{eq:branch_decoder_training_samples}
    \mathbf Y
    =
    \gamma_k(z)+\mathbf N,
    \qquad
    z\sim\mathcal U[0,1),
\end{equation}
where $\gamma_k(z)$ is the $k$th branch embedding and $\mathbf N$ is the
channel noise. The branch-conditioned decoder is therefore a regression map
\begin{equation}
    \label{eq:branch_decoder_map}
    d_k:\mathbb R^n\rightarrow [0,1),
    \qquad
    \widehat Z=d_k(\mathbf Y).
\end{equation}
The term branch-conditioned means that a separate decoder is trained for each
candidate branch and that the decoder indexed by the detected branch
$\widehat K$ is used at inference time. Thus, the overall receiver may be
viewed as a bank of decoders $\{d_1,d_2,\ldots,d_m\}$, together with the branch classifier from Section~\ref{ss:receiver}. Given $\widehat K=k$, the receiver applies $d_k$ to the same received vector
$\mathbf Y$ and then maps the estimated local coordinate back to the
corresponding source interval.

In the implementation used here, each branch decoder begins with a
Fourier-feature lifting of the ambient input. Specifically, for branch $k$, the
received vector is mapped to
\begin{equation}
    \label{eq:fourier_feature_lift}
    \Gamma_k(\mathbf Y)
    =
    \big[
    \sin(2\pi \mathbf Y\mathbf B_k),
    \cos(2\pi \mathbf Y\mathbf B_k)
    \big],
\end{equation}
where $\mathbf B_k$ is the Fourier-feature projection matrix used by the
branch-conditioned decoder. The lifted representation is then processed by a
neural network $F_k$ consisting of residual and dense layers, followed by a
scalar readout:
\begin{equation}
    \label{eq:decoder_composition}
    \widehat Z
    =
    d_k(\mathbf Y)
    =
    q_k\!\left(F_k(\Gamma_k(\mathbf Y))\right).
\end{equation}
The readout $q_k$ produces the normalized local-coordinate estimate
$\widehat Z\in[0,1)$. Thus, decoding is performed directly from the ambient
received vector through a branch-conditioned Fourier-feature representation,
rather than through a discretized likelihood field or an explicit
Laplace--Beltrami expansion.

This architecture is NORM-inspired because it uses spectral features to learn
branch-dependent maps associated with geometric signal sets~\cite{Chen2024}.
In the present implementation, the spectral representation is obtained by
Fourier-feature lifting of the received ambient vector and by training a
separate decoder for each branch. After branch classification, the selected
decoder only has to learn the inverse map associated with one branch geometry.
Thus, the method is a branch-conditioned spectral decoder inspired by NORM,
rather than a full Laplace--Beltrami neural-operator implementation. The
implemented receiver architecture is kept fixed across all reported C3M
experiments: the classifier maps $\mathbf y\in\mathbb R^n$ to a candidate
branch, and the selected branch-conditioned decoder applies the
Fourier-feature lifting in Eq.~\eqref{eq:fourier_feature_lift} before residual
and dense layers estimate the local coordinate. The decoder hyperparameters are
summarized in Table~\ref{tab:hyper}, and training uses the circular Huber loss
introduced in Section~\ref{ss:Huber}.
\begin{table}[t!]
\caption{Hyperparameters of the NORM-inspired branch-conditioned spectral decoder.}
\label{tab:hyper}
\centering
\begin{threeparttable}
\begin{tabular}{c|c}
Hyperparameter & Values \\
\hline
Epochs & $100$ \\
Batch size & $128$ \\
Learning rate (LR) & $1e{-}4$ \\
Minimum LR & $1e{-}6$ \\
LR scheduler & ReduceLROnPlateau \\
LR patience & $70$ \\
LR factor & $0.5$ \\
Loss function & Huber \\
Optimizer & AdamW \\
Weight Decay & $1e{-}4$ \\
Number of NORM training examples per curve & $200,000$ \\
Number of NORM evaluation examples per curve  & $50,000$\\
Number of training examples for curve classifier  & $15,000$ \\
Number of evaluation examples for curve classifier & $15,000$\\
\hline
\end{tabular}
\end{threeparttable}
\end{table}

\subsection{Huber loss on the angular error}
\label{ss:Huber}

For neural receivers trained for regression, the mean-squared error (MSE), or
$\ell_2$ loss, is often the default choice because it is smooth and performs
well when residuals are approximately Gaussian. In the present setting, however,
large residuals can be produced by Gaussian-mixture impulses, heavy-tailed
perturbations, branch-confusion events, or difficult regions of the branch
manifold. Since MSE grows quadratically with residual magnitude, such samples
can dominate the optimization and destabilize training.

Robust alternatives such as the mean absolute error (MAE), or $\ell_1$ loss,
reduce the influence of large residuals, but their nonsmooth behavior at the
origin is less attractive for gradient-based optimization. The Huber loss
combines a quadratic penalty for small errors with a linear penalty for large
errors. It therefore preserves local smoothness near the optimum while reducing
sensitivity to outliers. This makes it well matched to branch-conditioned
local-coordinate regression under impulsive or heavy-tailed noise.

The decoder estimates a normalized local coordinate in $[0,1)$, which is mapped
to an angular coordinate on the selected branch. The regression residual should
therefore respect the periodic geometry of the branch parameterization. Let
$\Delta$ denote the difference between the predicted and target angular
coordinates. We define the circular residual by projecting this difference onto
the unit circle and taking the principal angle,
\begin{equation}
\label{eq:circular_residual}
\Delta_{\mathrm{circ}}
=
\operatorname{atan2}\!\bigl(\sin(\Delta),\cos(\Delta)\bigr).
\end{equation}
This is equivalent, up to endpoint convention, to wrapping $\Delta$ into the
principal interval $(-\pi,\pi]$. The angular error magnitude is then
\[
\delta = |\Delta_{\mathrm{circ}}|.
\]

We define the Huber loss on this circular residual as
\begin{equation}
\label{eq:huber_loss}
L_\kappa(\delta)=
\begin{cases}
\frac{1}{2}\delta^2, & \delta \le \kappa,\\[3pt]
\kappa\left(\delta-\frac{1}{2}\kappa\right), & \delta > \kappa,
\end{cases}
\end{equation}
where $\kappa>0$ is the transition threshold. In the experiments, we use
$\kappa=1$. Equivalently, this special case can be implemented through
$q=\min(\delta,1)$, $\ell=\delta-q$, and
\[
L_1(\delta)=\frac{1}{2}q^2+\ell .
\]
The constant offset in Eq.~\eqref{eq:huber_loss} ensures continuity at the
transition point, and the loss is averaged over the training batch.

Together, circular residual wrapping and Huber regression define a robust
training objective for branch-conditioned local-coordinate estimation. This
objective is well matched to C3M because large residuals can arise from
branch-confusion events as well as from impulsive or heavy-tailed noise
realizations. The Gaussian-mixture noise model in Eq.~\eqref{eq:gmm_noise}
also connects directly to Huber's robust-estimation viewpoint
~\cite{Huber1964}. Most observations follow a nominal distribution, while a
small fraction behave as outliers. This motivates the use of robust loss
functions such as the Huber loss~\cite{Zhao2024}.

\section{Analysis}
\label{sec:analysis}

This section develops the analytical model and reference quantities used for
the proposed C3M construction. We first characterize the induced output
distribution under fixed isotropic AWGN and use the resulting posterior
representation to define MAP and MMSE estimators. We then express the mutual
information induced by the C3M input distribution and relate it to
rate--distortion theory and channel-capacity references.

The remainder of the section studies the main multi-curve effects. These
include branch-classification reliability, capacity-surrogate allocation,
geometry-only allocation, end-to-end and branch-known \SDR{} references,
non-uniform source allocation, geometric branch-selection behavior, and
heuristic upper bounds on the number of robustly admissible branch curves.

\subsection{Induced output distribution}
\label{ss:pdf}

Under the C3M encoder, the channel input is a random point on one of the $m$ branch curves. In the main Gaussian-noise model considered in this paper, the channel output is $\mathbf{Y}=\mathbf{X}+\mathbf{N}$,  where $\mathbf{N}\sim\mathcal{N}(\mathbf{0},\sigma_n^2\mathbf{I}_n)$, with $\sigma_n^2$ fixed independently of the branch index, radius parameters, and candidate curve. Since the encoder is memoryless and generally nonlinear, the proposed C3M construction induces a non-Gaussian mixture distribution at the receiver even under isotropic Gaussian channel noise.

Let
\begin{equation}
\label{eq:gaussian_density_iso}
    \phi_n\!\left(\mathbf{y};\boldsymbol{\mu},\sigma_n^2\mathbf{I}_n\right)
    =
    \frac{1}{(2\pi\sigma_n^2)^{n/2}}
    \exp\!\left(
        -\frac{\|\mathbf{y}-\boldsymbol{\mu}\|^2}{2\sigma_n^2}
    \right)
\end{equation}
denote the $n$-dimensional isotropic Gaussian density with mean $\boldsymbol{\mu}$ and covariance $\sigma_n^2\mathbf{I}_n$. The channel law conditioned on the branch index $K=k$ and local coordinate $Z=z$ is therefore $p_{\mathbf{Y}|K,Z}(\mathbf{y}|k,z)
    =
    \phi_n\!\left(\mathbf{y};\gamma_k(z),\sigma_n^2\mathbf{I}_n\right)$.
Because $S\sim\mathcal{U}(0,1)$ and the local normalization on each branch interval is affine, the conditional distribution of the local coordinate satisfies $p_{Z|K}(z|k)=1$, $z\in[0,1)$, for every $k=1,\ldots,m$ and zero otherwise.

For example, in the $n=4$ case, a branch with two radius pairs may be written as
\begin{equation}
\label{eq:gamma_4d_example}
\gamma_k(z)
=
\begin{bmatrix}
r_{k,1}\cos(2\pi\omega_{k,1}z)\\
r_{k,1}\sin(2\pi\omega_{k,1}z)\\
r_{k,2}\cos(2\pi\omega_{k,2}z)\\
r_{k,2}\sin(2\pi\omega_{k,2}z)
\end{bmatrix},
\qquad z\in[0,1).
\end{equation}
Writing $\mathbf y=[y_1,y_2,y_3,y_4]^T$, the fixed isotropic AWGN model implies that the received coordinates are conditionally independent given $(K,Z)=(k,z)$. Hence,
\begin{equation}
\label{eq:py_given_kz_4d_product}
\begin{aligned}
p_{\mathbf Y|K,Z}(\mathbf y|k,z)
&=
\prod_{i=1}^4
\frac{1}{\sqrt{2\pi\sigma_n^2}}
\exp\!\left(
-\frac{(y_i-\gamma_{k,i}(z))^2}{2\sigma_n^2}
\right) =
\frac{1}{(2\pi\sigma_n^2)^2}
\exp\!\left(
-\frac{\|\mathbf y-\gamma_k(z)\|^2}{2\sigma_n^2}
\right).
\end{aligned}
\end{equation}
After marginalizing over the latent coordinate $Z$, however, the received coordinates are generally dependent because they share the same underlying curve parameter.

Thus, conditioned on $K=k$, the source induces a uniform distribution on the normalized local parameter $Z$ along branch $k$. This does not imply that the ambient coordinates of $\gamma_k(Z)$ are uniformly distributed; rather, the uniformity holds in the branch parameter itself, and for constant-speed parametrizations equivalently in normalized arc length. In particular, the coordinate marginals are generally sinusoidal transformations of a uniform parameter and are therefore non-uniform.

Conditioning only on the branch index gives $p_{\mathbf{Y}|K}(\mathbf{y}|k)
    =
    \int_0^1
    \phi_n\!\left(\mathbf{y};\gamma_k(u),\sigma_n^2\mathbf{I}_n\right)\,du$,
where $u\in[0,1)$ is the normalized local coordinate on branch $k$. Averaging over the branch index yields the unconditional output density
\begin{equation}
\label{eq:py_mixture_rev}
\begin{aligned}
    p_{\mathbf{Y}}(\mathbf{y}) &=
    \sum_{k=1}^m p_k\,p_{\mathbf{Y}|K}(\mathbf{y}|k) =
    \sum_{k=1}^m p_k
    \int_0^1
    \phi_n\!\left(\mathbf{y};\gamma_k(u),\sigma_n^2\mathbf{I}_n\right)\,du.
\end{aligned}
\end{equation}
Thus, the receiver distribution is a mixture over the $m$ branch curves, with weights given by the interval-allocation probabilities $p_k$. This representation makes explicit the probabilistic structure seen by the curve classifier and provides the basis for the information-theoretic analysis that follows. A corresponding joint representation is $p_{K,Z,\mathbf{Y}}(k,z,\mathbf{y}) = p_k\,p_{Z|K}(z|k)\,p_{\mathbf{Y}|K,Z}(\mathbf{y}|k,z) =
p_k\,\phi_n\!\left(\mathbf{y};\gamma_k(z),\sigma_n^2\mathbf{I}_n\right)$, for $k=1,\ldots,m$ and $z\in[0,1)$, where we used the fact that $Z|K=k$ is uniform on $[0,1)$. For a single branch, the induced output density reduces to the corresponding one-manifold Gaussian mixture obtained by averaging the Gaussian channel law along that branch.

\begin{remark}[Unit-circle output law]
As a simple special case, consider a single unit-circle branch in $\mathbb{R}^2$ with
$\gamma(a)=[\cos a,\sin a]^T$, with angular parameter $a\sim\mathcal{U}[0,2\pi)$, and isotropic AWGN with variance
$\sigma_n^2$ per coordinate. Writing $\mathbf y=[y_1,y_2]^T$, the induced output density is
\begin{equation}
\label{eq:unit_circle_output_density}
\begin{aligned}
p_{\mathbf Y}(\mathbf y)
&=
\frac{1}{2\pi}\int_0^{2\pi}
\frac{1}{2\pi\sigma_n^2}
\exp\!\left(
-\frac{\|\mathbf y-\gamma(a)\|^2}{2\sigma_n^2}
\right)\,da =
\frac{1}{2\pi\sigma_n^2}
\exp\!\left(
-\frac{\|\mathbf y\|^2+1}{2\sigma_n^2}
\right)
I_0\!\left(\frac{\|\mathbf y\|}{\sigma_n^2}\right),
\end{aligned}
\end{equation}
where $I_0(\cdot)$ is the modified Bessel function of the first kind. This example illustrates that even a single smooth analog curve induces a non-Gaussian output law. Moreover, although the two output coordinates are uncorrelated by symmetry in this example, they are not independent, since both are coupled through the same latent curve parameter.
\hfill$\blacksquare$
\end{remark}

\subsection{Posterior representation and Bayesian estimators}
\label{ss:posterior}

By Bayes' rule, the posterior density of the branch index and local coordinate given the observation $\mathbf{y}$ is
\begin{equation}
\label{eq:posterior_kz}
\begin{aligned}
p_{K,Z|\mathbf{Y}}(k,z\mid \mathbf{y})
&=
\frac{p_k\,\phi_n\!\left(\mathbf{y};\gamma_k(z),\sigma_n^2\mathbf{I}_n\right)}
     {p_{\mathbf{Y}}(\mathbf{y})},\\
&\qquad k=1,\ldots,m,\quad z\in[0,1).
\end{aligned}
\end{equation}
Here $p_{\mathbf{Y}}(\mathbf{y})$ is the marginal likelihood of the observation $\mathbf{y}$, i.e., the normalizing constant that ensures the posterior density integrates to one over all admissible branch-index and local-coordinate pairs. Equivalently, for fixed $\mathbf y$,
\begin{equation}
\label{eq:posterior_proportional}
p_{K,Z|\mathbf Y}(k,z|\mathbf y)
\propto
p_k\,p_{\mathbf Y|K,Z}(\mathbf y|k,z),
\end{equation}
so the posterior is proportional to likelihood times prior, with the branch probabilities $\{p_k\}$ serving as priors over the discrete branch index. The normalization is over both the discrete branch index and the continuous local coordinate,
\begin{equation}
\label{eq:posterior_normalization}
p_{\mathbf Y}(\mathbf y)
=
\sum_{k=1}^m p_k\int_0^1
p_{\mathbf Y|K,Z}(\mathbf y|k,z)\,dz .
\end{equation}
This posterior representation makes explicit the Bayesian inference problem induced by the C3M encoder. In particular, the MAP estimator is 
\begin{equation}
\label{eq:map_kz}
(\widehat{K}_{\mathrm{MAP}},\widehat{Z}_{\mathrm{MAP}})
=
\arg\max_{k,\;z\in[0,1)}
p_{K,Z|\mathbf{Y}}(k,z\mid \mathbf{y}).
\end{equation}
For fixed isotropic AWGN, this is equivalent to minimizing a branch-dependent Euclidean-distance criterion with the branch probabilities $\{p_k\}$ entering as prior weights:
\begin{equation}
\label{eq:map_kz_isotropic}
(\widehat{K}_{\mathrm{MAP}},\widehat{Z}_{\mathrm{MAP}})
=
\arg\min_{k,\;z\in[0,1)}
\left\{
\|\mathbf{y}-\gamma_k(z)\|^2
-2\sigma_n^2\log p_k
\right\}.
\end{equation}
When the branch probabilities are equal, the prior term is constant across branches and the MAP rule reduces to nearest-curve decoding in Euclidean distance.

The MMSE source estimate is
\begin{equation}
\label{eq:mmse_est}
\widehat{S}_{\mathrm{MMSE}}(\mathbf{y})
=
\mathbb{E}[S\mid \mathbf{Y}=\mathbf{y}].
\end{equation}
The estimator in Eq.~\eqref{eq:mmse_est} is optimal in the mean-squared-error sense, i.e., it minimizes $\mathbb{E}[(S-g(\mathbf{Y}))^2]$ over all measurable estimators $g(\mathbf{Y})$. Its optimality under mean-squared error is recalled in Appendix~\ref{app:MMSE}.

Since $S$ is in one-to-one correspondence with $(K,Z)$ through the interval map $f$, both estimators may be interpreted equivalently in the source domain or in the branch-coordinate domain. For a Gaussian model with general covariance $\mathbf{\Sigma}_n$, the same posterior representation applies after replacing $\sigma_n^2\mathbf{I}_n$ by $\mathbf{\Sigma}_n$, and thus the MAP distance term becomes the Mahalanobis distance $(\mathbf y-\gamma_k(z))^T\mathbf{\Sigma}_n^{-1}(\mathbf y-\gamma_k(z))$.

\subsection{Mutual information induced by C3M}
\label{ss:mi}

Since the source symbol $S$ is in one-to-one correspondence with the pair $(K,Z)$ through the bijective interval map $f$, the mutual information between the source and the received vector satisfies
\begin{equation}
\label{eq:mi_decomposition_rev}
    I(S;\mathbf{Y})
    =
    I(K,Z;\mathbf{Y})
    =
    I(K;\mathbf{Y}) + I(Z;\mathbf{Y}\mid K).
\end{equation}
The first term, $I(K;\mathbf{Y})$, quantifies the coarse branch-identification information, whereas the second term, $I(Z;\mathbf{Y}\mid K)$, quantifies the fine along-curve information once the transmitted branch is known. This decomposition mirrors the proposed receiver structure, which first performs curve classification and then applies a branch-conditioned decoder.

For the fixed isotropic AWGN channel, the mutual information induced by the C3M encoder may also be written as $I(S;\mathbf{Y}) = h(\mathbf{Y}) - h(\mathbf{N})$, 
where $h(\cdot)$ denotes differential entropy. Since $\mathbf{N}\sim\mathcal{N}(\mathbf{0},\sigma_n^2\mathbf{I}_n)$, the noise entropy is $h(\mathbf{N})
    =
    \frac{n}{2}\log\!\left(2\pi e\sigma_n^2\right)$.
\begin{remark}[Diagonal Gaussian extension]
As a Gaussian extension, if the $n=4$ noise covariance is diagonal with unequal variances across coordinate pairs,
$\mathbf{\Sigma}_n
=
\operatorname{diag}(\sigma_1^2,\sigma_1^2,\sigma_2^2,\sigma_2^2)$, then the noise entropy becomes
\begin{equation}
\label{eq:noise_entropy_diag_4d}
\begin{aligned}
h(\mathbf N)
&=
\frac{1}{2}\log\!\left((2\pi e)^4\det\mathbf{\Sigma}_n\right) 
=
2\log(2\pi e)+2\log\sigma_1+2\log\sigma_2 .
\end{aligned}
\end{equation}
Equivalently, $h(\mathbf N)
=
\log(2\pi e\sigma_1^2)+\log(2\pi e\sigma_2^2)$.
\hfill$\blacksquare$
\end{remark}
In either case, the remaining task is to evaluate the output entropy induced by the mixture density in Eq.~\eqref{eq:py_mixture_rev}. Equivalently,
\begin{equation}
\label{eq:mi_integral}
\begin{aligned}
I(S;\mathbf{Y})
&=
\sum_{k=1}^m p_k \int_0^1 \int_{\mathbb{R}^n}
p_{\mathbf{Y}|K,Z}(\mathbf{y}|k,z) 
\log
\frac{p_{\mathbf{Y}|K,Z}(\mathbf{y}|k,z)}
     {p_{\mathbf{Y}}(\mathbf{y})}
\,d\mathbf{y}\,dz.
\end{aligned}
\end{equation}
Unless otherwise stated, logarithms in this subsection are natural and the resulting mutual information is measured in nats; conversion to bits is obtained by dividing by $\log 2$.

For numerical evaluation, the mutual information in Eq.~\eqref{eq:mi_integral} may be approximated by Monte Carlo sampling from the joint distribution $p_{K,Z,\mathbf{Y}}$. Specifically, one may draw samples $(K_i,Z_i,\mathbf{Y}_i)$ from the encoder--channel model and estimate $I(S;\mathbf{Y})$ by averaging the corresponding log-likelihood ratios
$\log p_{\mathbf{Y}|K,Z}(\mathbf{Y}_i\mid K_i,Z_i)
-
\log p_{\mathbf{Y}}(\mathbf{Y}_i)$.
\begin{remark}[Single-branch continuous-parameter interpretation]
Conditioned on branch $K=k$, the mutual information between the local coordinate and the received vector can also be written in entropy form as
\begin{equation}
\label{eq:single_branch_mi_entropy}
I(Z;\mathbf Y|K=k)
=
h(\mathbf Y|K=k)-h(\mathbf N),
\end{equation}
where
\begin{equation}
\label{eq:single_branch_output_entropy}
h(\mathbf Y|K=k)
=
-\int_{\mathbb R^n}
p_{\mathbf Y|K}(\mathbf y|k)
\log p_{\mathbf Y|K}(\mathbf y|k)\,d\mathbf y .
\end{equation}
For fixed isotropic AWGN, $h(\mathbf N)=\frac{n}{2}\log(2\pi e\sigma_n^2)$, so for $n=4$ this becomes $h(\mathbf N)=2\log(2\pi e\sigma_n^2)$. The output entropy term must generally be evaluated numerically from the branch-induced mixture density. In the small-noise limit, the posterior over the continuous curve parameter concentrates and $I(Z;\mathbf Y|K=k)$ can grow without bound; in the high-noise limit, the observation becomes independent of $Z$ and $I(Z;\mathbf Y|K=k)\to0$.
\hfill$\blacksquare$
\end{remark}
\begin{remark}[Scalar projection bound]
As a scalar projection example, let $Y_{\mathrm{sc}}=X+N$, where
$X=\cos(2\pi S)$, $S\sim\mathcal{U}(0,1)$, and
$N\sim\mathcal{N}(0,\sigma^2)$ is independent of $S$. Equivalently, with
$A=2\pi S$, one has $A\sim\mathcal{U}[0,2\pi)$ and $X=\cos A$. Since
$Y_{\mathrm{sc}}$ depends on $S$ only through $X$, 
$I(S;Y_{\mathrm{sc}})=I(X;Y_{\mathrm{sc}})=h(Y_{\mathrm{sc}})
-\frac{1}{2}\log(2\pi e\sigma^2)$. This remark uses natural logarithms;
conversion to bits is obtained by dividing by $\log 2$.

The induced scalar output density is
\begin{equation}
\label{eq:scalar_cos_output_density}
p_{Y_{\mathrm{sc}}}(y)
=
\frac{1}{2\pi}\int_0^{2\pi}
\phi_1(y;\cos a,\sigma^2)\,da .
\end{equation}
Thus, equivalently,
\begin{equation}
\label{eq:scalar_cos_mi}
I(S;Y_{\mathrm{sc}})
=
-\int_{\mathbb R}
p_{Y_{\mathrm{sc}}}(y)\log p_{Y_{\mathrm{sc}}}(y)\,dy
-
\frac{1}{2}\log(2\pi e\sigma^2),
\end{equation}
where the entropy integral generally has to be evaluated numerically.

The scalar entropy power inequality~\cite{Stam1959,Dembo1991,Verdu2006}
gives
\begin{equation}
\label{eq:epi_scalar}
e^{2h(Y_{\mathrm{sc}})}
\ge
e^{2h(X)} + e^{2h(N)} .
\end{equation}
The random variable $X=\cos(2\pi S)$ has the arcsine distribution on $[-1,1]$
and differential entropy $h(X)=\log(\pi/2)$, while
$h(N)=\frac{1}{2}\log(2\pi e\sigma^2)$. Substituting these quantities into
Eq.~\eqref{eq:epi_scalar} yields
\begin{equation}
\label{eq:epi_scalar_cos_bound}
I(S;Y_{\mathrm{sc}})
\geq
\frac{1}{2}\log\!\left(1+\frac{\pi}{8e\sigma^2}\right).
\end{equation}

In this scalar example, the limiting behavior is also explicit. As $\sigma^2\to 0$,
the output density converges to the arcsine law induced by $X=\cos(2\pi S)$,
whose differential entropy is $h(X)=\log(\pi/2)$. Hence
\begin{equation}
\label{eq:scalar_low_noise_limit}
I(S;Y_{\mathrm{sc}})
=
h(Y_{\mathrm{sc}})-h(N)
\approx
\log\!\left(\frac{\pi}{2}\right)
-
\frac{1}{2}\log(2\pi e\sigma^2),
\end{equation}
which diverges as $-\frac{1}{2}\log\sigma^2$ in the vanishing-noise limit. This
does not contradict the two-to-one nature of the cosine map: the noiseless
observation does not fully determine $S$, but it does determine the continuous
variable $X=\cos(2\pi S)$ exactly. In the high-noise limit, the mutual
information vanishes, with first-order behavior
\begin{equation}
\label{eq:scalar_high_noise_limit}
I(S;Y_{\mathrm{sc}})
=
\frac{\operatorname{Var}(X)}{2\sigma^2}
+
o(\sigma^{-2})
=
\frac{1}{4\sigma^2}
+
o(\sigma^{-2}),
\end{equation}
since $\operatorname{Var}(\cos(2\pi S))=1/2$. This bound and these limiting
expressions are only scalar projection examples, while the C3M analysis uses the full
vector output distribution in Eq.~\eqref{eq:py_mixture_rev}.
\hfill$\blacksquare$
\end{remark}

It is important to distinguish this quantity from channel capacity. The capacity of the Gaussian channel depends only on the channel law and the input power constraint, whereas Eqs.~\eqref{eq:mi_decomposition_rev}--\eqref{eq:mi_integral} characterize the mutual information induced by the specific distribution generated by the proposed C3M encoder. In what follows, we therefore refer to this quantity as the mutual information induced by the C3M encoder, or equivalently, the achievable information rate under the C3M input distribution.

\subsection{Rate--distortion and channel-capacity references}
\label{ss:rd}

The information-theoretic reference for distortion-limited communication is
obtained by combining the source rate--distortion function with the channel
capacity constraint~\cite{Gallager1968,Cover2006}. Since one source symbol is
transmitted using one $n$-dimensional channel use, Shannon separation implies
that reliable communication with distortion $D$ is achievable whenever
$R(D)\le C_n$, where $R(D)$ is measured in bits per source symbol and $C_n$ is
the channel capacity per $n$-dimensional channel use.

For the fixed isotropic AWGN channel with total input power constraint
$P=\mathrm{E}[\|\mathbf{X}\|^2]=1$ and noise covariance
$\sigma_n^2\mathbf{I}_n$, the capacity of one $n$-dimensional channel use is
\begin{equation}
\label{eq:capacity_n}
C_n
=
\frac{n}{2}\log_2\!\left(1+\frac{P}{n\sigma_n^2}\right)
=
\frac{n}{2}\log_2\!\left(1+\frac{\SNR}{n}\right),
\end{equation}
where, under the normalization adopted in Section~\ref{sec:sys},
$\SNR=P/\sigma_n^2=1/\sigma_n^2$. This capacity serves as an AWGN capacity
reference against which the mutual information induced by the C3M input
distribution can be compared. For the four-dimensional case, the fixed
isotropic AWGN capacity reference specializes to
\[
C_4
=
2\log_2\!\left(1+\frac{P}{4\sigma_n^2}\right).
\]
Under the normalization $P=1$, this becomes
\[
C_4
=
2\log_2\!\left(1+\frac{1}{4\sigma_n^2}\right)
=
2\log_2\!\left(1+\frac{\SNR}{4}\right).
\]
Equivalently, if $\sigma^2=4\sigma_n^2$ denotes the total noise variance across
the four channel dimensions, then $C_4=2\log_2(1+1/\sigma^2)$.

Because the C3M encoder $e\triangleq g\circ f$ is deterministic and bijective
between the source interval $[0,1)$ and the admissible code set $\mathcal X$,
the source symbol $S$ and the transmitted vector $\mathbf X=e(S)$ are in
one-to-one correspondence. Consequently, $I(S;\mathbf Y)=I(\mathbf X;\mathbf
Y)$, since $\mathbf X=e(S)$ is an invertible reparameterization of $S$ on the
C3M codebook. This distinction matters because nonlinearity itself does not
reduce mutual information. Rather, the induced input distribution and its
support determine how close the mapping can come to the AWGN capacity
reference. Under an average-power constraint, the capacity-achieving input for
the fixed isotropic AWGN channel is an $n$-dimensional Gaussian distribution,
whereas the C3M input distribution is supported on a union of one-dimensional
branch curves in $\mathbb{R}^n$. Thus, $C_n$ should be interpreted as an AWGN
capacity reference, while $I(S;\mathbf Y)$ is the achievable information rate
induced by the specific C3M source--channel mapping.

For a Gaussian source with variance $\sigma_\mathbb S^2$, the rate--distortion function
under MSE distortion is
\begin{equation}
\label{eq:gaussian_rd}
R_G(D)
=
\frac{1}{2}\log_2\!\left(\frac{\sigma_S^2}{D}\right).
\end{equation}
Equating Eq.~\eqref{eq:gaussian_rd} with the channel-capacity reference $C_n$
yields the Gaussian-source reference distortion
\begin{equation}
\label{eq:gaussian_rd_distortion}
D_G^*
=
\sigma_S^2\,2^{-2C_n}.
\end{equation}
For a uniform source, the true rate--distortion function is not given by
Eq.~\eqref{eq:gaussian_rd}. At high rate, however, the uniform-source
distortion reference is lower than the Gaussian-source reference by the
asymptotic shaping factor $6/(\pi e)$:
\begin{equation}
\label{eq:Du_rev}
D_U^{\mathrm{asy}}
=
D_G^*\frac{6}{\pi e}
=
\sigma_S^2\,2^{-2C_n}\frac{6}{\pi e}.
\end{equation}
Equivalently, the corresponding high-rate \SDR{} reference is larger by the
factor $\pi e/6$, or approximately \dBval{1.53}. This refined reference is
therefore best interpreted as a high-rate uniform-source reference, not as an
exact low-\SNR{} rate--distortion limit.

The branch decomposition can also be used to define conditional diagnostic
references. If the true branch index $K$ were provided to the decoder, then a
conditional branch-known rate constraint could be written as
\[
\sum_{k=1}^m p_k R_k(D_k)\le C_n,
\]
where $R_k(D_k)$ is the rate--distortion function associated with the source
interval assigned to branch $k$. This is a branch-known reference, not an
ordinary end-to-end reference for the original source. In the practical C3M
receiver, the branch index must be inferred from $\mathbf Y$, and any
information needed to identify the branch is part of the end-to-end source
description. The corresponding end-to-end and branch-known \SDR{} references
are separated explicitly in Section~\ref{ss:end_to_end_branch_known_sdr}.

Structured-code constructions provide an important asymptotic reference for
Gaussian-source settings, showing that separation-based references can be
approached arbitrarily closely in high-dimensional regimes~\cite{Erez2004}.
The C3M construction studied here is instead finite-dimensional and
low-blocklength. Its performance is governed by curve geometry, branch
separation, branch-classification reliability, and decoder accuracy rather than
by claims of asymptotic optimality.

It should be emphasized that the separation-based rate--distortion and capacity
references are asymptotic information-theoretic limits. In practical
finite-blocklength and low-\SNR{} regimes, separated systems may operate far
from these limits~\cite{Kostina2012}, whereas analog joint source--channel
mappings can provide graceful degradation under changing channel
conditions~\cite{Hekland2009}. Throughout the numerical comparisons in this
paper, these references are evaluated under the fixed isotropic AWGN model.
For a Gaussian channel model with general covariance $\mathbf{\Sigma}_n$, the
corresponding capacity and entropy expressions are obtained by replacing
$\sigma_n^2\mathbf{I}_n$ with $\mathbf{\Sigma}_n$.

\subsection{Classification reliability}
\label{ss:class_reli}

To interpret the branch-classification reliabilities appearing in the interval-allocation rule in Eq.~\eqref{eq:interval_alloc}, it is useful to relate them to the local separation geometry of the branch set. Under fixed isotropic AWGN, a convenient nearest-neighbor approximation is obtained from the Euclidean inter-branch distances
\begin{equation}
\label{eq:branch_distance_euclidean}
d_{kk'}
\triangleq
\min_{z,z'\in[0,1)}
\left\|
\gamma_k(z)-\gamma_{k'}(z')
\right\|.
\end{equation}
In the moderate- to high-\SNR{} regime, the conditional error probability for branch $k$ is then dominated by nearby competing branches and may be approximated as
\begin{equation}
\label{eq:pairwise_q_branch}
\Pr(\widehat{K}\neq k \mid K=k)
\;\approx\;
\sum_{k'\neq k}
Q\!\left(\frac{d_{kk'}}{2\sigma_n}\right).
\end{equation}
Thus, the quantities $q_k=1-\epsilon_k$ used in the interval-allocation rule are closely tied to the local separation structure of the branch set: larger inter-branch distances improve classification reliability, whereas closely spaced branches become the dominant source of confusion. 
For a Gaussian extension with general covariance $\mathbf{\Sigma}_n$, the Euclidean distance in Eq.~\eqref{eq:branch_distance_euclidean} is replaced by the corresponding Mahalanobis distance.
This reliability viewpoint will also be useful in interpreting the \SDR{} behavior of multi-curve designs under different interval-allocation rules.

The nearest-neighbor approximation further suggests that branch-classification reliability may be strongly asymmetric across the branch set. Although the multi-curve design is chosen to maximize global separation, the conditional error for branch $k$ is controlled mainly by its closest competitors. Hence branches with multiple nearby neighbors, or with smaller admissible tube radii, may exhibit substantially lower values of
\[
q_k=\Pr(\widehat{K}=k\mid K=k)
\]
than more isolated branches. This phenomenon is particularly relevant in low-dimensional settings, where geometric packing constraints limit the number of strongly separated branches. In such cases, one or more curves may become bottlenecks for classification performance, even when the overall design remains near-optimal in a global sense.

A useful conceptual decomposition of end-to-end distortion separates branch-index errors from within-branch reconstruction error. In particular, let $E_K$ denote the event that the branch index is decoded incorrectly. Then the total mean-squared error may be decomposed as
\begin{equation}
\begin{aligned}
\label{eq:mse_decomp_branch}
\mathbb{E}\!\left[(S-\widehat{S})^2\right]
&=
\Pr(E_K)\,\mathbb{E}\!\left[(S-\widehat{S})^2 \mid E_K\right]\\
&\quad+
\bigl(1-\Pr(E_K)\bigr)\,\mathbb{E}\!\left[(S-\widehat{S})^2 \mid E_K^c\right].
\end{aligned}
\end{equation}
This viewpoint clarifies why both inter-branch separation and along-branch resolution matter simultaneously in the multi-curve setting.

A more refined decomposition is obtained by conditioning on the transmitted branch. Let $D_k^{\mathrm{corr}}$ denote the conditional distortion when branch $k$ is correctly identified, and let $D_k^{\mathrm{err}}$ denote the corresponding distortion under branch misclassification. Then
\begin{equation}
\label{eq:branchwise_mse_decomp}
D_{\mathrm{total}}
=
\sum_{k=1}^m p_k\Bigl[(1-\epsilon_k)D_k^{\mathrm{corr}}+\epsilon_k D_k^{\mathrm{err}}\Bigr],
\end{equation}
where $\epsilon_k=\Pr(\widehat K\neq k\mid K=k)$. This form makes explicit that end-to-end performance depends jointly on within-branch reconstruction quality and branch-classification reliability.

The decomposition in Eq.~\eqref{eq:branchwise_mse_decomp} also highlights a useful optimistic reference obtained by assuming perfect branch identification, in which case only the within-branch reconstruction error remains. Practical receivers must infer the transmitted branch from noisy observations, so the gap to this reference measures the additional distortion associated with branch-identification errors.

In multi-branch systems, branch misclassification can induce a large distortion contribution because the reconstructed source value may be forced into an interval associated with a different branch. Consequently, even modest classification error probabilities may significantly affect end-to-end SDR when the branch intervals are well separated in source space.

\subsection{Capacity-surrogate viewpoint}
\label{ss:cap_surrogate}

A useful high-\SNR{} perspective is obtained by viewing multi-curve design
through a geometric capacity surrogate. Let $G_k$ denote the geometric score
associated with branch $k$. In a small-noise approximation, the induced mutual
information increases with the aggregate geometric score $\sum_{k=1}^m G_k$,
so that the corresponding objective may be viewed as increasing monotonically
with
\[
I^\star \propto \log\!\left(\sum_{k=1}^m G_k\right).
\]
From this perspective, maximizing the sum of the per-branch geometric scores
provides a natural high-\SNR{} design principle for the branch set.

A related geometric viewpoint is to balance total usable branch length against
effective union thickness, which captures the tradeoff between resolution and
robustness. This viewpoint separates the C3M design problem into two stages.
The geometry stage seeks a branch set with a large aggregate geometric score
while maintaining sufficient robustness and inter-branch separation. The
allocation stage divides the source interval among the selected branches
according to the resulting per-branch utilities. For constant-speed branches,
the geometric scores satisfy $G_k=L_k/(2\pi)$, so the capacity-surrogate
viewpoint connects directly to a geometry-only, length-based allocation rule.

\begin{remark}
In the constant-speed even-dimensional case, the capacity-surrogate viewpoint also suggests a direct algorithmic route to branch-set design. A natural formulation is
\begin{equation}
\label{eq:geom_opt_problem}
\max_{\{\mathbf r_k\}}
\;
\log\!\Big(\sum_{k=1}^m G_k\Big)
-\lambda\,[\rho_{\min}-\rho_U]_+^2,
\end{equation}
where $\rho_U$ denotes an overall robustness measure for the multi-curve construction, $\rho_{\min}$ is a target robustness threshold, $\lambda\ge 0$ is a penalty parameter, and $[x]_+\triangleq \max\{x,0\}$. Here the first term promotes a large aggregate geometric score, while the penalty term discourages branch sets whose usable tube radius falls below the target level. Writing
\[
G_k
=
\sqrt{\sum_{j=1}^{n/2} j^2 r_{j,k}^2},
\qquad
S=\sum_{\ell=1}^m G_\ell,
\]
the gradient of the objective term $\log S$ with respect to the radii parameters is
\begin{equation}
\label{eq:geom_gain_gradient}
\frac{\partial}{\partial r_{j,k}} \log\!\Big(\sum_{\ell=1}^m G_\ell\Big)
=
\frac{1}{S}\,\frac{j^2 r_{j,k}}{G_k}.
\end{equation}
Hence the geometry stage may be treated algorithmically, for example through projected-gradient or stochastic-approximation methods, with robustness and inter-branch separation imposed as additional constraints or penalties.
\hfill$\blacksquare$
\end{remark}

\subsection{Geometry-only allocation perspective}
\label{ss:geo_alloc}

A complementary perspective is obtained by considering source allocation based solely on branch geometry. In a high-\SNR{} small-noise approximation, a natural geometric score for branch $k$ is
\begin{equation}
    \label{eq:geom_score}
    G_k
    =
    \exp\!\left(
    \frac{1}{2\pi}\int_{0}^{2\pi}\log v_k(\theta)\,\mathrm{d}\theta
    \right),
    \end{equation}
    where $v_k(\theta)$ denotes the local branch speed, as in Eq.~\eqref{eq:speed}. This quantity is the geometric mean of the local speed over one period. A geometry-only allocation rule would then assign interval fractions proportionally to these scores,
    \begin{equation}
    \label{eq:geom_only_alloc}
    p_k
    =
    \frac{G_k}{\sum_{\ell=1}^m G_\ell},
    \qquad k=1,\ldots,m.
\end{equation}

\begin{remark}[Local Fisher-information interpretation]
    Another local interpretation is obtained from the Fisher information of the
    curve parameter. For a single branch observed through fixed isotropic AWGN,
    $\mathbf Y=\gamma_k(z)+\mathbf N$,
    $\mathbf N\sim\mathcal N(\mathbf 0,\sigma_n^2\mathbf I_n)$, the Fisher
    information for the normalized local coordinate $z$ is $J_k(z)
    =
    \frac{1}{\sigma_n^2}
    \left\|\gamma_k'(z)\right\|^2$. Thus, branches with larger local speed provide greater local sensitivity of
    the observation to changes in the source coordinate. For constant-speed
    branches, $J_k(z)$ is constant along the branch, which supports the use of
    curve length or speed-based geometric gains in high-\SNR{} allocation
    heuristics. This Fisher-information viewpoint is local and does not by
    itself provide a global mutual-information, rate--distortion, or
    branch-identification bound.
    \hfill$\blacksquare$
\end{remark}

Other geometry-only surrogates, such as the tube-density factors $\{\delta_k\}$, may be used in the same way. The classification-aware rule in Eq.~\eqref{eq:interval_alloc} extends this geometry-only viewpoint by weighting each branch by its reliability $q_k=1-\epsilon_k$, thereby reducing the allocation to branches that are geometrically efficient but unreliable to classify.

A branch parametrization is said to be constant-speed if the local speed
$v_k(\theta)=\|\gamma_k'(\theta)\|$ does not depend on the parameter $\theta$.
In that case, the branch parameter is proportional to arc length, so the
natural parametrization is already compatible with normalized arc-length
coordinates. This simplification is useful in the present setting because it
allows the geometric score to be expressed directly in terms of branch length
and avoids additional Jacobian factors in the corresponding decoder
expressions. For the branch families considered here, the natural
parametrizations are constant-speed. Hence $G_k=v_k=\frac{L_k}{2\pi}$, 
since the branch length over one period is $L_k=2\pi v_k$. Thus, for
constant-speed branches, the geometry-only rule reduces to an allocation
proportional to branch length. For non-constant-speed branch parametrizations,
related decoder formulations may also include a Jacobian compensation term
based on the local branch speed.

Since the full branch of length $L_k$ is traversed over a source interval of
length $p_k$, the effective encoder slope scales like $L_k/p_k$. Hence, in the
high-\SNR{} regime, the local tangential distortion is expected to scale like
$\sigma^2 p_k^2/L_k^2$, and after averaging over the interval mass this yields
a contribution of order $\sigma^2 p_k^3/L_k^2$.

\begin{remark}[Heuristic high-\SNR{} distortion decomposition]
A useful high-\SNR{} interpretation of multi-curve performance is obtained by separating tangential distortion from tube-escape effects. Heuristically, the mean-squared error may be approximated as
\begin{equation}
\label{eq:high_snr_multicurve_mse}
D(\sigma^2)
\approx
\sigma^2 \sum_{k=1}^m c_k \frac{p_k^3}{L_k^2}
+
C\sum_{k=1}^m p_k \exp\!\left(-\frac{r_k^2}{2\sigma^2}\right),
\end{equation}
where $L_k$ denotes the length of branch $k$, $r_k$ its admissible tube radius, and $c_k,C>0$ are branch-dependent and global constants, respectively. The first term represents tangential distortion when the received point remains in the correct tube, whereas the second term captures threshold effects associated with tube escapes. This decomposition helps explain why geometry-only allocation tends to favor long branches in the high-\SNR{} regime, while branches with larger admissible tube radii become increasingly important when tube escapes are non-negligible.
\hfill$\blacksquare$
\end{remark}

When the admissible tube radii vary significantly across branches and the noise level is such that tube escapes are no longer negligible, the escape term penalizes branches with small $r_k$. This motivates radius-aware geometric weighting in addition to length-based allocation. When tube escapes are negligible, the leading term in \eqref{eq:high_snr_multicurve_mse} decreases as the branch length increases, which provides a high-SNR justification for geometry-only allocation rules that favor longer branches. This viewpoint is useful because it clarifies the role of geometry in source allocation. However, in the present multi-curve setting, end-to-end performance is influenced not only by geometric efficiency but also by branch-classification reliability. At low or moderate SNR, performance may be dominated not by along-curve resolution alone, but by tube escapes and branch-classification errors. This provides further motivation for the classification-aware rule in Section~\ref{ss:allocation}, where the geometric contribution is weighted by the branch-reliability factors $\{q_k\}_{k=1}^m$.

\subsection{End-to-end and branch-known SDR references}
\label{ss:end_to_end_branch_known_sdr}

We distinguish between an end-to-end \SDR{} reference, in which the branch
index is not available as side information, and a branch-known reference, in
which the decoder is given the true branch index. This distinction is important
because a multi-curve code represents the source through both a discrete branch
index and a continuous coordinate along the selected branch. In the practical
C3M receiver, the branch index must be inferred from the received channel
output and is therefore not free side information.

Let $S\sim\mathcal U[0,1)$, so that $\sigma_S^2=1/12$, and let $C_n$ denote
the channel-capacity reference per $n$-dimensional channel use from
Eq.~\eqref{eq:capacity_n}. For the ordinary end-to-end reference, the
rate--distortion calculation is applied to the original scalar source using the classical Gaussian
rate--distortion and AWGN capacity formulas~\cite{CoverThomas2006,Berger1971}. The
Gaussian-source distortion and \SDR{} references are
\begin{equation}
\label{eq:sdr_g_end_to_end}
D_{\mathrm{G,end}}
=
\frac{1}{12}2^{-2C_n},
\qquad
\mathrm{SDR}_{\mathrm{G,end}}
=
2^{2C_n}.
\end{equation}
Using the high-rate uniform-source correction gives
\begin{equation}
\label{eq:sdr_u_end_to_end}
D_{\mathrm{U,end}}^{\mathrm{asy}}
=
\frac{1}{2\pi e}2^{-2C_n},
\qquad
\mathrm{SDR}_{\mathrm{U,end}}^{\mathrm{asy}}
=
2^{2C_n}\frac{\pi e}{6}.
\end{equation}
These end-to-end references do not contain an $m^2$ factor, because any
information needed to identify the branch is part of the transmitted source
description.

For comparison, we also define a genie-aided branch-known reference. In this
reference, the decoder is given the true branch index $K$ and only needs to
estimate the local coordinate within the corresponding source interval. Under
equal splitting into $m$ subintervals, each interval has length $1/m$ and
conditional variance $1/(12m^2)$. The branch-known Gaussian distortion and
\SDR{} references are therefore
\begin{equation}
\label{eq:sdr_g_branch_known}
D_{\mathrm{G,bk}}
=
\frac{1}{12m^2}2^{-2C_n},
\qquad
\mathrm{SDR}_{\mathrm{G,bk}}
=
m^2 2^{2C_n}.
\end{equation}
Similarly, the high-rate uniform-source branch-known reference is
\begin{equation}
\label{eq:sdr_u_branch_known}
D_{\mathrm{U,bk}}^{\mathrm{asy}}
=
\frac{1}{2\pi e\,m^2}2^{-2C_n},
\qquad
\mathrm{SDR}_{\mathrm{U,bk}}^{\mathrm{asy}}
=
m^2 2^{2C_n}\frac{\pi e}{6}.
\end{equation}
Thus, the familiar $m^2$ factor belongs to the branch-known reference. It is
not an ordinary end-to-end rate--distortion gain for the original source.

For non-uniform branch probabilities $p_1,\ldots,p_m$ with
$\sum_{k=1}^m p_k=1$, we also use a same-rate branch-known Gaussian diagnostic
reference. If branch $k$ is assigned interval fraction $p_k$, then its
conditional source variance is $p_k^2/12$. Applying the same Gaussian capacity
reference $C_n$ to each active branch gives
\begin{equation}
    \label{eq:dist_branch_known_nonuniform}
    D_{\mathrm{G,bk}}
    =
    \frac{1}{12}2^{-2C_n}
    \sum_{k=1}^m p_k^3,
\end{equation}
and the corresponding branch-known Gaussian \SDR{} reference is
\begin{equation}
    \label{eq:sdr_branch_known_nonuniform}
    \mathrm{SDR}_{\mathrm{G,bk}}
    =
    \frac{2^{2C_n}}{\sum_{k=1}^m p_k^3}.
\end{equation}
Equal splitting gives $\sum_{k=1}^m p_k^3=1/m^2$, which recovers
Eq.~\eqref{eq:sdr_g_branch_known}. This expression is used as a diagnostic
branch-known reference. It assumes the same effective conditional rate
reference for each active branch and does not optimize a separate rate
allocation across branches.

Using the fixed isotropic AWGN capacity expression in Eq.~\eqref{eq:capacity_n},
\begin{equation}
\label{eq:capacity_exp_factor}
2^{2C_n}
=
\left(1+\frac{P}{n\sigma_n^2}\right)^n
=
\left(1+\frac{\SNR}{n}\right)^n .
\end{equation}
Hence the end-to-end references can be written as
\begin{equation}
\label{eq:sdr_end_to_end_snr_form}
\begin{aligned}
\mathrm{SDR}_{\mathrm{G,end}}
=
\left(1+\frac{\SNR}{n}\right)^n,\qquad
\mathrm{SDR}_{\mathrm{U,end}}^{\mathrm{asy}}
=
\left(1+\frac{\SNR}{n}\right)^n\frac{\pi e}{6},
\end{aligned}
\end{equation}
while the equal-splitting branch-known references are
\begin{equation}
\label{eq:sdr_branch_known_snr_form}
\begin{aligned}
\mathrm{SDR}_{\mathrm{G,bk}}
=
m^2\left(1+\frac{\SNR}{n}\right)^n,
\qquad\mathrm{SDR}_{\mathrm{U,bk}}^{\mathrm{asy}}
=
m^2\left(1+\frac{\SNR}{n}\right)^n\frac{\pi e}{6}.
\end{aligned}
\end{equation}

The practical C3M performance reported in Section~\ref{sec:numexp} is neither
of these idealized references. The receiver observes only $\mathbf Y$, estimates
the branch index, and then estimates the source coordinate on the selected
branch. Its end-to-end distortion and \SDR{} are
\begin{equation}
\label{eq:c3m_practical_sdr}
D_{\mathrm{C3M}}
=
\mathbb E\!\left[(S-\widehat S)^2\right],
\qquad
\mathrm{SDR}_{\mathrm{C3M}}
=
\frac{\sigma_S^2}{D_{\mathrm{C3M}}}.
\end{equation}
This quantity includes both along-curve estimation error and
branch-identification error. The gap between a branch-known decoder and the
practical decoder therefore measures the penalty due to branch confusion.
Table~\ref{tab:sdr_references} summarizes the distinction between the
end-to-end Gaussian reference, the branch-known Gaussian reference, and the
practical C3M \SDR{} used in the numerical evaluation.

\begin{table}[t]
\centering
\caption{Interpretation of the \SDR{} quantities used in the C3M analysis and evaluation.}
\label{tab:sdr_references}
\begin{tabular}{lll}
\toprule
Quantity & Expression & Interpretation \\
\midrule
End-to-end Gaussian reference
&
$\left(1+\frac{\SNR}{n}\right)^n$
&
No branch side information \\

Branch-known Gaussian reference
&
$m^2\left(1+\frac{\SNR}{n}\right)^n$
&
Genie-aided diagnostic reference \\

Practical C3M \SDR{}
&
$\sigma_S^2/\mathbb{E}[(S-\widehat S)^2]$
&
Learned-decoder end-to-end performance \\
\bottomrule
\end{tabular}
\end{table}

A useful high-\SNR{} interpretation of multi-curve performance is obtained by
separating along-curve distortion from tube-escape effects. At a heuristic
level, the mean-squared error may be decomposed into a tangential term, which
decreases with increasing effective curve length, and a threshold term
associated with tube escapes or branch confusions. This viewpoint helps explain
why geometry-only allocation tends to favor long curves in the high-\SNR{}
regime, whereas more robust branches become increasingly important when the
noise level is high enough for tube escapes or branch confusions to play a
significant role.

These analytical references do not include finite-dimensional geometric
admissibility constraints, branch-overlap constraints, decoder training effects,
or classification errors. They should therefore be interpreted as idealized
information-theoretic references rather than finite-dimensional performance
guarantees. The Gaussian reference is not the exact Shannon limit for a uniform
source. It is a convenient reference obtained from the Gaussian
rate--distortion function. The expression $\mathrm{SDR}_{\mathrm{U}}^{\mathrm{asy}}$ is best interpreted
as a high-rate uniform-source reference, not as an exact low-\SNR{}
rate--distortion limit.

\subsection{Non-uniform source allocation}
\label{ss:non_uniform_src}

Finally, when the source allocation is highly non-uniform, a more refined design criterion is obtained by protecting the branches with largest probabilities $\{p_k\}$. In that case, the standard max--min inter-branch separation heuristic may be replaced by a weighted geometric surrogate $\max_{\{\mathbf r_k\}}
\;
\min_{k\neq k'}
w_{kk'}\,\|\mathbf r_k-\mathbf r_{k'}\|$, where $w_{kk'}\ge 0$ are design weights. By assigning larger values of $w_{kk'}$ to pairs involving dominant branches, the resulting curve set encourages larger separation margins around the branches that contribute most strongly to the overall distortion. This is especially relevant in low-dimensional settings, where geometric packing constraints limit the number of well-separated branches and one or more curves may otherwise become bottlenecks for classification performance.

It is useful to distinguish non-uniform branch probabilities from unequal conditional variances. If two branches have probabilities $p_1$ and $p_2$ but identical conditional source variances and are assigned the same target distortion, then the weighted rate constraint $p_1 R_1(D)+p_2 R_2(D)\le C_n$ reduces to $R(D)\le C_n$ whenever $R_1(D)=R_2(D)$ and $p_1+p_2=1$. In that special symmetric case, changing the probabilities alone does not change the per-branch reference distortion. For a uniform source partitioned into intervals, however, the branch probability is tied to the interval length. Thus, non-uniform interval fractions generally imply different conditional variances; for a contiguous interval of length $p_k$, $\sigma_{S|K=k}^2=\frac{p_k^2}{12}$, and the allocation problem becomes genuinely weighted.

For Gaussian rate--distortion references, one may write the weighted constraint as
\begin{equation}
\label{eq:weighted_rd_constraint}
    \sum_{k=1}^m p_k R_k(D_k)\le C_n,
    \qquad
    R_k(D_k)=\frac{1}{2}\log_2\!\left(\frac{\sigma_{S|K=k}^2}{D_k}\right).
\end{equation}
This form makes explicit that non-uniform source allocation affects the
branch-known reference through both the probabilities $p_k$ and the conditional
variances $\sigma_{S|K=k}^2$.

A complementary way to interpret non-uniform allocation is through a decoder-switching model. For simplicity, consider two branches with probabilities $p_1$ and $p_2$, and let
$\alpha=\Pr(\widehat K=2|K=1)$ and $\beta=\Pr(\widehat K=1|K=2)$ denote the two branch-confusion probabilities. If $D_{k\to \ell}$ denotes the conditional distortion incurred when branch $k$ was transmitted but the branch-conditioned decoder for branch $\ell$ is used, then the total distortion can be written as
\begin{align}
\label{eq:two_branch_switching_distortion}
D_{\mathrm{total}} &= p_1\bigl[(1-\alpha)D_{1\to 1}+\alpha D_{1\to 2}\bigr] 
+ p_2\bigl[\beta D_{2\to 1}+(1-\beta)D_{2\to 2}\bigr].
\end{align}
This expression separates the within-branch reconstruction distortion from the penalty associated with decoder-switching errors. In particular, when the branch probabilities are non-uniform, switching errors involving high-probability branches contribute more strongly to the total distortion and should therefore be weighted more heavily in both branch placement and receiver training.

\begin{remark}[Switching-aware numerical design objective]
    More generally, one can define a switching-aware numerical design objective
    by optimizing over branch curves, a branch-selection rule, and
    branch-conditioned decoders,
    \begin{align}
        \label{eq:switching_aware_design}
        &\min_{\{\gamma_k\},\,h,\,\{d_k\}}
        \;
        \mathbb E\!\left[(S-d_{h(\mathbf Y)}(\mathbf Y))^2\right]
        \quad
        \text{s.t.} \nonumber \\
        &\sum_{k=1}^m p_k\,
        \mathbb E\!\left[\|\gamma_k(Z)\|^2\mid K=k\right]\le P .
    \end{align}
    Here $h(\mathbf Y)\in\{1,\ldots,m\}$ denotes the branch-selection rule, and
    $d_k(\cdot)$ denotes the decoder used when branch $k$ is selected. If the
    functions $\{\gamma_k\}$, $h$, and $\{d_k\}$ are restricted to a
    finite-dimensional parameterized family, such as C3M curves with a learned
    branch classifier and branch-conditioned neural decoders,
    Eq.~\eqref{eq:switching_aware_design} yields a numerical design objective
rather than a closed-form analytical performance limit.
    \hfill$\blacksquare$
\end{remark}

When the branch-selection probabilities $\{p_k\}$ are non-uniform, these priors should also be reflected in the decoder. In a MAP formulation under fixed isotropic AWGN, the branch decision is no longer determined solely by Euclidean proximity, but must include the prior contribution associated with each branch. Thus, for an observation $\mathbf y$, a branch-conditioned MAP rule takes the form
$(\widehat{k},\widehat{z})
=
\arg\min_{k,\;z\in[0,1)}
\left\{
\|\mathbf y-\gamma_k(z)\|^2
-2\sigma_n^2\log p_k
\right\}$.
Hence, highly probable branches receive the appropriate prior weighting through the term $-2\sigma_n^2\log p_k$, while the geometric mismatch is measured by Euclidean distance under the fixed isotropic AWGN model. For a Gaussian extension with general covariance $\mathbf{\Sigma}_n$, the Euclidean distance is replaced by the corresponding Mahalanobis distance.

The same principle also applies to learning-based receivers. When $\{p_k\}$ are non-uniform, classifier and decoder training should account for the resulting class imbalance, for example through class-weighted losses or probability-aware sampling. Otherwise, dominant branches may be underprotected during training even though they contribute most strongly to the overall distortion. In this sense, the source allocation $\{p_k\}$ influences not only the encoder-side interval division, but also the statistically consistent design of the receiver.

\subsection{Geometric interpretation of multi-curve branch selection}
\label{ss:geo_interp}

A useful geometric interpretation is obtained by viewing multi-curve branch selection in terms of spherical codes in radius-vector space. For even ambient dimension $n$, one has $N=n/2$, so the admissible radii vectors lie on $\mathbb{S}^{N-1}$. In this way, multi-curve branch selection may be viewed as a packing problem on $\mathbb{S}^{N-1}$, where one seeks sufficiently well-separated radii vectors to obtain robust and reliably distinguishable branch curves.
In low dimensions, this immediately explains some of the observed branch-count limitations. For $n=4$, the radii vectors lie on $\mathbb{S}^{1}$, where only a very small number of strongly separated points can be placed; this is consistent with the practical difficulty of using many robustly distinguishable curves in four dimensions. For $n=6$, the design problem moves to $\mathbb{S}^{2}$, where larger and more symmetric point sets become feasible, for example six points arranged like an octahedron. For $n=8$, the corresponding design space is $\mathbb{S}^{3}$, where substantially richer spherical configurations are possible, for example the $24$ vertices of the $24$-cell. These examples should be interpreted as geometric illustrations rather than strict optimality statements, but they provide useful intuition for why the admissible number of robust branch curves tends to increase with dimension.

\subsection{Heuristic upper bounds on the number of branch curves}
\label{ss:num_branches}

A further geometric question is whether one can derive explicit upper bounds on
the number of robustly admissible branch curves. Dimension alone does not impose
such a bound, since arbitrarily many distinct curves may in principle be placed
on the ambient sphere if tube thickness and inter-branch separation are allowed
to shrink. A meaningful finite bound arises only after geometric robustness
constraints are imposed, such as a minimum admissible tube radius or a minimum
inter-branch separation. A natural robustness metric for the multi-curve
construction combines intra-branch tube admissibility with inter-branch
separation, so that the overall usable tube radius is limited by the smaller of
the individual branch thicknesses and half the minimum inter-branch distance.

At a heuristic level, let $\mathcal U_k(r_{\min})$ denote the radius-$r_{\min}$
tube around branch $k$, i.e., the tubular neighborhood generated by all points
within Euclidean distance $r_{\min}$ of the curve $\gamma_k$. If each branch
admits a disjoint tube of radius at least $r_{\min}$, then the total volume
occupied by the branch tubes must remain within the available bounding region.
For a one-dimensional curve of length $L_k$ embedded in $\mathbb{R}^n$, the
corresponding tube volume scales locally like $\operatorname{Vol}\!\big(\mathcal U_k(r_{\min})\big)
\approx
\kappa_{n-1}\,L_k\,r_{\min}^{\,n-1}$,  where $\kappa_{n-1}$ is the volume of the unit ball in the normal
$(n-1)$-dimensional cross-section. Consequently, if all admissible branch tubes
are pairwise disjoint and satisfy a common lower bound $L_k\ge L_{\min}$, one
obtains the heuristic packing estimate
\begin{equation}
\label{eq:heuristic_M_bound}
M
\;\lesssim\;
\frac{\operatorname{Vol}(\text{bounding region})}
     {\kappa_{n-1}\,L_{\min}\,r_{\min}^{\,n-1}}.
\end{equation}
Although the precise constants and admissibility conditions depend on the
detailed geometry of the construction, this scaling suggests that the maximum
number of robustly usable branches is governed jointly by ambient dimension,
minimum tube radius, and curve length, rather than by dimension alone.

\begin{figure*}[t!]
  \centering
  \subfloat[\label{fig:sdr_vs_snr_n4_m1}]{
    \includegraphics[width=.32\textwidth]{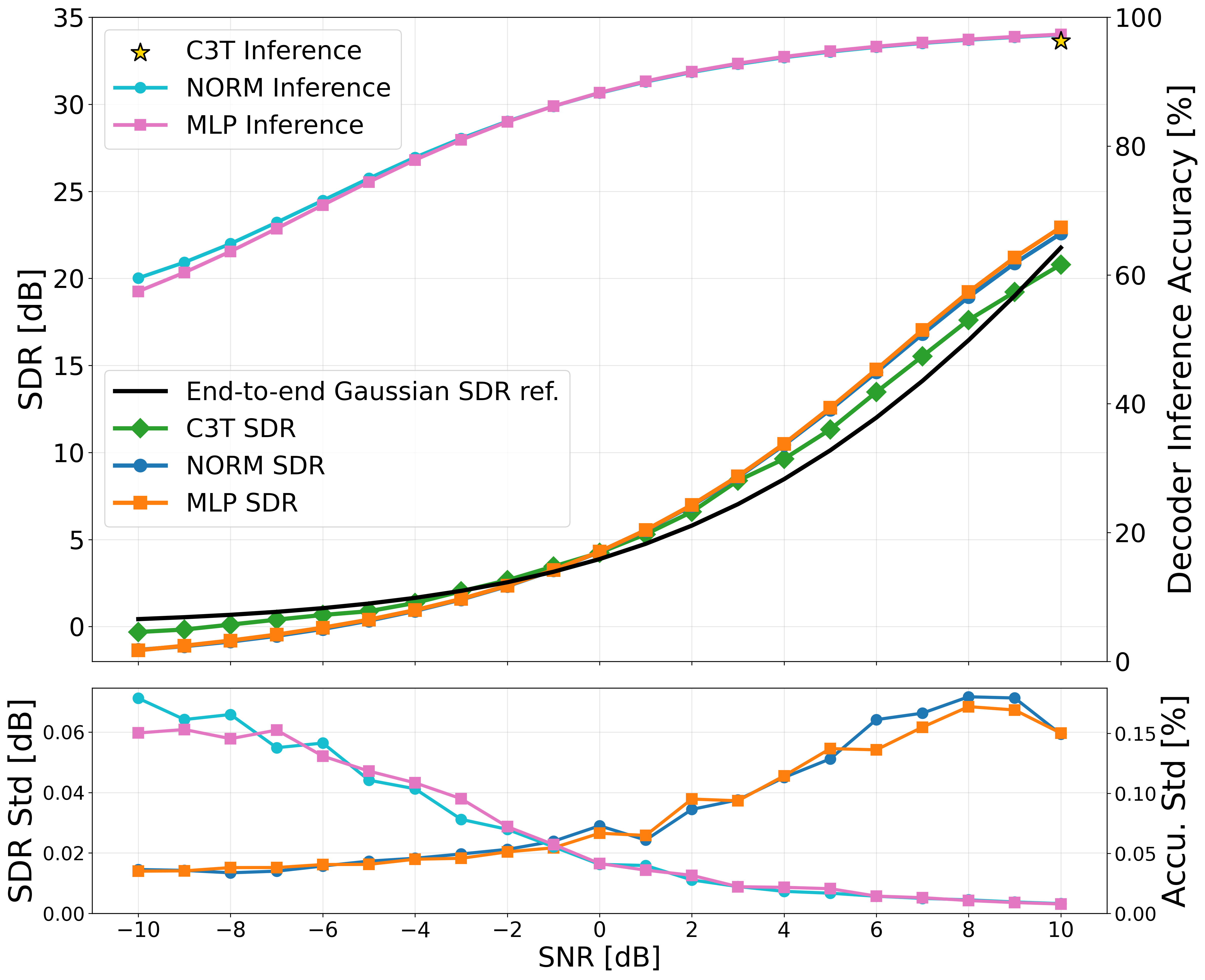}}\hfil
  \subfloat[\label{fig:sdr_vs_snr_n4_m2}]{
    \includegraphics[width=.32\textwidth]{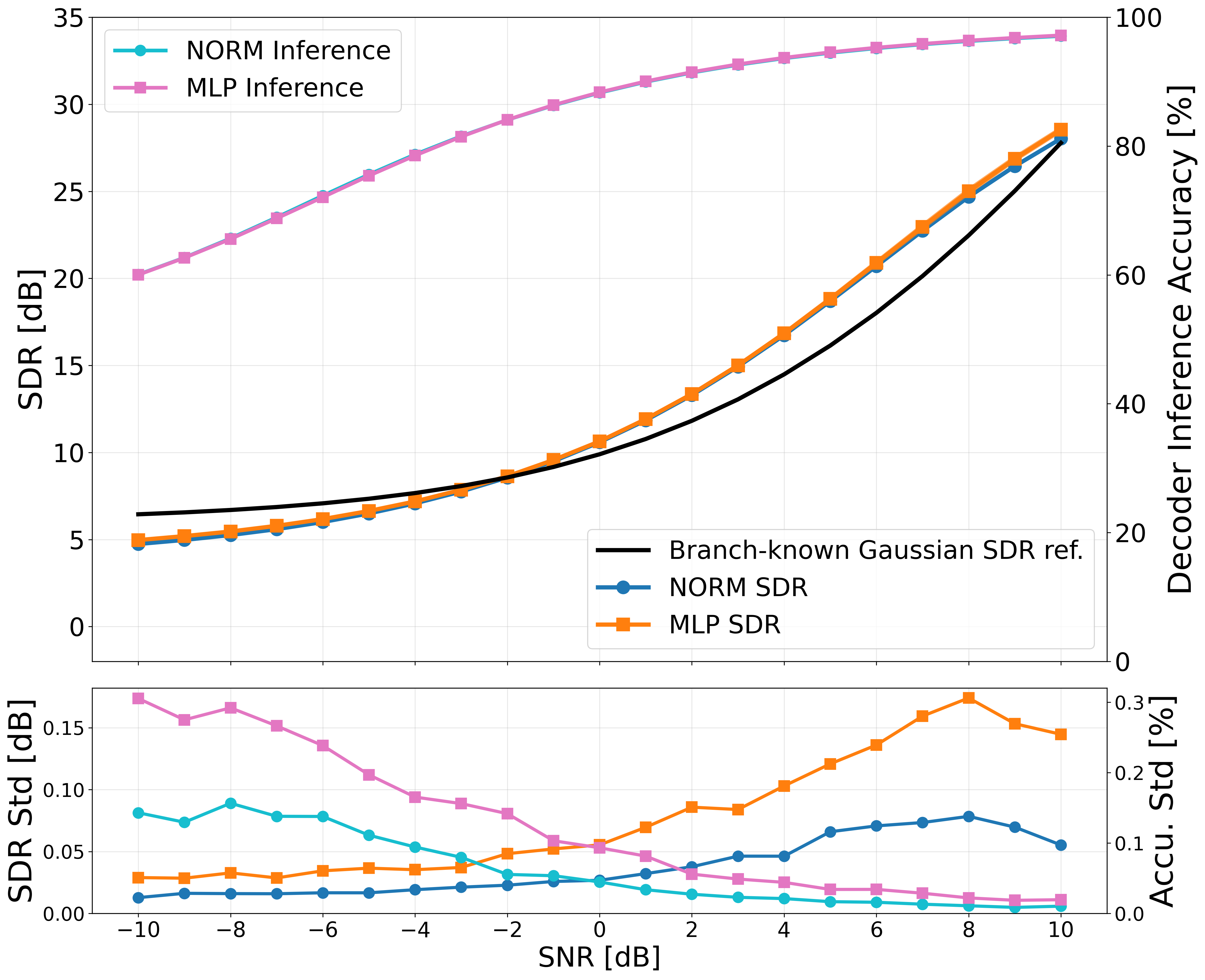}}\hfil
  \subfloat[\label{fig:sdr_vs_snr_n4_m3}]{
    \includegraphics[width=.32\textwidth]{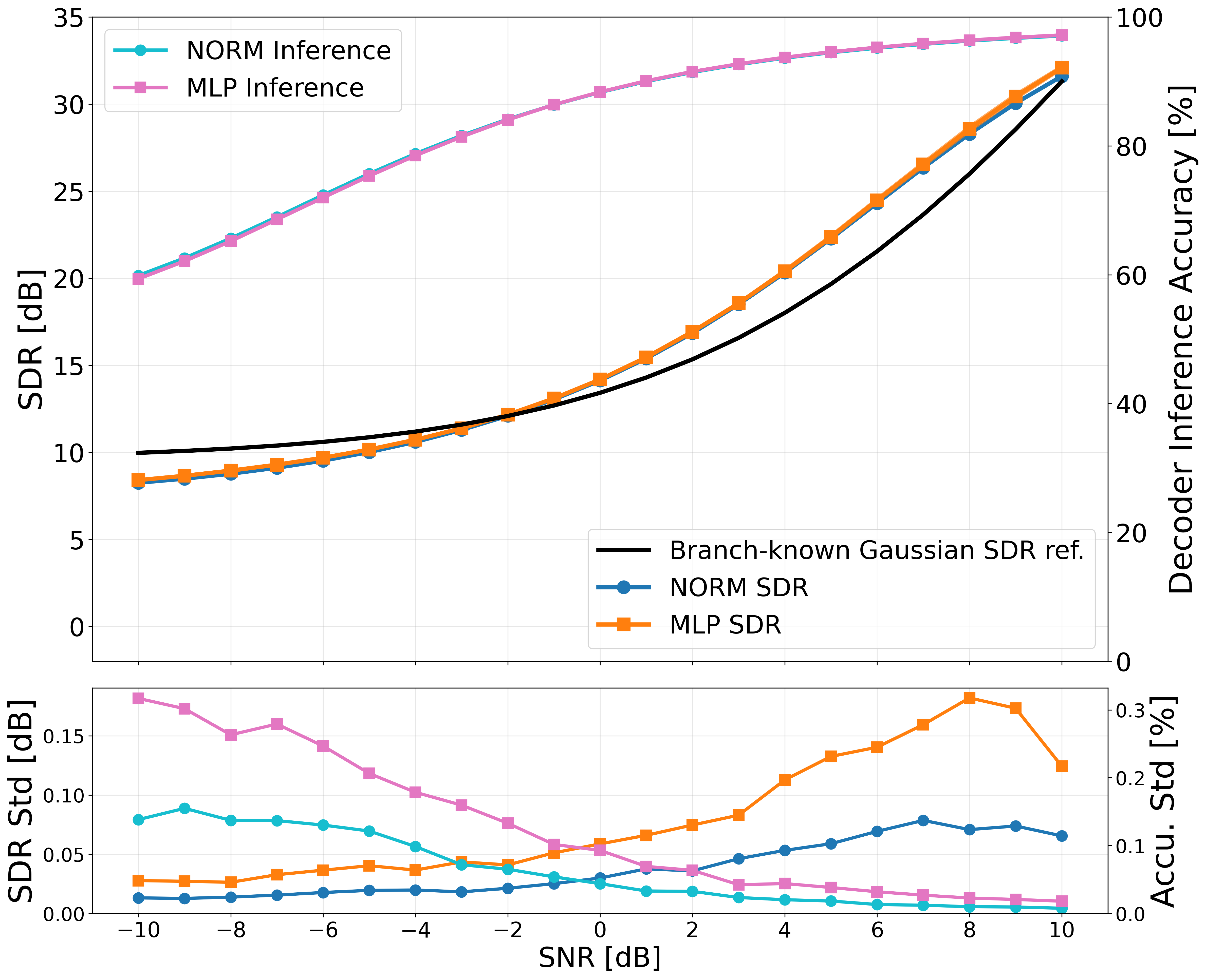}} \\
  \subfloat[\label{fig:sdr_vs_snr_n6_m1}]{
    \includegraphics[width=.32\textwidth]{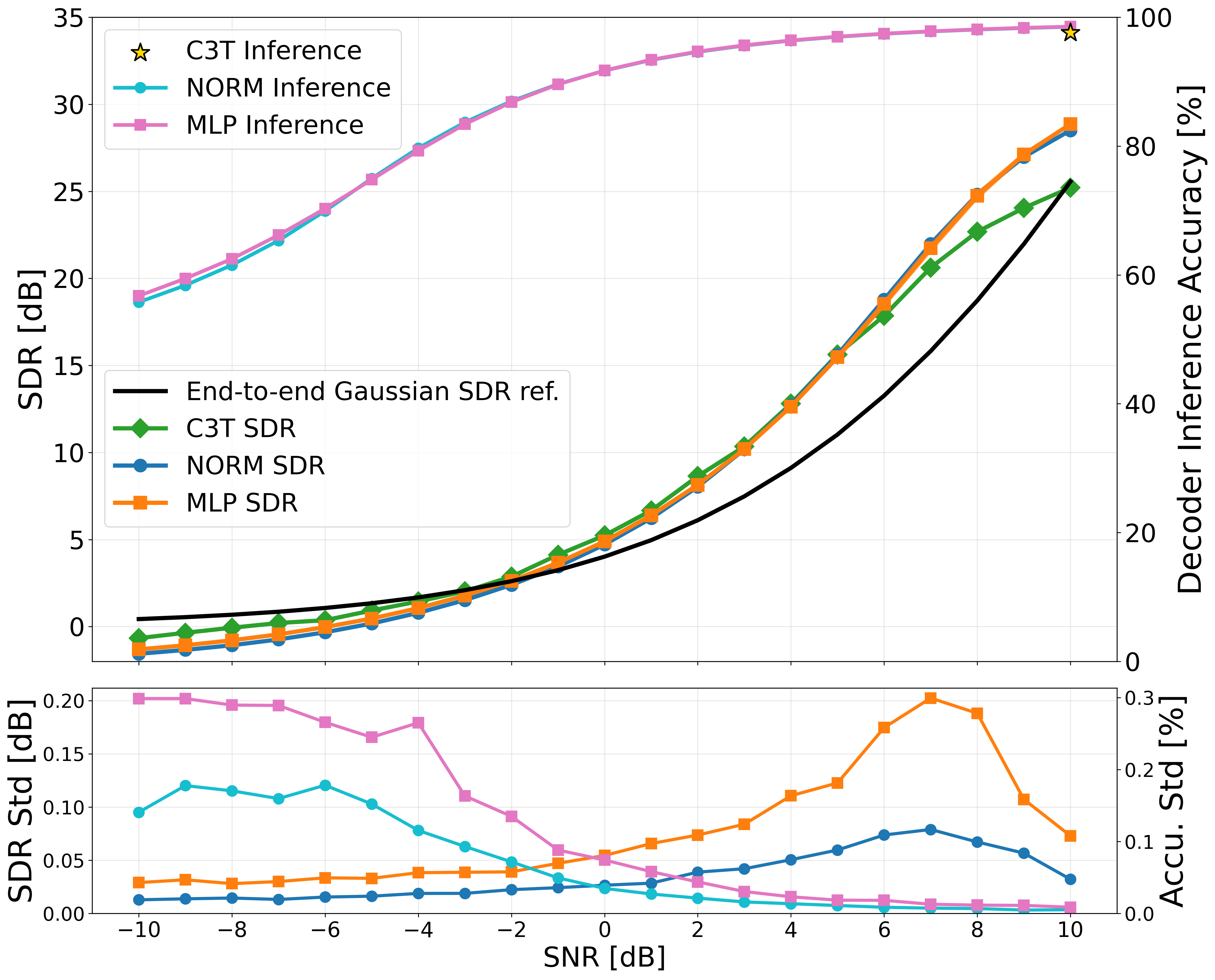}}\hfil
  \subfloat[\label{fig:sdr_vs_snr_n6_m2}]{
    \includegraphics[width=.32\textwidth]{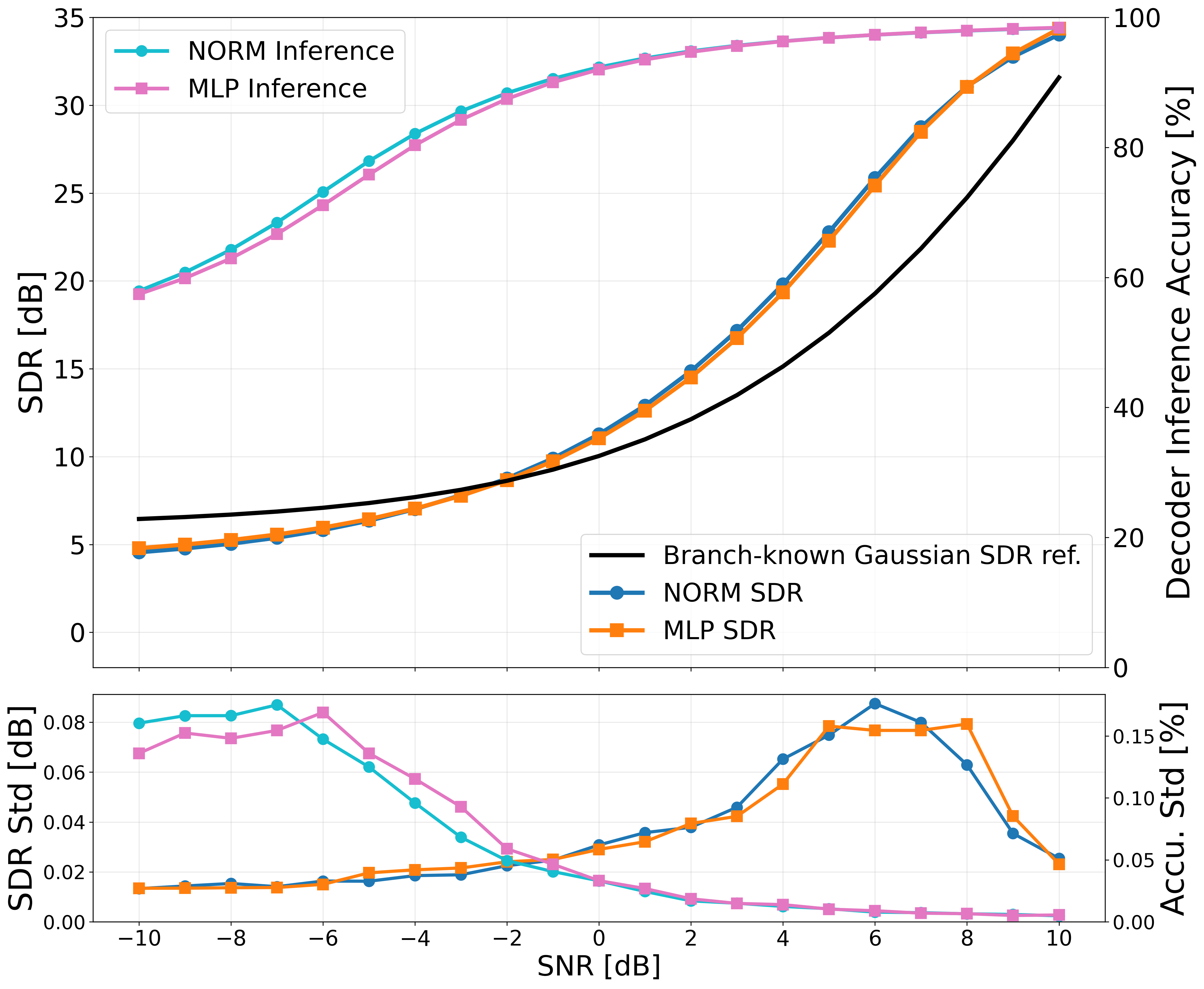}}\hfil
  \subfloat[\label{fig:sdr_vs_snr_n6_m3}]{
    \includegraphics[width=.32\textwidth]{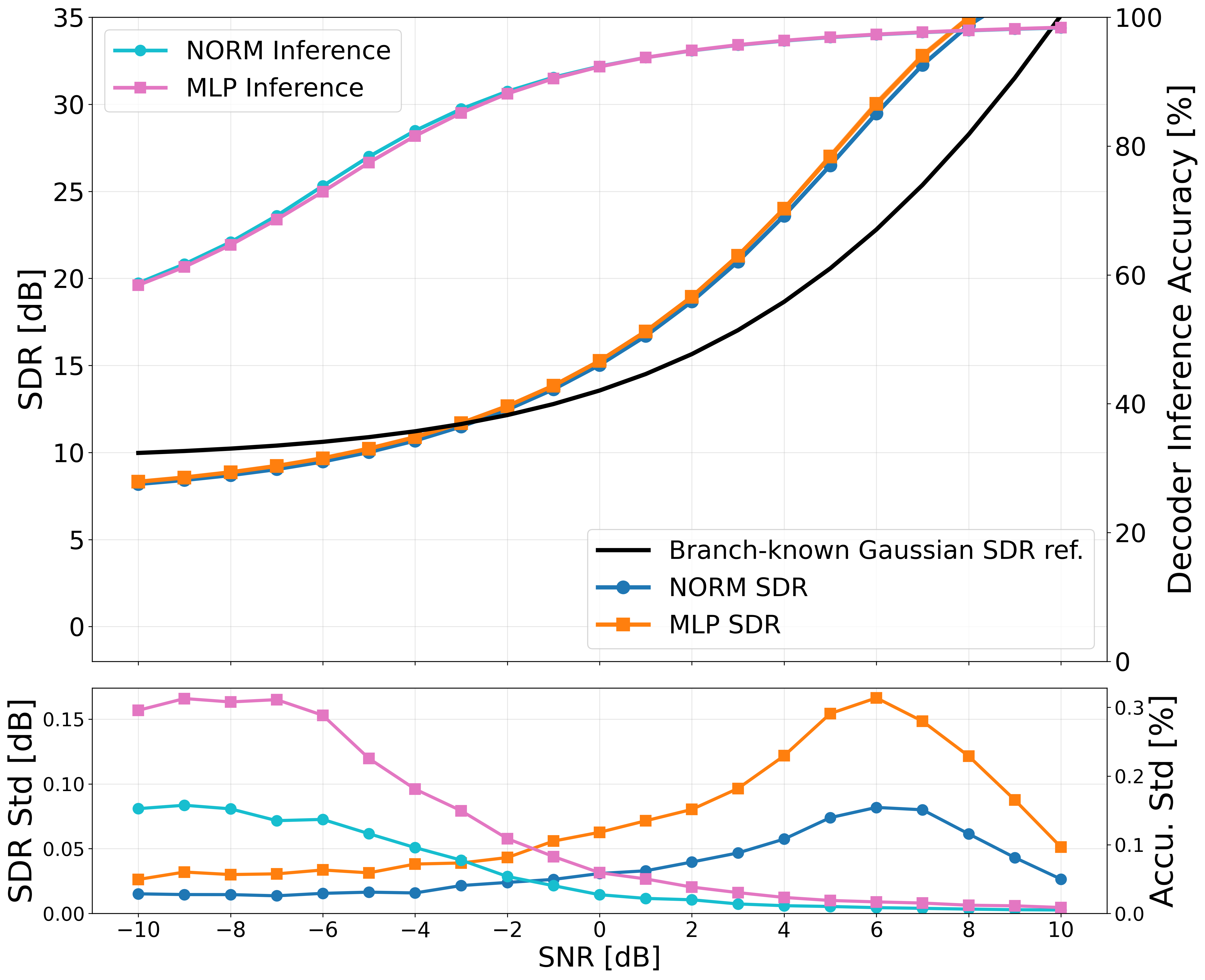}}
  \caption{\SDR{} and decoder-inference accuracy versus \SNR{} for C3M systems with different ambient dimensions and numbers of branch curves. The top row shows the four-dimensional systems and the bottom row shows the six-dimensional systems. The columns correspond to $m=1$, $m=2$, and $m=3$ branch curves, respectively. In each panel, the upper plot reports \SDR{} together with decoder-inference accuracy, while the lower plot reports the corresponding bootstrap standard deviation. The one-curve panels show the end-to-end Gaussian \SDR{} reference from Eq.~\eqref{eq:sdr_g_end_to_end} and include the C3T baseline~\cite{AndersC3T}. The multi-curve panels show the branch-known Gaussian \SDR{} reference from Eq.~\eqref{eq:sdr_g_branch_known} and compare NORM-inspired C3M decoding with the available MLP baselines. These reference curves are included for interpretation and are not finite-dimensional upper bounds on the learned decoders.}
  \label{fig:sdr_vs_snr}
\end{figure*}
\begin{table*}[t!]
\caption{Decoder \SDR{} and inference-accuracy statistics at \dBval{+10} \SNR{}.}
\label{tab:accu}
\centering
\begin{threeparttable}
\begin{tabular}{lcllcccc}
\toprule
Decoder & $n$ & $m$ & Selected/new curve\tnote{f} & \SDR{} [dB] & Inference accuracy [\%]\tnote{g} & $\sigma$\tnote{c} & MAD\tnote{d} \\
\midrule
C3T\tnote{a} & $4$ & $1$ & -- & $20.8$ & $96.3$ & $N/A$\tnote{e} & $N/A$\tnote{e} \\
MLP\tnote{b} & $4$ & $1$ & $\phi_1=47^\circ$ & $22.9$ & $97.3$ & $0.00788$ & $0.00550$ \\
MLP\tnote{b} & $4$ & $2$ & $\phi_2=46^\circ$ & $28.5$ & $97.2$ & $0.00884$ & $0.00594$ \\
MLP\tnote{b} & $4$ & $3$ & $\phi_3=44^\circ$ & $32.1$ & $97.2$ & $0.0180$ & $0.0117$ \\
NORM & $4$ & $1$ & $\phi_1=47^\circ$ & $22.6$ & $97.3$ & $0.0164$ & $0.0110$ \\
NORM & $4$ & $2$ & $\phi_2=47^\circ$ & $28.0$ & $97.1$ & $0.0182$ & $0.0118$ \\
NORM & $4$ & $3$ & $\phi_3=44^\circ$ & $31.6$ & $97.1$ & $0.00868$ & $0.00594$ \\
\midrule
C3T\tnote{a} & $6$ & $1$ & -- & $25.2$ & $97.6$ & $N/A$\tnote{e} & $N/A$\tnote{e} \\
MLP\tnote{b} & $6$ & $1$ & $(45^\circ,45^\circ)$ & $28.9$ & $98.6$ & $0.00947$ & $0.00609$ \\
MLP\tnote{b} & $6$ & $2$ & $(50^\circ,65^\circ)$ & $34.4$ & $98.4$ & $0.00414$ & $0.00339$ \\
MLP\tnote{b} & $6$ & $3$ & $(50^\circ,50^\circ)$ & $37.8$ & $98.4$ & $0.00997$ & $0.00687$ \\
NORM & $6$ & $1$ & $(50^\circ,60^\circ)$ & $28.5$ & $98.5$ & $0.00463$ & $0.00317$ \\
NORM & $6$ & $2$ & $(50^\circ,60^\circ)$ & $34.0$ & $98.4$ & $0.00574$ & $0.00342$ \\
NORM & $6$ & $3$ & $(50^\circ,55^\circ)$ & $37.5$ & $98.4$ & $0.00510$ & $0.00345$ \\
\bottomrule
\end{tabular}
\begin{tablenotes}[flushleft]
    \item[a] C3T \SDR{} from~\cite[Fig. 6]{AndersC3T} and accuracy from~\cite[Fig. 9]{AndersC3T}.
    \item[b] The MLP rows are learned-decoder reference points trained for
comparison. They should not be interpreted as a controlled same-geometry
architectural comparison with the NORM-inspired decoder.
    \item[c] Bootstrap standard deviation of the median inference accuracy.
    \item[d] Bootstrap median absolute deviation of the median inference accuracy.
    \item[e] Not measured for the data in~\cite[Fig. 9]{AndersC3T}.
    \item[f] For $m=1$, the entry gives the selected single curve. For $m=2$
    and $m=3$, the entry gives the newly added curve in the sequential design,
    with previously selected curves fixed. The complete NORM candidate-branch
    sets are listed in Tables~\ref{tab:curves_n4} and~\ref{tab:curves_n6},
    and the corresponding interval allocations and branch-classification
    reliabilities are listed in Table~\ref{tab:alloc_reliability}.
    \item[g] Inference accuracy is the relative source-coordinate reconstruction
accuracy computed after branch selection and decoding.
\end{tablenotes}
\end{threeparttable}
\end{table*}
\section{Numerical Experiments}
\label{sec:numexp}

We evaluate the proposed C3M designs under the fixed isotropic AWGN model,
$\mathbf{N}\sim\mathcal{N}(\mathbf{0},\sigma_n^2\mathbf{I}_n)$, using the
power normalization in Eq.~\eqref{eq:power} and the \SNR{} convention in
Eq.~\eqref{eq:snr}. When useful for interpretation, the corresponding
per-dimension SNR is $\SNR_{\mathrm{dim}}
=
\frac{P/n}{\sigma_n^2}
=
\frac{\SNR}{n}$. Unless otherwise stated, the reported \SDR{} values are evaluated under this
nominal AWGN setting. In Fig.~\ref{fig:sdr_vs_snr}, the one-curve panels use
the end-to-end Gaussian \SDR{} reference in
Eq.~\eqref{eq:sdr_g_end_to_end}. The multi-curve panels use the branch-known
Gaussian \SDR{} reference in Eq.~\eqref{eq:sdr_g_branch_known}, which includes
the equal-splitting $m^2$ factor and is used as a diagnostic reference. The
hyperparameters of the NORM-inspired decoder are shown in Table~\ref{tab:hyper}.

Table~\ref{tab:accu} and Fig.~\ref{fig:sdr_vs_snr} summarize the main AWGN
performance results. The figure compares \SDR{}, decoder-inference accuracy,
and bootstrap variability across $n\in\{4,6\}$ and $m\in\{1,2,3\}$, while the
table reports the selected or newly added curve parameters, \SDR{} values,
decoder-inference accuracies, and bootstrap statistics. In
Fig.~\ref{fig:sdr_vs_snr}, the one-curve panels use the end-to-end Gaussian
\SDR{} reference, whereas the multi-curve panels use the branch-known Gaussian
\SDR{} reference.
The results show an increase in \SDR{} as candidate branches are added. At
\dBval{+10} \SNR{}, the selected $m=3$ NORM-inspired systems reach
\dBval{31.6} for $n=4$ and \dBval{37.5} for $n=6$. The $n=6$, $m=3$ case is a
three-candidate branch set with one candidate effectively pruned by the
allocation rule. These multi-curve gains should be interpreted as
finite-dimensional C3M performance gains, with the branch-known reference used
only as a diagnostic comparison. They are not end-to-end rate--distortion gains
for the original scalar source. The MLP baselines trained for comparison follow
the same increasing-\SDR{} trend and provide learned-decoder reference points.
Since the MLP rows are slightly higher than the NORM-inspired rows in some
cases, the main performance claim is the multi-branch C3M construction rather
than architectural superiority of the NORM-inspired decoder.
\begin{table*}[b!]
\caption{Allocation-rule ablation at \dBval{+10} \SNR{} for the selected C3M
candidate branch sets. All rules are evaluated on the same selected branch
geometries. Entries report the mean \SDR{} over $100$ Monte Carlo trials, with
one standard deviation.}
\label{tab:allocation_ablation}
\centering
\begin{tabular}{lcccc}
\toprule
Configuration & Equal split & Length-only & Tube-density-only
& Density--reliability $\delta_k q_k$ \\
\midrule
$n=4,m=2$ & $28.03\pm0.07$ & $28.08\pm0.07$ & $28.01\pm0.06$ & $28.02\pm0.07$ \\
$n=4,m=3$ & $31.72\pm0.05$ & $31.75\pm0.07$ & $31.70\pm0.06$ & $31.60\pm0.06$ \\
$n=6,m=2$ & $34.06\pm0.02$ & $34.09\pm0.02$ & $34.02\pm0.03$ & $34.02\pm0.03$ \\
$n=6,m=3$ & $37.68\pm0.03$ & $37.71\pm0.03$ & $37.65\pm0.03$ & $37.52\pm0.03$ \\
\bottomrule
\end{tabular}
\end{table*}

Table~\ref{tab:allocation_ablation} compares equal splitting, length-only
allocation, tube-density-only allocation, and density--reliability allocation
on the same selected C3M branch geometries. This ablation separates the effect
of interval allocation from the effect of branch-geometry selection. For the
two-branch cases, the allocation rule has only a small effect on \SDR{}.
For the three-candidate cases, length-only allocation gives the largest mean
\SDR{} among the tested rules. The density--reliability rule instead favors
branches with higher classifier reliability; in the $n=6$, $m=3$ case, it prunes
the second candidate branch because its measured reliability is zero. Thus, the
density--reliability allocation should be interpreted as a reliability-aware
allocation diagnostic rather than as an \SDR{}-optimal rule for these fixed
trained receivers. Across all tested allocations, the main multi-branch
performance trend remains unchanged.

For the selected NORM-inspired C3M systems, the total system curve length and density increase from $19.23$ and $0.7062$ for the four-dimensional $m=2$ design to $28.95$ and $1.0656$ for the four-dimensional $m=3$ design. In six dimensions, the corresponding values increase from $27.58$ and $0.1862$ for $m=2$ to $41.88$ and $0.2733$ for $m=3$. These trends show that adding branch curves increases the usable geometric resolution, while larger ambient dimension provides additional degrees of freedom for branch placement and along-curve reconstruction.

MAP and MMSE estimators are characterized analytically in Section~\ref{ss:posterior}; numerical MAP/MMSE comparisons for the single-curve C3T case were studied in~\cite{AndersC3T}. The present experiments therefore focus on scalable learned decoders for the multi-curve setting.

\subsection{Computational complexity}
\label{ss:complexity}

The proposed C3M design has two distinct computational regimes: (a) offline
code design and training, and (b) online inference. The offline stage includes
candidate branch selection, classifier training, branch-conditioned decoder
training, and Monte Carlo evaluation. The online stage consists only of applying
the trained classifier and the selected branch decoder to each received vector.

Let $G$ denote the number of candidate branches considered at each sequential
selection step. In the four-dimensional one-parameter sweep, $G=|\Phi|$, while
in the six-dimensional two-parameter sweep, $G=|\Theta||\Phi|$. For a final
$m$-branch design initialized by a fixed C3T branch, the sequential search
requires approximately $(m-1)G$ candidate evaluations. By contrast, an
unordered exhaustive search over $G$ possible branches would require
$\binom{G}{m}$ candidate sets. Thus, the sequential design replaces a
combinatorial global search with a linear sequence of marginal branch-selection
problems.

For each candidate, let $N_{\mathrm{tr}}$ denote the number of training
samples, $E_{\mathcal C}$ and $E_h$ the numbers of classifier and decoder
training epochs, and let $P_{\mathcal C}(m)$ and $P_h$ denote the numbers of
trainable parameters in the branch classifier and in one branch-conditioned
decoder, respectively. Up to architecture-dependent constants, the offline
training cost per candidate scales as $\mathcal O\!\left(
E_{\mathcal C}N_{\mathrm{tr}}P_{\mathcal C}(m)
+
mE_hN_{\mathrm{tr}}P_h
\right)$, with an additional Monte Carlo evaluation cost proportional to
$N_{\mathrm{MC}}\big(P_{\mathcal C}(m)+P_h\big)$ for $N_{\mathrm{MC}}$ test
samples when one selected decoder is evaluated per received vector. This cost
is incurred offline and is therefore not part of the online inference cost.
At inference time, the trained receiver first estimates the branch index and
then applies the corresponding branch-conditioned decoder. Hence the per-symbol
online complexity is $\mathcal O\!\left(P_{\mathcal C}(m)+P_h\right)$, or simply $\mathcal O(P_h)$ in the single-branch case. The dependence on the
number of active branches enters primarily through the classifier output
dimension and through the number of decoders stored in the decoder bank.
Importantly, the learned receiver does not perform an online search over the
continuous curve parameter.

This distinguishes the learned C3M receiver from a discretized MAP or MMSE
implementation. If each branch were discretized using $N_z$ points, then a
nearest-curve MAP approximation would require $\mathcal O(mN_z n)$ distance
evaluations per received vector, and an MMSE approximation would require a
comparable summation over the branch-coordinate grid. The NORM-inspired
branch-conditioned spectral receiver replaces this online grid search with
fixed forward passes through the classifier and the selected branch decoder.
Thus, the online cost is determined by the classifier and selected-decoder
network sizes rather than by the branch-coordinate grid resolution.

\begin{figure}[t!]
  \centering
  \subfloat[Gaussian-mixture noise.\label{fig:robust_gmm}]{
  \includegraphics[width=.48\columnwidth]{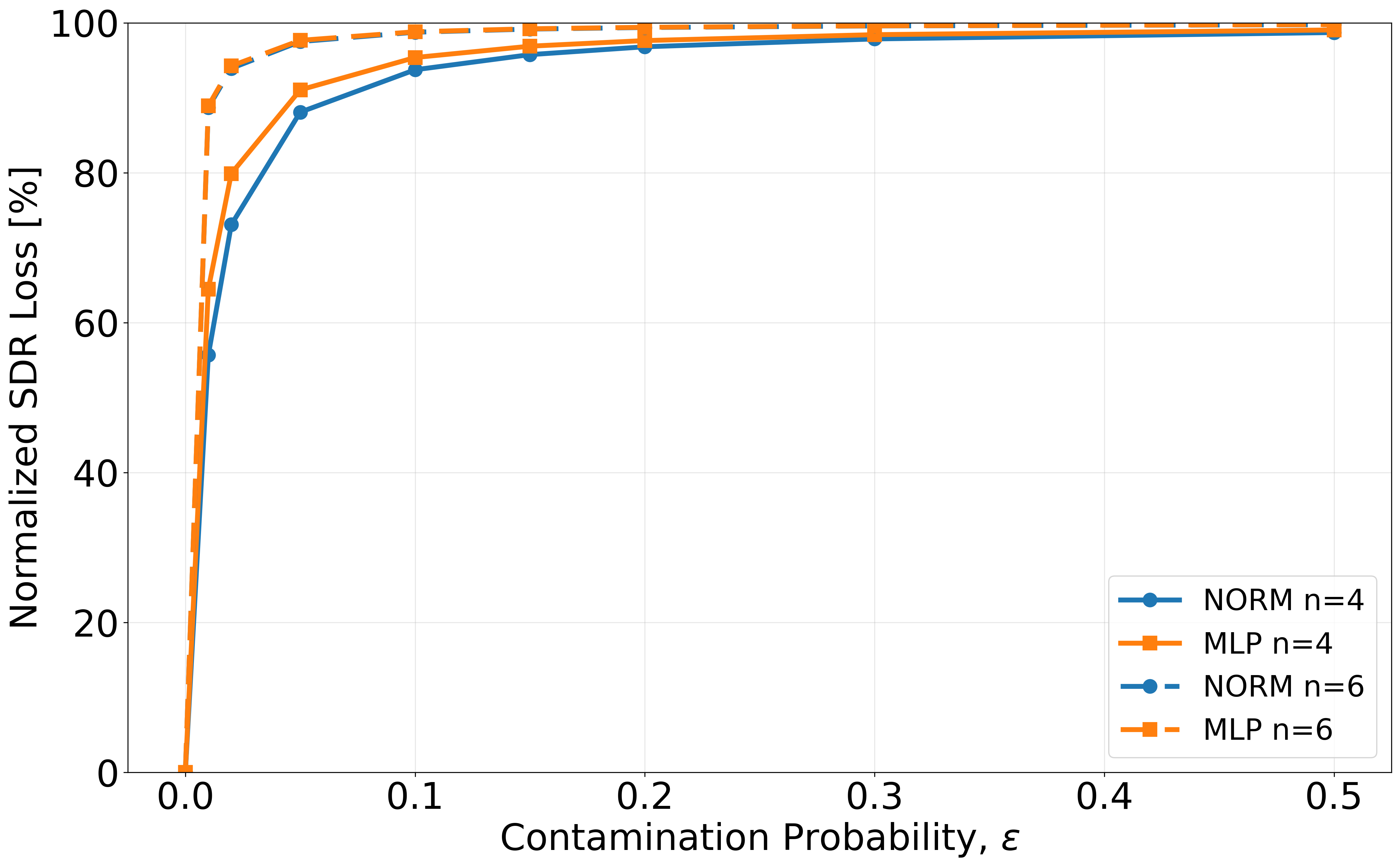}}\hfil
  \subfloat[$S\alpha S$ noise.\label{fig:robust_sas}]{
  \includegraphics[width=.48\columnwidth]{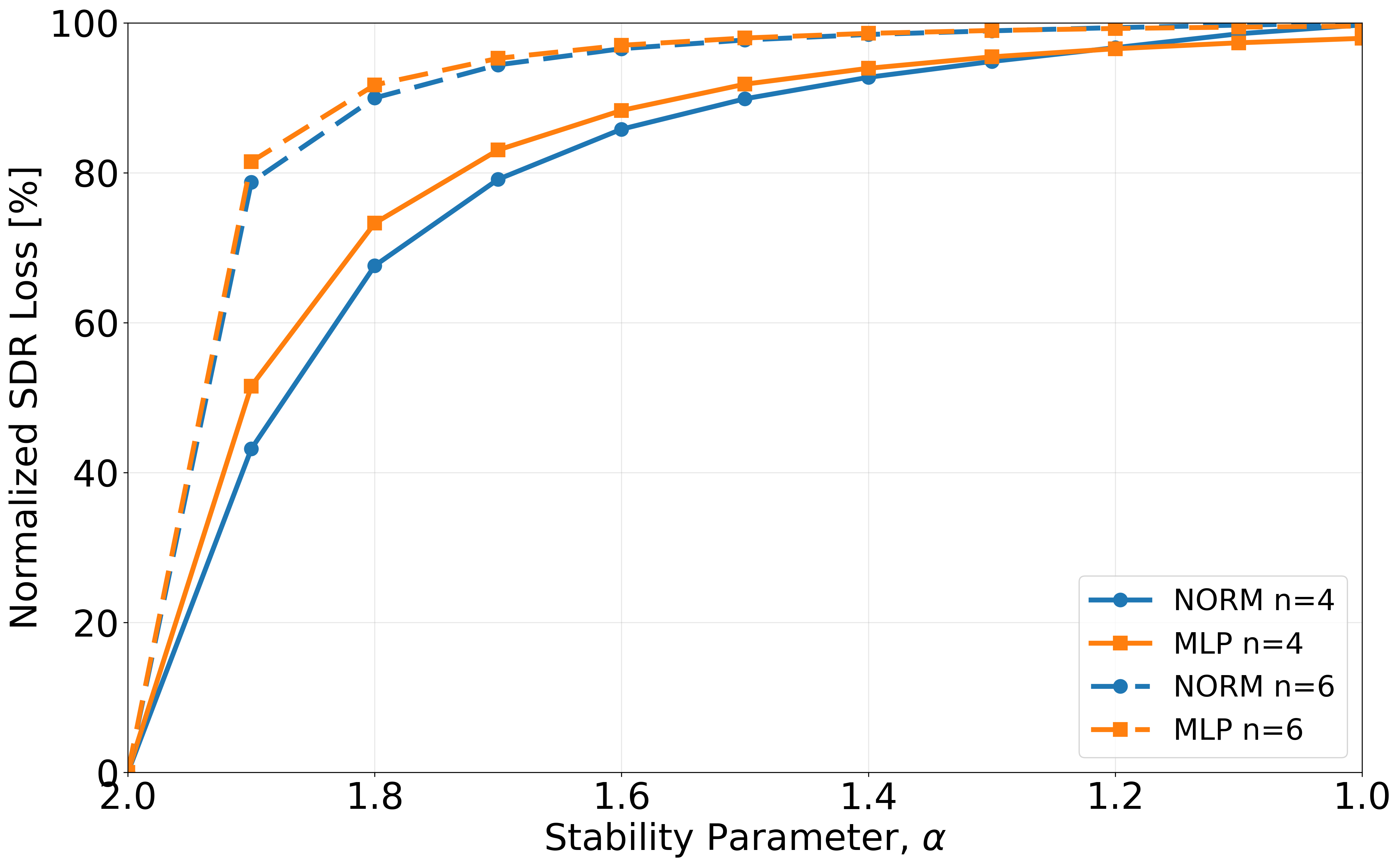}}
  \caption{Robustness of the single-curve decoders under non-Gaussian noise.
Panel (a) sweeps the Gaussian-mixture contamination probability $\epsilon$ in
Eq.~\eqref{eq:gmm_noise}, and panel (b) sweeps the symmetric alpha-stable
stability parameter $\alpha$ in Eq.~\eqref{eq:sas_noise}. Both panels report
the AWGN-relative normalized \SDR{} loss $\Delta_{\mathrm{rel}}\SDR{}$ in
percent, where smaller values indicate greater robustness. The curves compare
the $n=4$ and $n=6$ single-curve systems for the NORM-inspired decoder with circular
Huber loss and the MLP baseline. Curves are estimated from $100$ Monte Carlo
runs.}
  \label{fig:robust_noise}
\end{figure}

\subsection{Single-curve decoder robustness under impulsive non-Gaussian noise}
\label{ss:robust_noise}

To isolate decoder robustness from branch-classification effects, we evaluate
the single-curve setting under the non-Gaussian noise models in
Eqs.~\eqref{eq:gmm_noise} and~\eqref{eq:sas_noise}. For the Gaussian-mixture
experiments, we set $\mathbf{\Sigma}_w=\sigma_w^2\mathbf I_n$ and
$\mathbf{\Sigma}_v=\sigma_v^2\mathbf I_n$, and fix the impulsive-to-nominal
variance ratio to $\sigma_v^2/\sigma_w^2=100$. At the nominal \dBval{+10}
\SNR{} operating point, the corresponding linear SNR is
$\SNR_{\mathrm{lin}}=10$. With $P=1$, this gives
$\sigma_w^2=P/\SNR_{\mathrm{lin}}=0.1$. Hence,
$\sigma_v^2=10$, or $\sigma_v=\sqrt{10}\approx3.1623$, so impulsive samples
have ten times larger standard deviation than the nominal AWGN component. We
then sweep the contamination probability over the range $\epsilon\in[0,0.5]$.
For the symmetric alpha-stable model, we decrease the stability parameter from
$\alpha=2$ to $\alpha=1$, where $\alpha=2$ corresponds to the Gaussian endpoint
and $\alpha=1$ corresponds to the Cauchy endpoint of the sweep. These single-curve experiments assess how the circular Huber loss used by the NORM-inspired spectral decoder affects stability as the noise departs from the
nominal AWGN model. They do not evaluate multi-branch operation or branch-classification robustness under non-Gaussian noise.

We report robustness using the normalized \SDR{} loss relative to the nominal
AWGN case,
\begin{equation}
\label{eq:delta_sdr_rel}
\Delta_{\mathrm{rel}}\SDR{}
=
\frac{
\SDR_{\mathrm{AWGN}}-\SDR_{\mathrm{nonGaussian}}
}{
\SDR_{\mathrm{AWGN}}
},
\end{equation}
where the \SDR{} values in Eq.~\eqref{eq:delta_sdr_rel} are evaluated on a
linear scale. Thus, smaller values of $\Delta_{\mathrm{rel}}\SDR{}$ indicate
greater robustness, with $\Delta_{\mathrm{rel}}\SDR{}=0$ corresponding to no
loss relative to the nominal AWGN setting. In Fig.~\ref{fig:robust_noise}, this
quantity is plotted in percent.
The curves in Fig.~\ref{fig:robust_noise} are estimated from $100$ Monte Carlo
runs. For Gaussian-mixture noise at $\epsilon=0.10$, the NORM-inspired spectral
decoder has a smaller normalized \SDR{} loss than the MLP baseline by $1.64$
percentage points for $n=4$ and $0.06$ percentage points for $n=6$. For
$S\alpha S$ noise at $\alpha=1.8$, the corresponding reductions are $5.45$
percentage points for $n=4$ and $1.47$ percentage points for $n=6$. In these
single-curve tests, the Huber-trained decoder is therefore competitive with the
MLP baseline and gives lower normalized \SDR{} loss in several impulsive and
heavy-tailed regimes. In the single-curve setting, the Gaussian-mixture results are consistent with
the bounded-influence behavior of the Huber objective. The NORM-inspired
decoder generally exhibits smaller normalized \SDR{} loss than the MLP baseline
under low-to-moderate contamination.

\section{Conclusion and Future Work}
\label{sec:conclusion}

This paper introduced constant-curvature curve multicode (C3M), a multi-branch
extension of C3T for finite-dimensional analog joint source--channel coding
over fixed isotropic AWGN channels. The construction partitions a scalar source
interval across multiple constant-speed branch curves in even-dimensional
Euclidean space, thereby increasing the available geometric design degrees of
freedom while preserving a one-shot analog encoding structure. We derived the
induced mixture output distribution, posterior representation, MAP and MMSE
estimators, mutual-information expressions, and end-to-end and branch-known
\SDR{} reference curves. The analysis separates along-curve reconstruction from
branch-identification effects, which is essential for interpreting multi-curve
performance.

The numerical results show that adding candidate branches increases the
finite-dimensional \SDR{} of the learned C3M receivers under the nominal AWGN
model. At \dBval{+10} \SNR{}, the selected $m=3$ NORM-inspired systems reach
\dBval{31.6} for $n=4$ and \dBval{37.5} for $n=6$. The $n=6$, $m=3$ case is a
three-candidate branch set with one candidate effectively pruned by the
allocation rule. These gains should be interpreted as finite-dimensional C3M
performance gains, not as rate--distortion gains for the original scalar
source. The branch-known Gaussian reference is used only as a diagnostic
comparison for the multi-curve panels.

Several extensions follow directly from the present formulation. The first is
automated branch-set design. In the present work, branch candidates are selected
by finite sweeps over radii parameters. A natural next step is to treat the
radii vectors and branch orientations as variables in a constrained geometric
optimization problem. For constant-speed even-dimensional constructions, the
geometry-only objectives admit simple derivatives with respect to the radii
parameters, which suggests projected-gradient, stochastic-approximation, and
simultaneous-perturbation methods for automated C3M branch design
\cite{RobbinsMonro1951,KieferWolfowitz1952,Spall1992}. This direction would
also allow branch geometry and interval allocation to be optimized jointly,
rather than sequentially. The fixed-geometry ablation in
Table~\ref{tab:allocation_ablation} provides a first comparison of allocation
rules; a natural extension is to include the allocation variables directly in
the branch-selection optimization.

A second direction is to develop receivers that more directly exploit the
intrinsic geometry of the C3M signal set. One approach is to lift each noisy
channel observation to a likelihood or log-likelihood function over the
circular branch coordinate and then process this intrinsic signal using
$\mathbb S^1$-equivariant or manifold-aware neural architectures
\cite{Cohen2016,Masci2015,Monti2017,Cohen2019}. Such decoders could exploit the
harmonic structure of the C3T/C3M encoder, refine multimodal posteriors caused
by curve folds, and produce calibrated directional uncertainty through a von
Mises or related circular-statistics readout \cite{Mardia2000}. Kolmogorov--Arnold networks provide another possible class of decoders and branch classifiers and should be compared with MLP and Fourier-feature
baselines on the same selected branch geometries~\cite{Liu2025}.

A third direction is to separate branch identification from along-curve
estimation more explicitly. A hybrid receiver could first preselect a small set
of candidate branches using coarse geometry-aware scores and then apply refined
branch-conditioned projection, MAP estimation, MMSE estimation, or learned
local-coordinate decoding only to the selected candidates. This coarse-to-fine
structure may reduce online complexity when the number of branches is large.
Low-complexity branch classifiers are also natural in this setting. In four
dimensions, the features $u_1(\mathbf y)=\sqrt{y_1^2+y_2^2}$ and $u_2(\mathbf y)=\sqrt{y_3^2+y_4^2}$
are invariant to phase within each harmonic pair. In the noiseless limit, these
features coincide with the corresponding branch radii and therefore provide a
low-dimensional representation for coarse branch classification in
radius-vector space. Such magnitude-based classifiers may be useful as front
ends for hybrid receivers, followed by refined branch-conditioned projection,
MAP estimation, MMSE estimation, or learned local-coordinate decoding.
Gray-style assignment of neighboring source intervals to geometrically nearby
branches is a related design option, since it can reduce the source-domain
distortion caused by occasional branch-index errors~\cite{Gray1953}. For
non-uniform branch probabilities, receiver training should also incorporate
source priors through class-weighted losses, probability-aware sampling, or
prior-weighted decision rules so that high-probability branches are not
underprotected~\cite{Cui2019}.

Finally, algebraic constructions may provide systematic tools for generating
structured branch families. One exploratory direction is to use number-field
embeddings to generate torus curves whose winding directions and relative
separations are controlled algebraically. CM subfields of
$K=\mathbb Q(i,\sqrt{-2})$ may provide candidate curve families with distinct
torus directions, while related structures such as Minkowski embeddings,
Eisenstein-integer lattices \cite{AndersMBM}, and Hurwitz-quaternion
constructions may provide additional tools for multi-curve design
\cite{Souza2024,Natarajan2015}. These algebraic constructions could complement
numerical branch optimization by providing structured candidate families with
analyzable separation and symmetry properties. A longer-term direction is to
use C3T/C3M mappings as one-shot analog source--channel interfaces in
distributed sensing, edge inference, or federated-learning systems, connecting
the finite-dimensional geometric approach studied here with learned JSCC
systems \cite{Bourtsoulatze2019}.
\appendices

\section{Proof of Optimal Bayesian Estimator}
\label{app:MMSE}

\begin{proposition}
\label{prop:mmse_optimality}
The MMSE estimator $\widehat{S}_{\mathrm{MMSE}}(\mathbf{Y})=\mathbb{E}[S\mid \mathbf{Y}]$ minimizes $\mathbb{E}[(S-g(\mathbf{Y}))^2]$ over all measurable estimators $g(\mathbf{Y})$.
\end{proposition}
\begin{proof}
For any estimator $g(\mathbf{Y})$,
\begin{equation}
\begin{aligned}
\mathbb{E}\!\left[(S-g(\mathbf{Y}))^2\right] &= 
\mathbb{E}\!\left[\bigl(S-\mathbb{E}[S\mid \mathbf{Y}]
+\mathbb{E}[S\mid \mathbf{Y}]-g(\mathbf{Y})\bigr)^2\right] \\
&=
\mathbb{E}\!\left[\bigl(S-\mathbb{E}[S\mid \mathbf{Y}]\bigr)^2\right]
+
\mathbb{E}\!\left[\bigl(\mathbb{E}[S\mid \mathbf{Y}]-g(\mathbf{Y})\bigr)^2\right] \nonumber\\
&\quad+
2\,\mathbb{E}\!\left[
\bigl(S-\mathbb{E}[S\mid \mathbf{Y}]\bigr)
\bigl(\mathbb{E}[S\mid \mathbf{Y}]-g(\mathbf{Y})\bigr)
\right].
\end{aligned}
\end{equation}
The cross-term vanishes by the tower property: $\mathbb{E}\!\left[
\bigl(S-\mathbb{E}[S\mid \mathbf{Y}]\bigr)
\bigl(\mathbb{E}[S\mid \mathbf{Y}]-g(\mathbf{Y})\bigr)
\right]=0$. Hence $\mathbb{E}\!\left[(S-g(\mathbf{Y}))^2\right]
=
\mathbb{E}\!\left[\bigl(S-\mathbb{E}[S\mid \mathbf{Y}]\bigr)^2\right]
+
\mathbb{E}\!\left[\bigl(\mathbb{E}[S\mid \mathbf{Y}]-g(\mathbf{Y})\bigr)^2\right]$, 
and the second term is nonnegative, with equality if and only if $g(\mathbf{Y})=\mathbb{E}[S\mid \mathbf{Y}]$ almost surely.
\end{proof}

\bibliographystyle{IEEEtran}
\bibliography{IEEEabrv,c3m}

\end{document}